\documentclass[aps,pra,superscriptaddress,twocolumn,nofootonbib,floatfix,longbibliography]{revtex4-2} 

\usepackage[final]{graphicx}    
\graphicspath{{figures_paper/}}
\usepackage{soul}
\usepackage[utf8]{inputenc}
\usepackage{natbib}
\usepackage{xcolor}
\usepackage{amsfonts}
\usepackage{amsmath}
\usepackage{hyperref}
\hypersetup{colorlinks=true}

\definecolor{asparagus}{rgb}{0.53, 0.66, 0.42}

\newcommand{\Ket}[1]{\left|#1\right>}
\newcommand{\Bra}[1]{\left<#1\right|}
\newcommand{\BraKet}[2]{\left<#1|#2\right>}
\newcommand{\KetBra}[2]{|#1\rangle\langle#2|}

\newcommand{\mbf}[1]{\mathbf{#1}}

\newtheorem{theorem}{Theorem}

\begin{document}

\title{Phase space geometry and optimal state preparation in quantum metrology with collective spins}

\author{Manuel H. Muñoz-Arias}
\email{munm2002@usherbrooke.ca}
\affiliation{Center for Quantum Information and Control, Department of Physics and Astronomy, University of New Mexico, Albuquerque, New Mexico 87131, USA}
\affiliation{Institut Quantique and Département de Physique, Université de Sherbrooke, Sherbrooke, Quebec, J1K 2R1, Canada}
\author{Ivan H. Deutsch}
\affiliation{Center for Quantum Information and Control, Department of Physics and Astronomy, University of New Mexico, Albuquerque, New Mexico 87131, USA}
\author{Pablo M. Poggi}
\email{ppoggi@unm.edu}
\affiliation{Center for Quantum Information and Control, Department of Physics and Astronomy, University of New Mexico, Albuquerque, New Mexico 87131, USA}
\begin{abstract}
We revisit well-known protocols in quantum metrology using collective spins and propose a unifying picture for optimal state preparation based on a semiclassical description in phase space. We show how this framework allows for quantitative predictions of the timescales required to prepare various metrologically useful states, and that these predictions remain accurate even for moderate system sizes, surprisingly far from the classical limit. Furthermore, this framework allows us to build a geometric picture that relates optimal (exponentially fast) entangled probe preparation to the existence of separatrices connecting saddle points in phase space. We illustrate our results with the paradigmatic examples of the two-axis counter-twisting and twisting-and-turning Hamiltonians, where we provide analytical expressions for all the relevant optimal time scales. Finally, we propose a generalization of these models to include $p$-body collective interaction (or $p$-order twisting), beyond the usual case of $p=2$. Using our geometric framework, we prove a no-go theorem for the local optimality of these models for $p>2$.

\end{abstract}

\date{\today}
\maketitle


\section{Introduction}
\label{sec:intro}



Quantum metrology employs nonclassical states as a resource for estimating unknown parameters with a sensitivity beyond that allowed by the standard quantum limit~\cite{Giovannetti2006,Giovannetti2011,Pezze2018,Degen2017}. Of particular interest are ensembles of qubits (real or pseudo-spins) for use in clocks and magnetometers.  Typically one considers collective spins of total angular momentum components $\hat{J}_\alpha$. Examples of metrologically useful quantum states of collective spins include spin squeezed states of atoms~\cite{KitagawaUeda1993,Sorensen2001,Ma2011,Gross2012,Wineland1992}, cat-states or GHZ states~\cite{Toth2014,Huang2015}, and Dicke states~\cite{Dicke1954}, among others. The aforementioned examples share the common property of being entangled, and it is well-known that entanglement is essential, but not sufficient, to enable sensitivity going beyond what can be achieved with classical resources (i.e., with product states)~\cite{Giovannetti2006,Pezze2009}.

The problem of local quantum metrology~\footnote{Here we emphasise that the types of quantum metrology protocol which exploit the quantum states we are about to mention is often referred as \emph{local quantum metrology}. This with the aim of making clear the fact that we would not be thinking about Bayesian quantum metrology.} can be regarded as a problem of optimal state preparation whereby one seeks to prepare a metrologically useful state in the shortest amount of time possible. For example, in their pioneering work Kitagawa and Ueda studied the preparation of spin squeezed states introducing one axis twisting (OAT) and the two-axis counter-twisting (2ACT) Hamiltonians~\cite{KitagawaUeda1993}. Here, the term ``twisting" refers to generators which are quadratic in the collective spin operators $(\hat{J}_x,\hat{J}_y,\hat{J}_z)$. While both Hamiltonians successfully generate squeezing, the OAT ($\hat{H}\sim \hat{J}_x^2$) does not saturate the fundamental bound of sensitivity dictated by quantum metrology and the maximum squeezing is reached at a time which scales as $J^{\frac{1}{\eta}}$, where $J$ is the size of the collective spin and $\eta > 1$.  On the other hand, the 2ACT ($\hat{H}\sim \hat{J}_x^2-\hat{J}_y^2$) reaches peak spin squeezing in a time that is logarithmic in the size of the collective spin, and saturates the fundamental limit of sensitivity imposed by quantum mechanics. Later, Micheli \textit{et al.}~\cite{Micheli2003} studied a model in which the OAT generator is combined with a transverse field, leading to Twisting and Turning (TaT) dynamics. They showed that this new generator permits the preparation of highly entangled metrologically relevant states, \textit{i.e.}, cat-like states, at times which are logarithmic in the size of the collective spin, and that the best performance is obtained when the ratio of twisting and external field strengths assumes a critical value. Related work by Sorelli \textit{et al.}~\cite{Sorelli2019} investigated the twist and turn generator in the short time regime (complementary to \cite{Micheli2003}), and argued for the optimality of this dynamics in the generation of spin squeezing. Furthermore, previous works by Yukawa \textit{et al.}~\cite{Yukawa2014} and Kajtoch \textit{et al.}~\cite{Kajtoch2015} have explored in detail the dynamics generated by the 2ACT Hamiltonian, emphasizing the different types of metrologically relevant states which are generated and verifying, either numerically or semianalytically, the logarithmic dependence of their preparation times with the size of the collective spin.

In this work we revisit these protocols and present a unified description of optimal state preparation for quantum metrology in collective spin systems based on their semiclassical dynamics. In particular, we discuss how optimal (i.e., exponentially fast) state preparation is allowed by the existence of saddle points in the classical flow associated with a collective spin Hamiltonian. Using this picture, we develop a comprehensive analysis of the geometry of the separatrix and identify the key properties which guarantee the local and global optimality for state preparation in generic collective spin models. We revisit the problems of 2ACT and TaT and use this semiclassical approach to derive new expressions for the timescales required to achieve several types of useful states for metrology. This gives a solid theoretical foundation to the time scales that were previously found numerically~\cite{Yukawa2014,Kajtoch2015}. We then apply this framework to the previously unexplored problem of using higher-order twisting generated by Hamiltonians of the form $J_{\hat{n}}^p$, with $p \geq 2$, to generate spin squeezing and metrologically useful states. We show that, while $p$-body 2ACT is optimal only for $p = 2$, a notion of optimality can be redefined for $p$-body TaT in the cases of $p = 3$ and $p = 4$.


The remainder of the manuscript is organized as follows. In Sec.~\ref{sec:qmetro_cspins} we give an overview of the semiclassical description of collective spin systems and of basic aspects of quantum metrology. In Sec.~\ref{sec:all_counter_twistins} we study the two-axis counter-twisting Hamiltonian (2ACT), its phase space structure and dynamics, and derive analytical results for the preparation times of several entangled states useful for quantum metrology. From this analysis we distill the conditions of the separatrix geometry which guarantee the local and global optimality of a given collective spin Hamiltonian for state preparation. In Sec.~\ref{sec:all_twist_turn} we turn our attention to the twist and turn (TaT) Hamiltonian and provide three different physical interpretations for the ``critical coupling'' regime first identified in Ref. \cite{Micheli2003}. 
In Sec.~\ref{sec:all_p_spin} we introduce generalizations of the 2ACT and TaT Hamiltonians by adding higher-order twisting terms and we prove several results which point at the optimality of the case of second order-twisting.
Finally, in Sec.~\ref{sec:outlook} we present some concluding remarks and discuss potential avenues for future work.

\section{Quantum metrology with collective spin systems} 
\label{sec:qmetro_cspins}

\subsection{Collective spin systems and their classical limit}

We consider systems of $N$ spin-$\frac{1}{2}$ particles  described by the set of collective angular momentum operators $\hat{\mbf{J}}=(\hat{J}_x,\hat{J}_y,\hat{J}_z)$, where $\hat{J}_\alpha = \frac{1}{2}\sum_{i=1}^N \hat{\sigma}_\alpha^{(i)}$, and $\hat{\sigma}_\alpha^{(i)}$ is a Pauli operator acting on particle $i$, $\alpha=x,y,z$. Any Hamiltonian of the form
\begin{equation}
    \hat{H} = \sum\limits_\alpha a_\alpha \hat{J}_\alpha + \sum\limits_{\alpha \beta} b_{ \alpha \beta} \hat{J}_\alpha \hat{J}_\beta + \sum_{\alpha \beta \gamma} c_{\alpha \beta \gamma} \hat{J}_\alpha \hat{J}_\beta \hat{J}_\gamma \ldots + \mathrm{h.c.}
    \label{eq:collectiveH}
\end{equation}
describes a collective spin model, where particles interact uniformly among themselves and with external fields. By construction, the Hamiltonian $\hat{H}$ commutes with the total angular momentum operator $\hat{J}^2$, and so the total spin value $J$ is conserved. We will focus on the subspace of $J=N/2$, often called the symmetric subspace, which is composed of all pure states which are invariant under permutation of any two particles \cite{Stockton2003}. Two convenient choices of basis for this subspace are the Dicke states $\{\Ket{J,m}\}$, $m=-J,-J+1,\ldots,J$, which are the eigenvectors of $\hat{J}_z$, and the overcomplete basis of spin coherent states (SCS) $\Ket{\theta,\phi}=e^{-i\phi \hat{J}_z}e^{-i\theta \hat{J}_y}\Ket{J,J}=\Ket{\uparrow_{\theta,\phi}}^{\otimes N}$, where $\Ket{\uparrow_{\theta,\phi}}$ describes the state of a single qubit pointing up along the $(\theta,\phi)$ direction on the Bloch sphere, and $0 \leq \theta \leq \pi$, $0\leq \phi \leq 2\pi$. 

\begin{figure}[t!]
\centering{\includegraphics[width=1\linewidth]{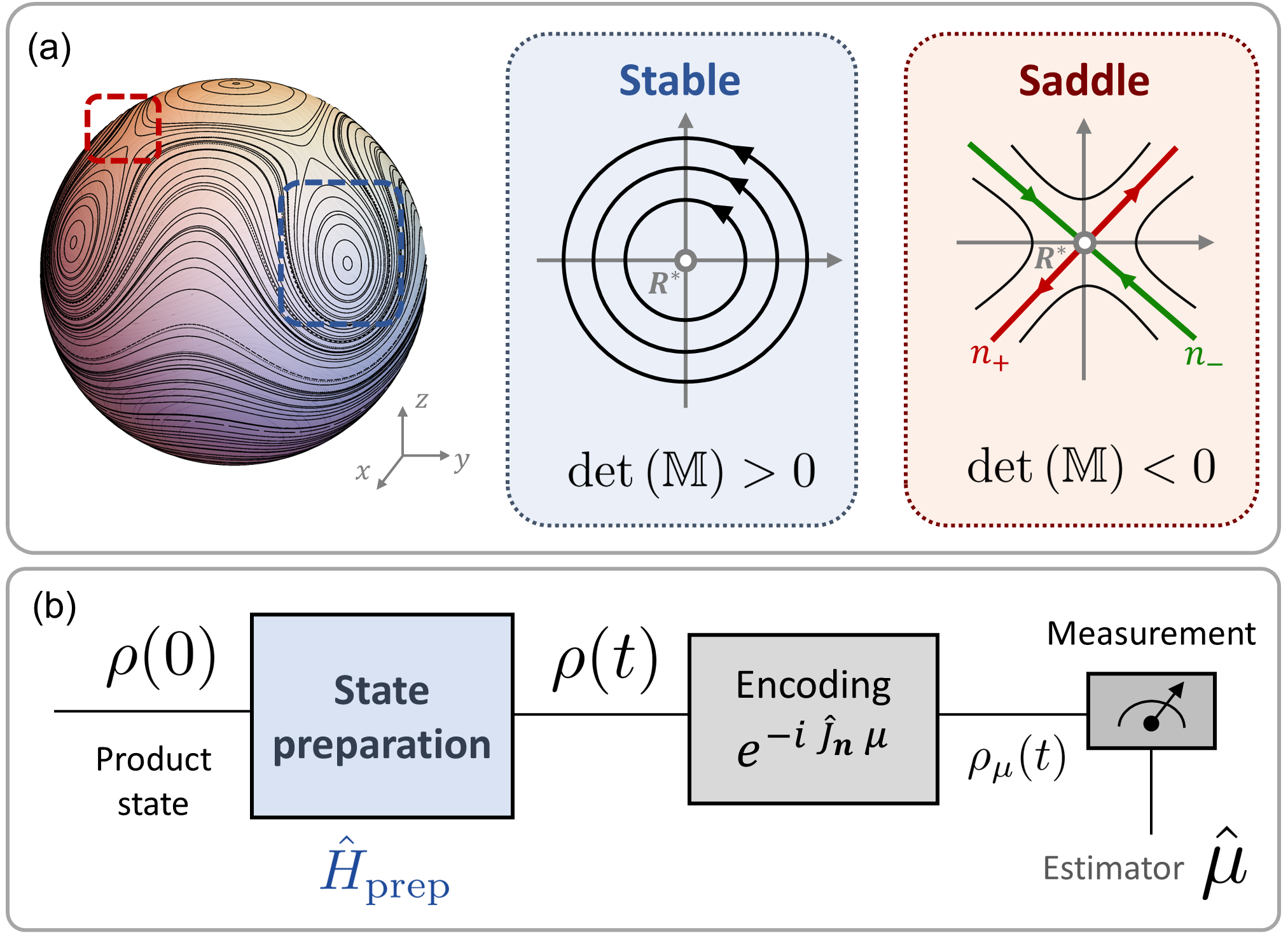}}
\caption{(a) Phase space portrait of the classical motion of a collective spin system. Prominent phase space structures include stable orbits and saddle points. (b) Interferometric approach to quantum metrology with collective spins. A product state $\rho(0)$ undergoes a state preparation procedure to yield the probe state $\rho(t)$, which is then used to sense an unknown parameter $\mu$. }
\label{fig:schematic}
\end{figure}

Any Hamiltonian of the form in Eq.~(\ref{eq:collectiveH}) admits a well-defined classical description in the thermodynamic limit $N\rightarrow \infty$. This description, in turn, coincides with the mean-field limit of the model and can be obtained from the equations of motion for the expected values of spin components, $\frac{d\langle\hat{J}_\alpha \rangle}{dt} = -i\langle[ \hat{J}_\alpha,H]\rangle, \ \mathrm{with}\ \alpha = x,y,z$ $(\hbar=1)$. In the thermodynamic limit this leads to equations of motion for a phase space flow for the classical variables $\mbf{R}\equiv (X,Y,Z)=\lim_{J \to \infty}\langle \hat{\mbf{J}}\rangle /J$ by neglecting correlations and setting $\langle A B\rangle = \langle A\rangle \langle B\rangle$. This flow has the form $\frac{d\mbf{R}}{dt} = \mbf{F}[\mbf{R}]$ and describes the nonlinear motion of a ``top'' on a unit sphere since $X^2 + Y^2 + Z^2 = 1$. The properties of the flow can be studied locally by identifying the fixed points, \textit{i.e.}, $\mathbf{R}^*$ such that $\mbf{F}[\mbf{R}^*]=0$, and analyzing their stability through calculation of the Jacobi matrix, $\mathbb{M}[\mbf{R}]=\frac{\partial \mbf{F}}{\partial \mbf{R}}$. If $\det \mathbb{M} >0$, the fixed point is stable or elliptic, and if $\det \mathbb{M} <0$ it is a saddle or an hyperbolic fixed point. 
In Fig. \ref{fig:schematic} (a) we illustrate the local motion of each case. In particular, the vicinity of a saddle point is described by two directions in phase space, one unstable $\mbf{n}_+$ and one stable $\mbf{n}_-$, along which the motion is of the form $\dot{R}_\pm(t) = \pm \lambda R_\pm(t)$ and so $R_\pm (t) \propto e^{\pm \lambda t} R_\pm(0)$. As a result, local motion along a unstable branch leads to an exponential stretching of trajectories in phase space. Globally, a given flow may possess many hyperbolic fixed points, and their stable and unstable branches are typically connected by an isolated trajectory in phase space which is called the \textit{separatrix}. In Sect. \ref{sec:all_counter_twistins} we will illustrate how the geometric properties of the separatrix in the classical phase space affect the motion of quantum states, and in particular we will study how to use this knowledge to  study the generation of metrologically-useful states.  

\subsection{Basic aspects of quantum metrology}
The goal of quantum metrology~\cite{Giovannetti2006} is to estimate the value of a weak signal encoded as an unknown physical parameter $\mu$ in a quantum system, and to do so with a precision beyond what is achievable with solely classical resources. There are many approaches to this problem, including quantum critical metrology~\cite{Salvatori2014,Frerot2018}, the use of chaotic sensors~\cite{Fiderer2018}, nonlinear metrology~\cite{Boixo2008a,Boixo2008b,Beau2017}, the use of interaction-based readouts~\cite{Macri2016,Davis2016,Anders2018,Nolan2017} or twist-and-untwist protocols~\cite{Volkoff2022}. 
In this work, we will focus on the interferometric approach~\cite{Pezze2018}, which is schematically depicted in Fig. \ref{fig:schematic} (b). In this scheme, a probe state is prepared through evolution under a  chosen Hamiltonian acting on an easy to prepare initial state, $\hat{\rho}(t)=e^{-i \hat{H}_{\mathrm{prep}} t}\hat{\rho}(0) e^{i \hat{H}_{\mathrm{prep}} t}$. The resulting probe state then undergoes another transformation which imprints the unknown parameter $\mu$ into the state, typically as in $\hat{\rho}_\mu(t)=e^{-i \mu \hat{G}} \hat{\rho}(t)e^{i \mu \hat{G}}$. Finally, measurements are performed on $\hat{\rho}_\mu(t)$ according to a well chosen POVM, from which an (unbiased) estimator $\hat{\mu}$ of the parameter $\mu$ is constructed. The variance of such estimator, denoted $\Delta \mu$, has a fundamental limit which is given by the  the Quantum Cramer-Rao bound,
\begin{equation}
 \label{eqn:quantum_cramer_rao_bound}
 \Delta \mu \ge \Delta\mu_{\rm QCR} = \frac{1}{\sqrt{\nu 
F_{\rm Q}[\hat{\rho}_\mu]}},
\end{equation}
where $\nu$ is the number of independent measurements and $F_{\rm Q}[\hat{\rho}_\mu]$ is the Quantum Fisher Information (QFI). For the present scheme, assuming that $\hat{\rho}(t)=\KetBra{\psi(t)}{\psi(t)}$ is pure and in absence of noise or decoherence, the QFI takes the simple form
\begin{equation}
    F_{\rm Q} [|\psi\rangle, \hat{G}] = 4 \langle \Delta \hat{G}(t)\rangle^2 = 4 \left(\langle G(t)^2\rangle - \langle G(t)\rangle^2\right) 
\end{equation}
where $\langle A(t)\rangle \equiv \Bra{\psi(t)}\hat{A}\Ket{\psi(t)}$. The maximum sensitivity predicted by the Quantum Cramer-Rao is achieved only if one can implement the optimal measurement (POVM), which is problem-dependent and often challenging in practice. Nevertheless, we will consider the QFI as one of the figures of merit to quantify metrological performance. \\

For the purposes of this work, we consider the unknown parameter $\mu$ to be a small angle of rotation of the collective spin around a known axis $\mbf{n}$ such that $\hat{G}=\hat{J}_{\mbf{n}}$. If the probe state $\Ket{\psi(t)}$ is a SCS, the maximum QFI gives $F_{\rm Q}[\hat{\rho}_{\mu}] \propto N$. This is often referred to as the standard quantum limit (SQL), and provides a reference point to gauge the usefulness of a proposed quantum metrological strategy. Any strategy making use of quantum resources which permits a sensitivity of the estimator going above the standard quantum limit is advantageous. When entangled states are allowed, the maximum possible QFI can scale as $F_{\rm Q}[\hat{\rho}_{\mu}] \propto N^2$, which is referred to as the Heisenberg limit.

In this work we will focus on how the semiclassical picture ascribed to $\hat{H}_{\rm prep}$ can be used to understand the preparation of metrologically useful states, and to make predictions about the times required to reach them. To quantify this, one figure of merit is the metrological gain based on the QFI, defined as 
\begin{equation}
\label{eqn:metrological_gain}
\zeta^2 = \frac{N}{F_Q[|\psi\rangle,\hat{J}_{\vec{n}}]},
\end{equation}
where any value of $\zeta^2<1$ will indicate improvement over the SQL.  In addition, we will also consider metrology protocols in which the POVM is not necessarily optimal, but restricted to a specific type. In particular, for a Ramsey-type interferometer as it is used in clocks and magnetometers, the metrological gain is quantified by the Wineland squeezing parameter~\cite{Wineland1992}
\begin{equation}
\label{eqn:metrological_squeezing}
\xi^2 \equiv \frac{\Delta \varphi^2}{\Delta \varphi^2_{\mathrm{SCS}}} = 2J \frac{\langle\Delta \hat{J}_\perp \rangle ^2}{|\langle \hat{\mbf{J}}\rangle|^2}. 
\end{equation}
Here, $\Delta \varphi$ is the uncertainty in the measurement of the desired phase in the interferometer and $\Delta \varphi_{\mathrm{SCS}}$ is the corresponding uncertainty when the input probe state is a SCS.  This can be reexpressed in terms of the collective spin variables, where $\langle\Delta \hat{J}_\perp\rangle$ is the projection noise of the collective spin perpendicular to the direction of rotation in the sensor, and $|\langle 
\mathbf{\hat{J}} \rangle|$ is the length of the collective spin.  For an input SCS, $\langle\Delta \hat{J}_\perp\rangle = \sqrt{J/2}$, $|\langle\mathbf{\hat{J}} \rangle|=J$, and $\xi^2=1$.  For a spin squeezed state of the sort originally considered by Kitagawa and Ueda, one can have $\xi^2 < 1$ and thus sensitivity in Ramsey interferometry beyond the SQL.
Furthermore, with sufficient squeezing one can attain scaling in improved sensitivity associated with the Heisenberg limit, $\xi^2 \propto 1/N$. In the next sections we will analyze these metrological figures of merit for different state preparation protocols by leveraging the semiclassical picture of collective spin systems. 



\section{Phase space geometry and quantum metrology with two-axis counter-twisting Hamiltonian}
\label{sec:all_counter_twistins}
Two-axis counter-twisting was introduced by Kitagawa and Ueda as the mechanism to obtain the optimal spin squeezed states, leading to Heisenberg scaling in the sensitivity~\cite{KitagawaUeda1993}. 
Our goal in this section is to understand this fact based on the geometrical properties of its classical phase space. The 2ACT Hamiltonian describes the dynamics of a collective spin $\mathbf{\hat{J}}$ under the action of two twisting operations, \textit{i.e.}, quadratic nonlinearities, along perpendicular axis. As originally written in~\cite{KitagawaUeda1993}, this Hamiltonian is given by 
\begin{equation}
\label{eqn:ct_hamil_1}
\hat{H}_{\rm CT} = \chi(\hat{J}^2_{\frac{\pi}{2},\frac{\pi}{4}} - \hat{J}^2_{\frac{\pi}{2},\frac{-\pi}{4}}) = \frac{\chi}{2i}(\hat{J}^2_+ - \hat{J}^2_-),
\end{equation}
where $\chi$ is the counter-twisting strength, and we have chosen the two axis as the $\pm 45 ^{\circ} $ directions on the $x-y$ plane, such that $\hat{J}_{\frac{\pi}{2},\frac{\pm\pi}{4}} = \frac{1}{\sqrt{2}}(\hat{J}_x \pm \hat{J}_y)$. Using the latter expression, Eq.~(\ref{eqn:ct_hamil_1}) can be written as 
\begin{equation}
\label{eqn:ct_hamil_2}
\hat{H}_{\rm CT} = \chi(\hat{J}_x\hat{J}_y + \hat{J}_y\hat{J}_x).
\end{equation}

To investigate the phase space geometry of the model in Eq.~(\ref{eqn:ct_hamil_2}) we proceed as described in Sec. \ref{sec:qmetro_cspins}.A to compute its associated classical flow (further details can be found in Appendix \ref{app:ct_stuff}). The equations of motion for the classical variables $\mbf{R}=(X,Y,Z)$ read
\begin{subequations}
\label{eqn:flow_ct}
\begin{align}
\frac{dX}{dt} &= -\tilde{\chi}XZ, \\
\frac{dY}{dt} &= -\tilde{\chi}YZ, \\
\frac{dZ}{dt} &= -\tilde{\chi}(X^2 - Y^2),
\end{align}
\end{subequations}

\noindent where $\tilde{\chi} = \chi/N$ and the fixed points are given by 
\begin{subequations}
\label{eqn:fixed_points_ct}
\begin{align}
(X,Y,Z) &= (0,0,\pm1), \\
(X,Y,Z) &= (\frac{1}{\sqrt{2}},\mp \frac{1}{\sqrt{2}},0), \\
(X,Y,Z) &= (-\frac{1}{\sqrt{2}}, \pm\frac{1}{\sqrt{2}},0),
\end{align}
\end{subequations}

\noindent which correspond to the north and south poles of the unit sphere, and the poles of each of the twisting axis. Their stability is deduced from the Jacobi matrix $\mathbb{M}[\mbf{R}]=\frac{\partial \mbf{F}}{\partial \mbf{R}}$, which in this case takes the form 

\begin{equation}
\label{eqn:tangent_map_ct}
\mathbb{M}[\mathbf{R}] = 
\begin{pmatrix}
\tilde{\chi}Z && 0 && \tilde{\chi}X \\ 
0 && -\tilde{\chi}Z && -\tilde{\chi}Y \\
-4\tilde{\chi}X && 4\tilde{\chi}Y && 0
\end{pmatrix}.
\end{equation}

\begin{figure*}[t!]
\centering{\includegraphics[width=\linewidth]{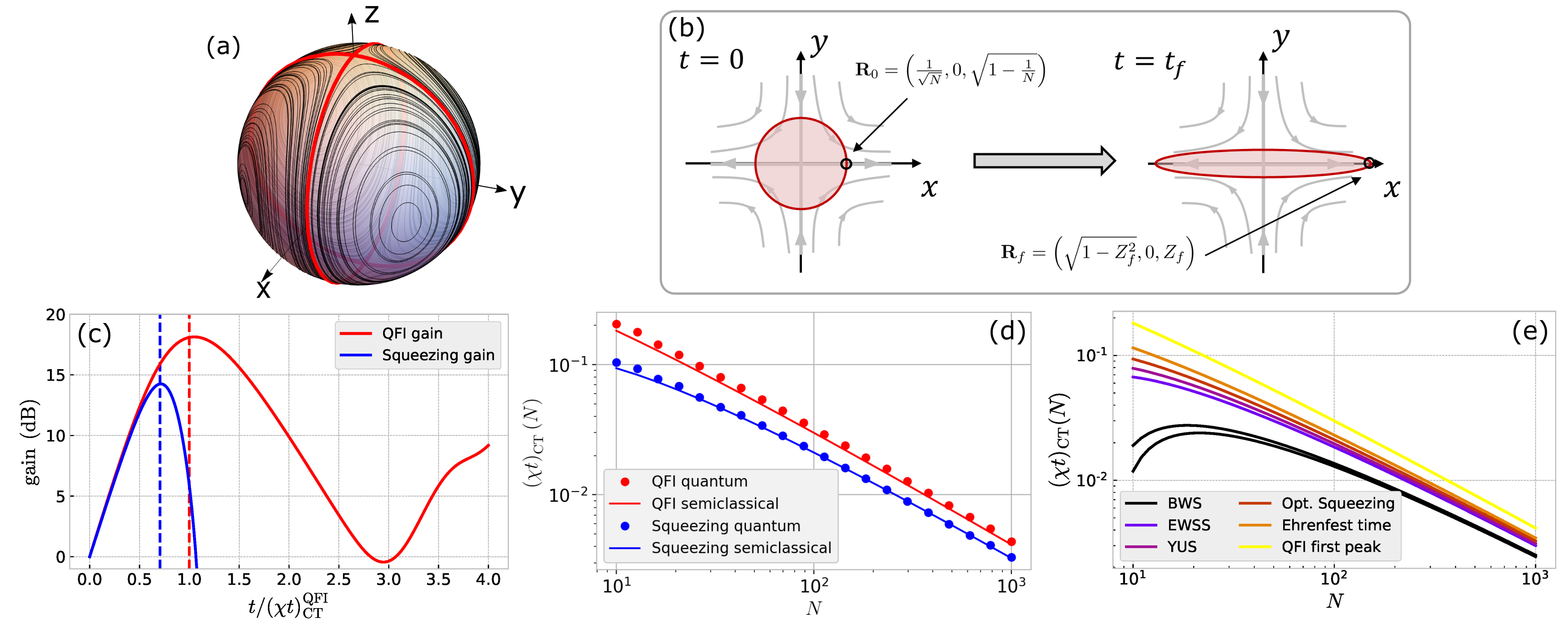}}
\caption{\textbf{(a)} Phase space portrait of the 2ACT phase space flow in the thermodynamic limit, c.f. Eqs. (\ref{eqn:flow_ct}). The red trajectory shows the separatrix. \textbf{(b)} Semiclassical picture for the dynamics of a spin coherent state located at the saddle point of the corresponding classical phase space flow.
\textbf{(c)} Metrological gains based on spin squeezing (solid blue) and QFI (solid red) for the 2ACT dynamics. The dashed lines show the time to the peak of each metrological gain computed in the main text. The results shown are for $N = 100$. \textbf{(c)} Comparison between analytical estimates and exact numerical simulations of the time required to peak QFI and spin-squeezing for the 2ACT dynamics. \textbf{(d)} Time scales to different metrologically relevant quantum states prepared with 2ACT dynamics as function of $N$. For the Yurke state, we considered $\alpha = 0.678$.}
\label{fig:ct_scales}
\end{figure*}

When evaluated at the fixed point $(X,Y,Z) = (0,0,\pm 1)$, Eq.~(\ref{eqn:tangent_map_ct}) is diagonal with eigenvalues $(\pm\tilde{\chi}, \mp\tilde{\chi},0)$. As the nonzero eigenvalues are real and come in pairs $\mathcal{M}_\pm = \pm\tilde{\chi}$, the fixed point is a saddle point, and here the principal directions of the separatrix curve are orthogonal and aligned with the $x$- and $y$-axis. In other words, they define great circles in the $x-z$ and $y-z$ planes, respectively. Furthermore, as discussed in the previous section, we know that an initial condition placed on one of the separatrix branches will evolve according to $\frac{dP_{\pm}}{dt} = \pm\tilde{\chi}P_\pm$, where $P_\pm = X,Y$. The other four fixed points in Eq.~(\ref{eqn:fixed_points_ct}) are stable centers. This can be easily verified by looking at the eigenvalues of the Jacobi matrix evaluated at the fixed points, which are given by $\lambda_{\pm} = \pm i2\tilde{\chi}$. This most basic information about the fixed points of 2ACT classical flow allows us to produce an accurate picture of the global structure of its phase space trajectories, whose exact form is shown in Fig.~\ref{fig:ct_scales} (a). \\



The local and global geometry of the separatrices provide useful information about the quantum evolution of an initial SCS placed at either of the two poles, $|J,\pm J\rangle$.  Without loss of generality, consider the case with $m=+J$. As depicted in Fig. \ref{fig:ct_scales} (b), while evolving under $\hat{H}_{\rm CT}$ the center of the distribution is fixed and the quantum projection noise will be initially squeezed exponentially fast, as the state will get stretched-out along the great circle on the $x$-$z$ plane and squeezed-in along the great circle on the $y$-$z$ plane. As a result, the motion along the separatrix branches leads to spin squeezing and thus governs the generation of metrologically useful quantum states. 

This qualitative picture can be made rigorous and used to estimate the timescales required to generate optimal spin-squeezing and other metrologically relevant quantum states. To do this we calculate the time required for points to travel along sections of the separatrix lines. When restricted to the unstable branch of the separatrix, $Y=0$ and thus $X^2 = 1-Z^2$.  From Eq.~(\ref{eqn:flow_ct}) we have that
\begin{equation}
\frac{dZ}{dt} = -\tilde{\chi}(1-Z^2),    
\end{equation}
which is solved by 
\begin{equation}
\label{eqn:int_time_scale_ct}
\tilde{\chi} t = -\int_{Z(0)}^{Z(t_f)}\frac{dZ}{1-Z^2}.    
\end{equation}
We can then compute how long it takes for a point starting at a contour of the SCS uncertainty patch, $Z(0) = \sqrt{1-\frac{1}{N}}$, to travel along the separatrix to some final point $Z(t_f) < Z(0)$ with $t_f>0$, 
\begin{equation}
\label{eqn:time_scale_ct}
\tilde{\chi}t = \ln\left[\frac{(1-Z(t_f))(\sqrt{N} + \sqrt{N-1})}{\sqrt{1 - Z^2(t_f)}} \right].
\end{equation}
Finding the time to the different metrologically relevant quantum states is then translated into the problem of finding the appropriate value of $Z(t_f)$. 



\subsection{Time to optimal spin squeezing and first peak of Quantum Fisher Information}

Given our choice of intial state, one can readily see that the length of the mean spin is given by $|\mathbf{\hat{J}}| = \langle \hat{J}_z \rangle_t$. Furthermore, we know that the anti-squeezed projection is $\hat{J}_x$, the direction determined by the unstable separatrix branch. Thus, at any later time $t>0$ we can write these two quantities as 
\begin{equation}
\label{eqn:vals_for_sq_ct}
|\mathbf{\hat{J}}| = J\cos(\theta), \quad \Delta J_x = J\sin(\theta),
\end{equation}
where $\theta = \theta(t)$, is the polar angle of spherical coordinates. In this setting the variance being squeezed is that of $\hat{J}_y$. To compute its value we proceed as follows. At the initial time, $t=0$, the area of the uncertainty patch of the SCS is that of a circle with radius $\Delta J_x = \Delta J_y = \sqrt{J/2}$. At any later time, $t>0$, this area is that of an ellipse with semi-major axis equal to $\Delta J_x$ and semi-minor axis equal to $\Delta J_y$. Hamiltonian dynamics preserves the areas and thus, after equating these two areas and solving for $\Delta J_y$, we obtain $\Delta \hat{J}_y = \frac{1}{2\sin(\theta)}$. The squeezing parameter in Eq.~(\ref{eqn:metrological_squeezing}) generated by this counter-twisting is thus 
\begin{equation}
\label{eqn:squeezing_ct}
\xi^2_{\rm CT} = \frac{4}{N\sin^2(2\theta)}.
\end{equation}
This expression has a minimum at $\theta = \pi/4$, the value at which the optimal spin squeezing is achieved, $\xi_{\rm CT}^2 = 4/N$~\footnote{This is the well know results of Heisenberg limited spin squeezing achieved with 2ACT~\cite{KitagawaUeda1993}, for $N=1024$, this gives an squeezing of $80\log_{10}(2)\approx 25$dB}. We can see immediately that $Z(t_f) = 1/\sqrt{2}$ gives the time scale to optimal squeezing. Plugging this value into Eq.~(\ref{eqn:time_scale_ct}) we obtain a principal result of this analysis,
\begin{equation}
\label{eqn:time_squeezing_ct}
(\tilde{\chi} t)_{\rm CT}^{\rm sq} = \ln\left[(\sqrt{2} - 1)(\sqrt{N} + \sqrt{N-1})\right].
\end{equation}
We note that this straightforward geometric analysis permits the exact identification of the numerical factor inside the logarithm, improving over the numerical or semi analytical approaches of Refs.~\cite{Yukawa2014,Kajtoch2015}.


Let us now turn our attention to the evolution of the QFI. In our setting, time evolution leads squeezing of  $\hat{J}_y$ and so we consider the QFI with respect to small rotations around the $x$-axis. This has a maximum whenever the variance of $\hat{J}_x$ has a maximum, and the latter is guaranteed to happen at the time at which the edges of the uncertainty patch are as far from each other as is allowed by the phase space geometry. This occurs when the state reaches points on the separatrix which are diametrically opposed. Given our choice of initial state, this happens when $Z(t_f) = 0$. Thus, substituting this value of $Z(t_f)$ into Eq.~(\ref{eqn:time_scale_ct}) we obtain the time to the first peak of the QFI, and consequently the time to the first peak of the metrological gain,  
\begin{equation}
\label{eqn:time_QFI}
(\tilde{\chi} t)_{\rm CT}^{\rm QFI} = \ln\left(\sqrt{N} + \sqrt{N-1}\right) \approx \frac{1}{2}\ln(4N),
\end{equation}
where the rightmost side of Eq.~(\ref{eqn:time_QFI}) is valid in the limit of $N\to\infty$.

The expressions in Eq.~(\ref{eqn:time_squeezing_ct}) and Eq.~(\ref{eqn:time_QFI}), albeit computed from a completely classical perspective, give quantitative predictions of the time required for a quantum system of system size $N$ evolving according to the 2ACT Hamiltonian to optimize metrological gain, be it measured by the QFI or spin squeezing. In order to test this prediction, we numerically simulate the quantum dynamics for various values of $N$ and compute the time-dependent metrological gains for each case. The case of $N=100$ is shown as an example in Fig. \ref{fig:ct_scales} (c), where it can be seen that the peaks of both quantities coincide with the predictions provided by our expressions. Furthermore, Fig. \ref{fig:ct_scales} (d) shows a systematic comparison between the exact numerical results and the analytical predictions as a function of the system size $N$. We find an excellent agreement between both, even for system sizes as small as $N \simeq 30$, which is far from the classical limit.

Finally, we point out that exploiting information about the classical motion of points along sections of the separatrix allows us to also write explicit expressions for the spin squeezing parameter and quantum Fisher information as a function of time. Further details of this analysis are discussed in Sec.~\ref{subsec:explicit_ct} of Appendix.~\ref{app:ct_stuff}.

\subsection{Time to other quantum states relevant for quantum metrology}
Previous works~\cite{Yukawa2014,Kajtoch2015} showed that starting from the stretched state, one can employ the dynamics of the 2ACT to prepare several other quantum states which are useful for quantum metrology with almost unit fidelity, that is, $|\langle \Psi|e^{-it\hat{H}_{{\rm CT}}}|J,J\rangle|^2 \approx 1$ , where $|\Psi\rangle$ is one of the following: the Berry-Wiseman state (BWS)~\cite{Berry2000}, Equally weighted superposition state (EWSS)~\cite{Yukawa2014}, some of Yurke states (YUS)~\cite{Combes2004}, and Twin-Fock state (TFS)~\cite{Yukawa2014}. Following our previous discussion, here we will use our semiclassical framework to derive expressions for the time scales required to prepare each of the first three states mentioned before, as these are the ones whose fidelity peaks before the first peak of the QFI. For the interested reader, we discuss the details in Sec.~\ref{subsec:other_states_ct} of Appendix ~\ref{app:ct_stuff}.

As before, the time to the peaks of fidelity to each of these three states can be estimated using Eq.~(\ref{eqn:time_scale_ct}), and thus we only need an appropriate value of $Z(t_f)$. Recognizing that under the current setting the variance of the time-evolved state goes as $\Delta \hat{J}_x = J\sin(\theta)$, then $Z(t_f)$ can be obtained from the variance of each of the states, discussed in detail in Appendix~\ref{app:ct_stuff}. Here we only list the final results.

For the BWS we can find a lower and upper bound for the time scale, they are given by
\begin{subequations}
\label{eqn:time_BWS}
\begin{align}
(\tilde{\chi} t)^{\rm BWS_<}_{\rm CT} &= \ln\left[(\sqrt{8} - \sqrt{7})(\sqrt{N} + \sqrt{N-1})\right], \\
(\tilde{\chi} t)^{\rm BWS_>}_{\rm CT} &=  \ln\left[(\sqrt{7} - \sqrt{6})(\sqrt{N} + \sqrt{N-1})\right].
\end{align}
\end{subequations}
For the EWSS we find
\begin{equation}
\label{eqn:time_EWSS}
(\tilde{\chi} t)_{\rm CT}^{\rm EWSS} = \ln\left[(\sqrt{3} - \sqrt{2})(\sqrt{N} + \sqrt{N - 1})\right],
\end{equation}
where the second line in Eq.~(\ref{eqn:time_EWSS}) holds in the limit of $N \gg 1$. For the family of Yurke states, we find 
\begin{equation}
\label{eqn:time_yurke}
(\tilde{\chi} t)_{\rm CT}^{\rm YUS} = \ln\left[\left( \frac{2 - \sqrt{2 + \sin^2(\alpha)}}{ \sqrt{2-\sin^2(\alpha) }} \right) (\sqrt{N} + \sqrt{N - 1})\right],
\end{equation}
expression which is valid in the limit of $N \gg 1$. The full expression is given in App.~\ref{app:ct_stuff}. Furthermore, we notice that only Yurke states with relatively small values of $\alpha$ are prepared by the 2ACT dynamics (see~\cite{Kajtoch2015} for a more detailed discussion).\\

From the results above, one might ask why the classical picture allows us to predict the time scales fairly well. To answer this, we can directly compare them with the Ehrenfest time. For systems with exponential instabilities originating at a saddle point, which can be characterized with a positive Lyapunov exponent~\cite{Schubert2012wave,Pappalardi2018}, this time is given by 
\begin{equation}
\label{eqn:ehrenfest_saddle}
t_{\rm Erfst} = \frac{1}{2\Lambda_{\rm sd}}\ln\left(\frac{1}{\hbar_{\rm eff}}\right),
\end{equation}
where $\Lambda_{\rm sd}$ is the Lyapunov exponent of the saddle and $\hbar_{\rm eff}$ is the effective Plank constant. For collective spin systems one has $\hbar_{\rm eff} = \frac{1}{N}$ and for the 2ACT Hamiltonian we know $\Lambda_{\rm sd}^{\rm CT} = \tilde{\chi}$, thus the Ehrenfest time is given by 
\begin{equation}
(\tilde{\chi} t)_{\rm CT}^{\rm Erfst} = \frac{\ln(N)}{2}.
\end{equation}
We show the different time scales computed in this section, including the Ehrenfest time, in Fig.~\ref{fig:ct_scales} (c). Excluding the time scale to peak QFI, all of the time scales with the 2ACT correspond to times shorter than the Ehrenfest time, and thus, one expects a semiclassical treatment to yield accurate predictions. Notice that our approach still provides accurate results for time scales which are longer than the Ehrenfest time, as illustrated by the validity of the expression for $(\tilde{\chi} t)_{\rm CT}^{\rm QFI}$. At this time scale the the support of the time evolved state is still concentrated on eigenstate which overlap considerably with the classical separatrix~\cite{Schubert2012wave}, and thus, our semiclassical approach based on the separatrix geometry retains validity.

\subsection{Summary}
The analysis presented in this section provides us with an intuitive picture of how collective spin Hamiltonians generate metrologically-useful states. The first important aspect is that one wants to construct a Hamiltonian having, in the classical limit, a saddle point, which is always accompanied by a separatrix line. Then the geometrical arrangement of the separatrix branches completely dictates the evolution and metrological utility of states generated from a SCS centered at the saddle point through Hamiltonian evolution.  It is in this sense that the two-axis counter-twisting represents the optimal choice of Hamiltonian. The geometry of its separatrix is locally optimal, since the branches are orthogonal, and it is globally optimal, since its branches defined great circles on the unit sphere, \textit{i.e.}, geodesics on the surface defining phase space. In the remainder of this work we will use these lessons to study other collective spin Hamiltonians and their potential for the generation of metrologically useful quantum states. 

\section{Phase space geometry and quantum metrology with a twisting and turning Hamiltonian}
\label{sec:all_twist_turn}
While locally and globally optimal for preparation of metrologically-useful quantum states, the 2ACT Hamiltonian requires the use of two twisting operations along perpendicular axis, which is difficult to implement experimentally. An alternative approach is to use a single twisting operation complemented with a linear term, \textit{i.e.}, a rotation term, resulting in a Hamiltonian of the form
\begin{equation}
\label{eqn:tat_hamil}
\hat{H}_{\rm TaT} = \Omega \hat{J}_x + \chi\hat{J}_z^2,
\end{equation}
where $\Omega$ is the rate of turning and $\chi$ the twisting strength. This model is often referred to as Twisting and Turning (TaT) in the quantum metrology literature~\cite{Micheli2003,Strobel2014,Muessel2015,Mirkhalaf2018,Pezze2018,Sorelli2019}, and it has been implemented in spinor BEC~\cite{Muessel2015}, and could be readily implemented in certain cavity QED setups~\cite{Braverman2019,Li2022}. One also recognizes this as the Lipkin-Meshkov-Glick (LMG) model, describing a transverse Ising model with all-to-all coupling~\cite{Lipkin1965,Vidal2004,Dusuel2004,Latorre2005,Dusuel2005,Heiss2005,Gutierrez2021}. 

In an early pioneering work~\cite{Micheli2003} it was recognized that the TaT dynamics generated by Eq.~(\ref{eqn:tat_hamil}) prepares highly entangled states at times which are logarithmic in the system size $N$ when starting from a spin coherent state centered at the point $(X,Y,Z) = (1,0,0)$, that is, $|\theta_0,\varphi_0\rangle = |\frac{\pi}{2},0\rangle$. This presents a significant improvement over what is achievable with the use of only a single twisting operation, the so called One Axis Twisting Hamiltonian (OAT)~\cite{Pezze2018}, and positions the TaT Hamiltonian on similar footing as the 2ACT Hamiltonian. Our goal is to use the phase space dynamics and separatrix geometry of this model to quantitatively explain this fact. We point out that the phase space description of this model has a direct connection to ground state, excited state, and dynamical quantum phase transitions of the LMG model. We will discuss this connection in Sec.~\ref{subsec:tat_and_lmg}.

Introducing $\tilde{\chi}=N \chi$ as before, the phase space flow associated with the classical TaT dynamics is given by (see Appendix~\ref{app:tat_stuff} for further details)
\begin{subequations}
\label{eqn:flow_tat}
\begin{align}
\frac{dX}{dt} &= \tilde{\chi}YZ, \\
\frac{dY}{dt} &= -\Omega Z - \tilde{\chi}XZ, \\
\frac{dZ}{dt} &= \Omega Y.
 \end{align}
\end{subequations}
The solutions of $\frac{d\mbf{X}}{dt}=0$, leading to the fixed points of the flow, depends on the system parameters. For $\frac{\Omega}{\tilde{\chi}}>1$, there are only two fixed points at $(X,Y,Z) = (\pm1,0,0)$, \textit{i.e.}, the poles with respect to the turning axis. These are stable, and so phase space is filed with trajectories representing Larmor precession of the mean spin. On the other hand, for $\frac{\Omega}{\tilde{\chi}}<1$ there are four fixed points, two given by $(X,Y,Z) = (\pm1,0,0)$, where the one at $X = -1$ is stable and the one at $X = 1$ is a saddle point. The two additional fixed points are at 
\begin{equation}
(X,Y,Z) = \left(\frac{\Omega}{\tilde{\chi}},0,\pm\sqrt{1-\left(\frac{\Omega}{\tilde{\chi}}\right)^2}\right).
\end{equation}
The two parameter regimes are connected through a pitchfork bifurcation of the stable point at $(X,Y,Z) = (1,0,0)$ which occurs at the critical point, $\frac{\Omega}{\tilde{\chi}}=1$.

As before, we focus on the motion around the saddle point. The separatrix branches emerge from this point, and conservation of energy guarantees that all points on the separatrix have the same energy as the saddle. By evaluating the eigenvalues of the Jacobi matrix at this saddle, one obtains its local Lyapunov exponent, 
\begin{equation}
\label{eqn:lyap_tat}
\Lambda^{\rm TaT}_{\rm sd} = \tilde{\chi} \sqrt{\frac{\Omega}{\tilde{\chi}}\left(1 - \frac{\Omega}{\tilde{\chi}} \right)}.
\end{equation}
Equation (\ref{eqn:lyap_tat}) dictates the exponential rate at which points move away from the saddle point. This rate is maximum when $\Omega/\tilde{\chi}=1/2$, and the corresponding Lyapunov exponent is given by $\left. \Lambda_{\rm sd}^{\rm TaT}\right|_{\rm CC} = \tilde{\chi}/2$. In~\cite{Micheli2003} this parameter regimes was denoted ``critical coupling,'' and it was shown to provide the fastest preparation time towards a cat-like state. Furthermore, \cite{Sorelli2019} studied  the dynamics of TaT in the short time regime and it was argued that the same parameter regime, \textit{i.e.}, critical coupling, was locally optimal. The maximum of Eq.~(\ref{eqn:lyap_tat}) already extends the results of Refs.~\cite{Micheli2003,Sorelli2019}, as it shows that at critical coupling, a point on the separatrix line travels with the maximum velocity allowed by $\hat{H}_{\rm TaT}$. We will present a formal result of this statement in Sec.~\ref{sec:speed_limit}. 

\subsection{Separatrix geometry and physical meaning of critical coupling}
\label{subsec:separatrix_geo_tat}
\begin{figure}[t!]
\centering{\includegraphics[width=0.40\textwidth]{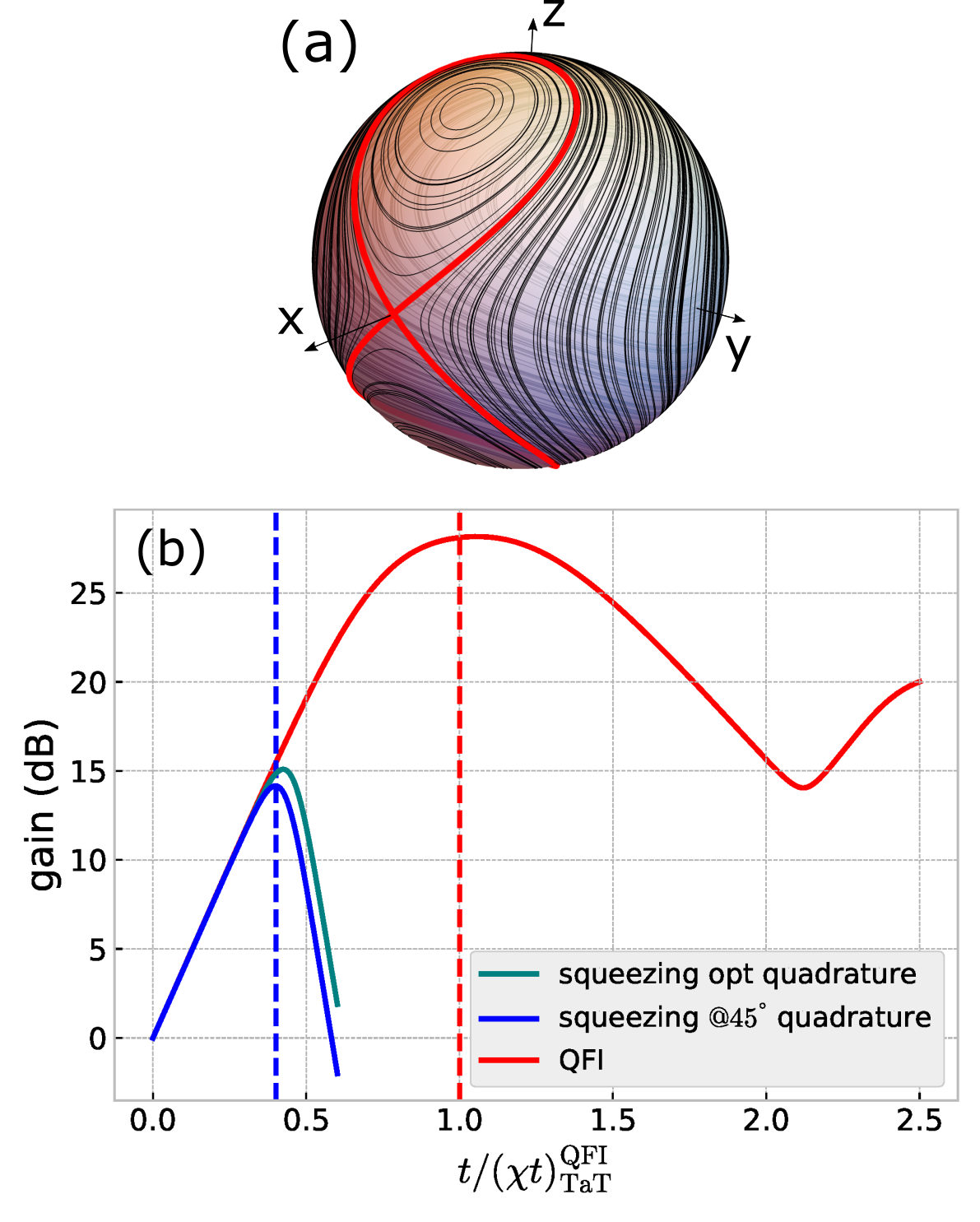}}
\caption{\textbf{(a)} Phase portrait of the phase space flow of the TaT Hamiltonian at critical coupling. The separatrix is shown in solid red. \textbf{(b)} Metrological gains, at critical coupling, based on the squeezing (solid blue) and the QFI (solid red), for an initial SCS along the positive $x$-direction. The dashed lines show our result for the time to peak of each metrological gain. Results are for a system with $N = 1024$.}
\label{fig:tat_gain}
\end{figure}

The importance of the separatrix geometry was already recognized in Ref.~\cite{Micheli2003}. In fact, they defined critical coupling as the parameter regime at which the distance between opposite ends of the separatrix was maximal, \textit{i.e.}, equal to one diameter of the unit sphere~\footnote{This fact implies that opposite ends of the separatrix touch the points $(X,Y,Z) = (0,0,\pm1)$. In other words, the state $|\psi\rangle = |J,J\rangle$ has energy equal to that of the separatrix line. This defines the dynamical critical point of the dynamical quantum phase transition of Hamiltonian in Eq.~(\ref{eqn:tat_hamil}). Interestingly, the metrological relevance of this dynamical quantum phase transition was investigated recently in Ref.~\cite{Guan2021}}. We illustrate the structure of the TaT phase space flow and separatrix line at critical coupling in Fig.~\ref{fig:tat_gain} (a).

From our previous discussion we learned that critical coupling is also the parameter regime at which the Lyapunov exponent of the saddle point is maximum. One can also show that at critical coupling the branches of the separatrix in the vicinity of the saddle point are orthogonal. In Appendix~\ref{app:tat_stuff} we show that for an arbitrary value of $\frac{\Omega}{\tilde{\chi}}$, the angle between separatrix branches is 
\begin{equation}
\label{eqn:angle_separatrix_tat}
\cos(\upsilon) = \frac{2 - \frac{\tilde{\chi}}{\Omega} + (\frac{\tilde{\chi}}{2\Omega})^2 Z^2}{\frac{\tilde{\chi}}{\Omega} - ( \frac{\tilde{\chi}}{2\Omega} )^2 Z^2} \approx \frac{2\Omega}{\tilde{\chi}} - 1,
\end{equation}
where $\upsilon$ is the angle between separatrix branches, and the far right holds in the vicinity of the saddle point, \textit{i.e.}, $Z,Y\to0$. Thus when $\tilde{\chi} = 2\Omega$, at critical coupling, we have $\upsilon = \frac{\pi}{2}$ and the separatrix branches are orthogonal.

Qualitatively, the result in Eq.~(\ref{eqn:angle_separatrix_tat}) tells us that, locally in the vicinity of the saddle point and at critical coupling, the TaT Hamiltonian is, effectively, a 2ACT Hamiltonian. One can formalize this statement by an appropriate choice of the collective spin axis. In particular, if we consider
\begin{equation}
\label{eqn:new_opes}
\hat{J}_1 = \frac{\hat{J}_y + \hat{J}_z}{\sqrt{2}}, \enspace \hat{J}_2 = \frac{\hat{J}_z -\hat{J}_y}{\sqrt{2}}, \enspace \hat{J}_x = J - \hat{o},
\end{equation}
with $\hat{J}_1$ and $\hat{J}_2$ representing two orthogonal directions in the $y$-$z$ plane at $45^\circ$ and $-45^\circ$, respectively. Furthermore, we are only interested in dynamics taking place withing the vicinity of the saddle point, and exploit this fact to write $\hat{J}_x = J - \hat{o}$, its mean-field minus fluctuations, with $\hat{o}$ an operator representing the fluctuations. Using Eq.~(\ref{eqn:new_opes}) we can rewrite Eq.~(\ref{eqn:tat_hamil}) as 
\begin{equation}
\label{eqn:tat_hamil_eff_1}
\hat{H}_{\rm TaT}^{\rm eff} = \Omega J - \left(\Omega - \frac{\chi}{2} \right)\hat{o} + \frac{\chi}{2}\left( \hat{J}_1 \hat{J}_2 + \hat{J}_2 \hat{J}_1 \right)
\end{equation}
where we have ignored terms $\mathcal{O}(\hat{o}^2)$; see Appendix~\ref{app:tat_stuff} for details.

At critical coupling Eq.~(\ref{eqn:tat_hamil_eff_1}) becomes
\begin{equation}
\label{eqn:tat_hamil_eff_2}
\hat{H}_{\rm TaT}^{\rm eff} = \frac{\chi}{2}\left( \hat{J}_1 \hat{J}_2 + \hat{J}_2 \hat{J}_1  \right),
\end{equation}
which is a 2ACT Hamiltonian as the one in Eq.~(\ref{eqn:ct_hamil_2}). The form of the Hamiltonian in Eq.~(\ref{eqn:tat_hamil_eff_2}) points at a general result. Any collective spin Hamiltonian whose associated phase space flow has a saddle point with orthogonal separatrix branches, can always be mapped locally to an effective 2ACT Hamiltonian under the appropriate choice of axis.


In summary, the different properties that define critical coupling, $\tilde{\chi} = 2\Omega$ for the TaT Hamiltonian are:
\begin{enumerate}
    \item The distance between the separatrix branches is maximal, and equal to the diameter in the unit sphere.
    \item The Lyapunov exponent of the saddle point is maximal, and equal to  $\left. \Lambda_{\rm sd}^{\rm TaT}\right|_{\rm CC} = \frac{\tilde{\chi}}{2}$.
    \item The separatrix branches are orthogonal in the vicinity of the saddle point, and thus, locally, the TaT Hamiltonian is effectively a 2ACT Hamiltonian.
    \item Also, the local wells of the classical energy surface to the left and right sides of the saddle point have the same depth, and thus time evolution happens symmetrically around the saddle point (see Appendix~\ref{app:tat_stuff}).
\end{enumerate}
Condition 4 above is fundamental to the validity of our semiclassical approach. Up to now we did not require it explicitly, as both the 2ACT and the TaT satisfy it for all parameter ranges. However, it will be essential in generalizations of the TaT, and we will further comment on its importance in Sec.~\ref{sec:all_p_spin}.

\subsection{Time scales to peak spin squeezing and to first peak of QFI}
In this subsection we focus on the special case of critical coupling and compute the time scales to peak spin squeezing and the first peak of the QFI using the traveling time of points along some sections of the separatrix branches. Following~\cite{Micheli2003,Sorelli2019} we consider an initial SCS centered at the positive $x$-direction, that is, $|\theta_0, \varphi_0 \rangle = |\frac{\pi}{2}, 0\rangle$. In a similar fashion to the 2ACT analysis, the center of the uncertainty patch is fixed, and the TaT dynamics only stretches and squeezes the uncertainty patch. As before, we can approximate the desired time scales by the time it takes for a point at the border of the uncertainty patch, which now takes the form $Z(0) = \frac{1}{\sqrt{2N}}$, to travel to some final location $Z(t_f)$ along the separatrix. This time scale is given by
\begin{equation}
\label{eqn:tat_time_scale}
(\tilde{\chi}t)_{\rm TaT} = 2\ln\left[ \frac{Z(t_f)(\sqrt{2N} + \sqrt{2N-1})}{1 + \sqrt{1 - Z^2(t_f)}} \right].
\end{equation}
The details of its derivation are given in Appendix~\ref{app:tat_stuff}. The problem becomes now how to find the appropriate value of $Z(t_f)$ for a given time scale. 

\subsubsection{Time scale to peak spin squeezing}
The third bullet point at the end of Sec.~\ref{subsec:separatrix_geo_tat} points at the fact that, locally, the separatrix branches of the TaT must be aligned with great circles. Here we will exploit this fact in order to estimate the appropriate value of $Z(t_f)$ for the peak squeezing time scale. The key ingredient in this analysis is the fundamental theorem of the local theory of parametric curves (see Chap.~1 of Ref.~\cite{Do1976}). This result tells us that, up to rigid motions (displacements and rotations), a curve is completely characterized by its values of curvature $\kappa$ and torsion $\tau$. Hence, given two curves of interest, one can investigate their local equivalence by finding the range of parameter values for which they both have the same curvature and torsion.

A great circle on the unit sphere has curvature and torsion given by  
\begin{equation}
\label{eqn:ct_curv_tor}
\kappa_{\rm CT} = 1, \enspace \tau_{\rm CT} = 0.
\end{equation}
We estimate $Z(t_f)$ as the limit value for which the curvature $\kappa_{\rm TaT}$ and torsion $\tau_{\rm TaT}$ of the TaT separatrix at critical coupling, give the same values as those in Eq.~(\ref{eqn:ct_curv_tor}). This value is given by $Z^*(t_f) = \cos(\theta^*) \approx 0.132684$ (see Appendix~\ref{app:tat_stuff}). In order to obtain an analytical expression of the value of $Z(t_f)$ recall that at critical coupling the two stable fixed points emerging at $\frac{\Omega}{\tilde{\chi}} = 0$, have $Z$ coordinate given by $Z_{\pm} = \pm\sqrt{1-\left(\frac{\Omega}{\tilde{\chi}}\right)} = \frac{\sqrt{3}}{2}$, where the last equality holds at critical coupling. It immediately follows that 
\begin{equation}
\label{eqn:z_squeezing_bound}
Z^*(t_f) \le 1 - Z_+,
\end{equation}
is a tight bound. We then approximate $Z(t_f)$ as $Z(t_f) = 1-Z_+ = 1 - \frac{\sqrt{3}}{2}$, and plugging the latter into Eq.~(\ref{eqn:tat_time_scale}), we obtain the time scale to peak spin sqeezing as 
\begin{equation}
\label{eqn:tat_time_scale_squeezing}
(\tilde{\chi}t)_{\rm TaT}^{\rm sq} = 2 \ln\left[ \left(\frac{2 - \sqrt{3}}{2 + \sqrt{\sqrt{48} - 3}}\right) (\sqrt{2N} + \sqrt{2N-1}) \right].
\end{equation}

\subsubsection{Time scale to the first peak of QFI}
As discussed in Ref.~\cite{Micheli2003}, the first peak of the QFI occurs when the quantum uncertainty patch has been stretched all the way to opposite ends of the separatrix. Due to our choice of axis in Eq.~(\ref{eqn:tat_hamil}) and our choice of initial state, we know that at the first peak of the QFI the state is stretched along the $z$-direction, thus we consider the QFI associated with the generator $\hat{G} = \hat{J}_z$. This conditions then tells us that the appropriate value of $Z(t_f)$ for this time scale is given by $Z(t_f) = 1$. Plugging this value into Eq.~(\ref{eqn:tat_time_scale}) we obtain 
\begin{equation}
\label{eqn:tat_time_scale_QFI}
(\tilde{\chi} t)_{\rm TaT}^{\rm QFI} = 2\ln\left[ \sqrt{2N} + \sqrt{2N-1} \right] \approx \ln[8N],
\end{equation}
where the rightmost side of Eq.~(\ref{eqn:tat_time_scale_QFI}) holds in the limit of $N\gg1$. We illustrate how well the expressions in Eq.~(\ref{eqn:tat_time_scale_squeezing}) and Eq.~(\ref{eqn:tat_time_scale_QFI}) give the time scales to peak spin squeezing and peak QFI, respectively, in Fig.~\ref{fig:tat_gain} (b).

\subsubsection{Time to first peak of QFI: 2ACT vs TaT at critical coupling}
The time scales to first peak of QFI in Eq.~(\ref{eqn:time_QFI}) and in Eq.~(\ref{eqn:tat_time_scale_QFI}) look deceptively similar. In fact, in the limit of large $N\gg1$, they only differ by a constant factor, whose origin is purely geometrical. In order to show this, we will assume that there is only a finite amount of twisting strength that can be achieved in a given implementation, and thus, if one is required to twist along two different axes, the strength of each twist will be halved compared to that of a Hamiltonian which  twists along a single axis. Given that the Lyapunov exponents of the 2ACT saddle and the TaT saddle at critical coupling are $\Lambda_{\rm sd}^{\rm CT} = \tilde{\chi}$ and $\Lambda_{\rm sd}^{\rm TaT}\left.\right|_{\rm CC} = \frac{\tilde{\chi}}{2}$, this assumption implies that the exponents are equal. In other words we take $\Lambda_{\rm sd}^{\rm CT} = \Lambda_{\rm sd}^{\rm TaT}\left.\right|_{\rm CC} = \Lambda_{\rm sd}$, and measure timescales in terms of this exponent.

Then, in the limit of large $N\gg1$, we can write 
\begin{equation}
(\Lambda_{\rm sd}t)_{\rm CT}^{\rm QFI} = \frac{1}{2}\ln[4N], \enspace (\Lambda_{\rm sd}t)_{\rm TaT}^{\rm QFI} = \frac{1}{2}\ln[8N].    
\end{equation}
The difference between this two time scales is given by 
\begin{equation}
(\Lambda_{\rm sd}t)_{\rm CT}^{\rm QFI} - (\Lambda_{\rm sd}t)_{\rm TaT}^{\rm QFI} = \frac{\ln(2)}{2}.
\end{equation}
The origin of this can be traced back to the additional length that a point has to travel along the separatrix between $Z(0)$ and $Z(t_f)$. At critical coupling, and in the limit $N\to\infty$, the length difference between the 2ACT and TaT separatrices is given by 
\begin{equation}
\label{eqn:length_diff}
l_{\rm TaT} - l_{\rm CT} = \mathrm{E}(i) - \frac{\pi}{2}\approx 0.3393 \approx \frac{\ln(2)}{2},
\end{equation}
where $\mathrm{E}(i)$ is the complete elliptic integral of the second kind. We give the details of this result in Appendix~\ref{app:tat_stuff}. We see then that the small difference in time scale in the limit of large $N\gg1$ is only due to the additional length of the TaT separatrix at critical coupling with respect to the length of a great circle.

\subsection{Local optimality of twisting and turning at critical coupling as a quantum speed limit}
\label{sec:speed_limit}
In the previous section we showed that the Lyapunov exponent of the saddle point of the TaT phase space flow is maximum at critical coupling. This implies that in the classical limit, a point traveling along the separatrix moves with the maximum allowed velocity.. This  has direct consequences for the quantum dynamics, and it can be recast as the saturation of a quantum speed limit~\cite{Pires2016,Deffner2017}. To this end, we use recent results derived in~\cite{Poggi2021} for quantum speed limits in Gaussian-preserving bosonic dynamics. 

We are interested in a quantum speed limit constrained to local dynamics, \textit{i.e.}, times such that the quantum state remains mostly in the vicinity of the saddle point. The first step to show that at critical coupling the speed of evolution saturates the quantum speed limit, is to write the TaT Hamiltonian using the Holsten-Primakoff approximation~\cite{Holstein1940}. We define it with the fixed component of angular momentum relative to the position of the saddle point, \textit{i.e.}, the positive $x$-direction, and fluctuations in the orthoginal directions, and thus $\hat{J}_x \approx J$,  $\hat{J}_y \approx \sqrt{J}\hat{q}$,  $\hat{J}_z \approx \sqrt{J}\hat{p}$, where $\hat{q}$, $\hat{p}$, are two bosonic quadrature operators. After this procedure the TaT Hamiltonian takes the form
\begin{equation}
\label{eqn:tat_hamil_hp}
\hat{H}_{\rm TaT} = -\frac{\Omega}{2}\hat{q}^2 + \frac{1}{2}(\tilde{\chi} - \Omega)\hat{p}^2,
\end{equation}
Details of this derivation are discussed in Appendix~\ref{app:speed_limit}. 

\begin{figure}[t!]
\centering{\includegraphics[width=0.43\textwidth]{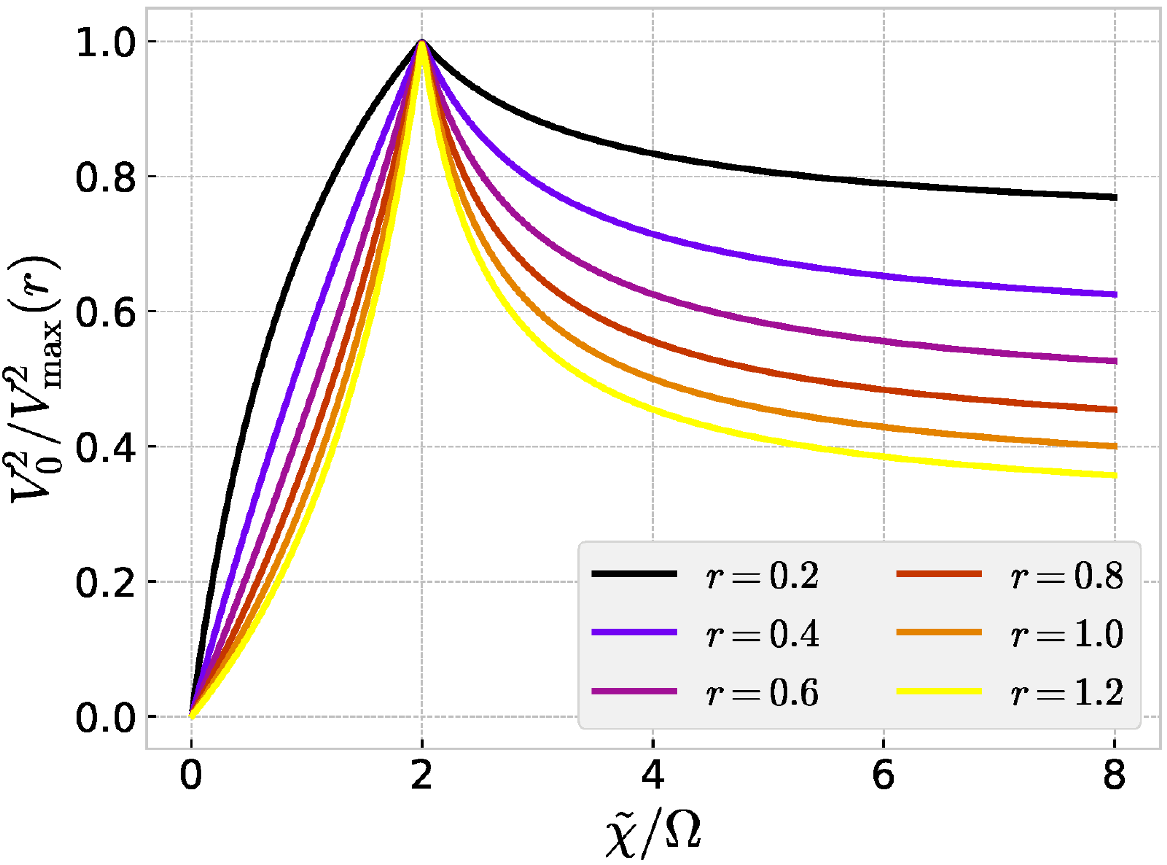}}
\caption{Behavior of the ratio in Eq.~(\ref{eqn:ratio_speed_limit}) between the actual quantum speed of the initial state and the maximum possible one, as a function of $\frac{\tilde{\chi}}{\Omega}$, for different values of the squeezing $r$. Notice how the quantum speed limit is saturated at critical coupling, independently of the amount of squeezing.}
\label{fig:qsl_ratio}
\end{figure}

Equation~(\ref{eqn:tat_hamil_hp}) can be written as 
\begin{equation}
\label{eqn:tat_hamil_gauss}
\hat{H}_{\rm TaT} = \frac{1}{2}\mathcal{Z}^{T} \mathbb{G} \mathcal{Z},
\end{equation}
where $\mathcal{Z} = (\hat{q}, \hat{p})$ is a vector of quadratures and $\mathbb{G}$ is a $2\times2$ matrix given by 
\begin{equation}
\label{eqn:g_matrix}
\mathbb{G} = -\Omega\begin{pmatrix}
1 && 0 \\
0 && 1-\frac{\tilde{\chi}}{\Omega}
\end{pmatrix}.
\end{equation}
From this last expression we see that at critical coupling, locally the TaT Hamiltonian is a perfect squeezer, \textit{i.e.}, it is phase matched~\cite{Trail2010}. In this context we consider the quantum speed limit as the maximum speed of evolution allowed by this Hamiltonian. The quantum speed $V_t$ is defined as the rate of change of the fidelity between $\Ket{\psi_t}$ and $\Ket{\psi_{t+dt}}$, 
\begin{equation}
    F(\psi_t,\psi_{t+dt})=1-V_t^2 dt^2
\end{equation}
\noindent and for pure states $F(\psi_1,\psi_2)=\lvert \BraKet{\psi_1}{\psi_2}\rvert^2$. Using results from \cite{Poggi2021}, we can compute the speed for a generic Gaussian state with a fixed level of squeezing $r$ (see Appendix~\ref{app:speed_limit} for details). With these constraints the maximum speed of evolution given the TaT Hamiltonian is
\begin{equation}
V^2_{\rm max}(r) = \frac{\tilde{\chi}^2}{8} + \frac{1}{4}\tilde{\chi}\left| \tilde{\chi} - 2\Omega \right|r. 
\end{equation}
On the other hand, the speed of evolution for our initial state is given by 
\begin{equation}
V_0^2 =  \langle \Delta^2 \hat{H}_{\rm TaT} \rangle_0 = \frac{\tilde{\chi}^2}{8}.
\end{equation}
In order to explore if there is a parameter regime where the quantum speed limit is saturated, we study the ratio 
\begin{equation}
\label{eqn:ratio_speed_limit}
\frac{V_0^2}{V^2_{\rm max}(r)} = \frac{\frac{\tilde{\chi}^2}{\Omega^2}}{\frac{\tilde{\chi}^2}{\Omega^2} + 2\frac{\tilde{\chi}}{\Omega}\left| \frac{\tilde{\chi}}{\Omega} - 2\right|r}.
\end{equation}
This expression has a maximum equal to one at $\frac{\tilde{\chi}}{\Omega} = 2$, \textit{i.e.}, critical coupling. Notably, this fact is independent of the value of the squeezing $r$, which we illustrate in Fig.~\ref{fig:qsl_ratio}.

\subsection{Quantum metrology and phase diagrams in many-body systems: 
Twisting and Turning and the Lipkin-Meshkov-Glick model}
\label{subsec:tat_and_lmg}
\begin{figure}[t!]
\centering{\includegraphics[width=\linewidth]{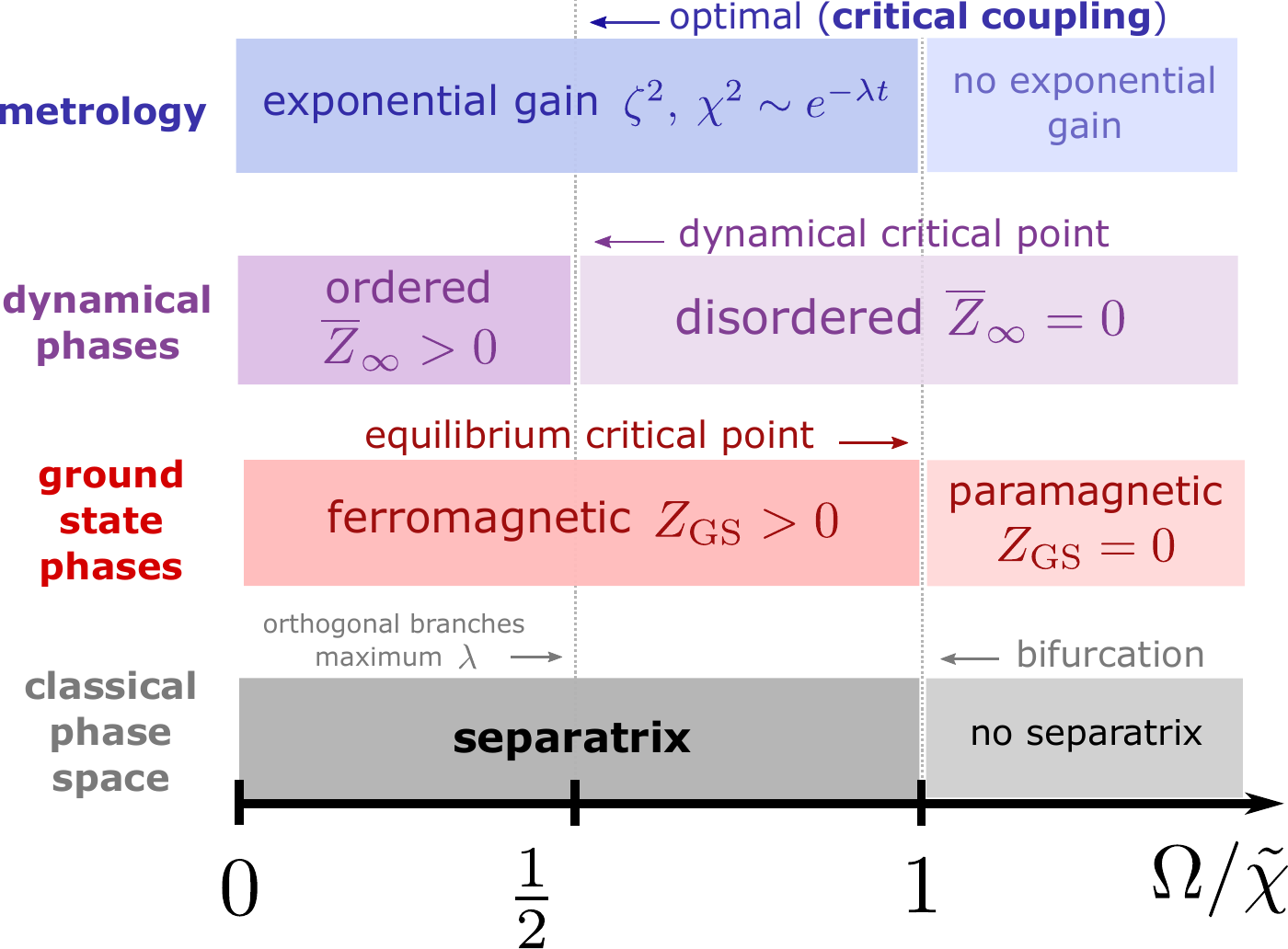}}
\caption{Schematic of the different parameter regimes of the TaT / LMG Hamiltonian of Eq. (\ref{eqn:tat_hamil}) as a function of $\Omega/\tilde{\chi}$. Depicted are aspects related to metrological state preparation, dynamical quantum phase transitions, ground-state phase quantum phase transitions and classical phase space structures.}
\label{fig:lmg_diag}
\end{figure}

Throughout this work we have discussed the prominent role played by the classical phase space structures of collective spin models in the preparation of probe states for quantum metrology. In particular, we have shown that the geometric and dynamical properties of the separatrix dictate the parameter regimes leading to the exponential increase of spin squeezing and QFI. In this context it is instructive to recall that the separatrix also plays a major role in the description of equilibrium and nonequilibrium phases in collective spin models. The case of the TaT Hamiltonian, Eq.~(\ref{eqn:tat_hamil}), is paradigmatic in this sense, since it also corresponds to a special case of the Lipkin-Meshov-Glick (LMG) model, which has been extensively investigated in the context of quantum phase  transitions~\cite{Lipkin1965,Latorre2005,Vidal2004,Dusuel2004,Dusuel2005,Wang2019,Gutierrez2021}. In fact, there is a one-to-one correspondence between the phase diagram of the LMG and the regimes of optimal probe state preparation with TaT dynamics, which are schematically depicted in Fig.~\ref{fig:lmg_diag} and analyzed in the following.

We can define an equilibrium order parameter describing the character of the ground state as
\begin{equation}
    \mathrm{Equilibrium}: Z_{\mathrm{GS}} = \frac{1}{J}\Bra{\phi_\mathrm{GS}}\hat{J}_z\Ket{\phi_\mathrm{GS}}
\end{equation}
\noindent where $\Ket{\phi_\mathrm{GS}}$ is the ground state of the Hamiltonian in Eq.~(\ref{eqn:tat_hamil}). Similarly, we define a nonequilibrium or dynamical order parameter
\begin{equation}
    \mathrm{Dynamical}: \overline{Z}_{\infty} = \lim \limits_{T\rightarrow \infty} \frac{1}{T}   \int\limits_0^T dt \frac{1}{J} \Bra{\psi(t)}\hat{J}_z\Ket{\psi(t)}
    \label{eqn:dqpt_op}
\end{equation}
\noindent where $\Ket{\psi(t)}= e^{-i \hat{H} t} \Ket{\psi(0)}$ and the initial state is the spin coherent state $\Ket{\psi(0)}=\Ket{\uparrow}^{\otimes N}$. As discussed previously, analysis of the TaT/LMG model classical flow reveals a bifurcation at $\Omega/\tilde{\chi}=1$. For $\Omega < \tilde{\chi}$, all trajectories revolve around the stable fixed points at $(\pm 1,0,0)$. As a result, in this regime we have both $Z_{\mathrm{GS}}=0$ and $\overline{Z}_{\infty}=0$. At the bifurcation (where the separatrix is born) the ground state of the system, identified with trajectories of minimal classical energy, begin shifting towards the $Z$ axis, and  $Z_{\mathrm{GS}}>0$. This then correspond to the ground-state critical point. However, the dynamical order parameter still vanishes, $\overline{Z}_{\infty}=0$, since the initial condition still revolves around the $x$-axis. As we decrease the value of the external field and $\Omega<\tilde{\chi}<1$, the stable fixed points get closer to the $z$-axis and the separatrix grows bigger. Eventually, this trajectory will pass through $(0,0,1)$. At this stage, the initial condition in Eq.~(\ref{eqn:dqpt_op}) gets trapped \textit{inside} the separatrix, and rotates around the stable fixed point, leading to $\overline{Z}_{\infty}>0$. This is the dynamical critical point, $\Omega/\tilde{\chi}=\frac{1}{2}$ which coincides with the critical coupling regime discussed previously.

Furthermore the separatrix line has been recognized as the classical origin of certain type of Excited State Quantum phase Transitions (ESQPTs)~\cite{Stransky2014,Cejnar2021}. In particular, for a system whose thermodynamic limit is described by a single degree of freedom, this ESQPT is given as a logarithmic divergence of the density of states as a consequence of the clustering of eigenstates whose energy is inside a small energy window centered at the separatrix energy. As such, as we sweep $\frac{\Omega}{\tilde{\chi}}$ towards zero and for $\frac{\Omega}{\tilde{\chi}}<1$, different excited states undergo ESQPT as their energy becomes equal to the energy of the separatrix. Since this type of quantum phase transitions refer to macroscopic changes in the structure of excited states, their consequences are observable in the dynamics of appropriately chosen initial states. In fact, there is a direct correspondence between ESQPT in a group of excited states and the type of DQPT discussed above (see for instance Refs.~\cite{corps2022a,Corps2022b}) which has implications
for quantum enhanced metrology~\cite{Zhou2022}. 

Finally, the connection depicted in Fig.~\ref{fig:lmg_diag} goes beyond just the TaT/LMG model, and its applicable to generic collective spin models. In the next section we will introduce and study two additional examples.

\section{Beyond Hamiltonians with two-body interactions: \texorpdfstring{$p$}{\textit{p}}-order twisting}
\label{sec:all_p_spin}
So far we have considered the paradigmatic examples of 2ACT and TaT to analyze the preparation of metrologically useful states of collective spins. However, the tools we have employed in their analysis are agnostic to the specific model under study, and the geometry of phase space flows and separatrices allows us to study any collective spin Hamiltonian. In this section we turn our attention to generalizations of the 2ACT and TaT models studied before, to account for higher-order many-body twisting operations. In particular, we consider interactions such as $ \hat{J}_\alpha^p$ associated to a $p$-body collective coupling between $p$ spin-1/2 particles. 

\subsection{Two-axis counter-twisting with a \texorpdfstring{$p$}{\textit{p}}-order twist}
Consider a generalization of the Hamiltonian in Eq. (\ref{eqn:ct_hamil_1}) to include $p$-order twisting ($p\geq 2$ throughout),
\begin{equation}
    \hat{H}_{\rm CT}^{(p)} = \chi \left(\hat{J}_x^p - \hat{J}_y^p\right)
\end{equation}
where, for convenience, we have rotated the axis so that the twisting directions correspond with $x$ and $y$. Proceeding as in Sec.~\ref{sec:all_counter_twistins} we can derive the equations of motion for the classical flow, which read
\begin{subequations}
\label{eqn:flow_ct_p}
\begin{align}
\frac{dX}{dt} &= -\tilde{\chi}Y^{p-1} Z, \\
\frac{dY}{dt} &= -\tilde{\chi}X^{p-1}Z, \\
\frac{dZ}{dt} &= \tilde{\chi}\left(X^{p-1} Y + Y^{p-1} X \right).
\end{align}
\end{subequations}
The fixed points of the flow can be readily computed from Eqs.~(\ref{eqn:flow_ct_p}) and turn out to be independent of $p$ and located, as before, on the north and south pole $(0,0,\pm 1)$ and along the twisting directions $(\pm 1,0,0)$ and $(0,\pm 1,0)$. The stability analysis (see Appendix \ref{app:ptat_stuff}) reveals that the equatorial fixed points are stable for all values of $p$. For the polar ones, however, we find that the eigenvalues of the Jacobi matrix $\mathbb{M}$ are $\pm 1$ (leading to saddle points) only for $p=2$. For $p>2$, we find that $\mathbb{M}=0$ when evaluated in these fixed points and thus the standard linear stability analysis is insufficient to classify the local motion of the system. We can still study the behavior in this regime by considering the dynamics near $Z\simeq 1$, and looking at the motion along the branches $x=y=v$ and $u=x=-y$, for which we obtain
\begin{equation}
    \frac{dv}{dt}=-\tilde{\chi} v^{p-1},\  \mathrm{and}\ \frac{du}{dt}=-\tilde{\chi} (-1)^{p-1} u^{p-1}.
\end{equation}
We focus on the case of $p$ even, for which the local motion clearly shows saddle-point-like behavior, with a stable ($v$) and an unstable ($u$) branch. For $p=2$, we get exponential motion $u_{p=2}(t) \sim e^{\tilde{\chi} t}$ as described in Sec.~\ref{sec:all_counter_twistins}, while for $p>2$ we obtain upon direct integration
\begin{equation}
    u_p(t) = \left( \frac{1}{u_0^{p-2}} - (p-2)\tilde{\chi} t\right)^{-\frac{1}{p-2}},
\end{equation}
which grows more slowly. Thus, while the $p$-2ACT Hamiltonian will produce spin squeezing, it will not do so exponentially in time unless $p=2$.



\subsection{Twisting and turning with a \texorpdfstring{$p$}{\textit{p}}-spin}
Consider now a generalization of the TaT Hamiltonian, which we refer to as $p$TaT, given by
\begin{equation}
\label{eqn:ptat_hamil}
\hat{H}_{p{\rm TaT}} = \Omega \hat{J}_x + \chi\hat{J}_z^p,
\end{equation}
with $\Omega$ the turning rate and $\chi$ the $p$-twisting strength. The Hamiltonian in Eq.~(\ref{eqn:ptat_hamil}) is an unnormalized version of the $p$-spin models, largely studied in the context of quantum annealing~\cite{Jorg2010,Bapst2012,Matsuura2017}, and equilibrium and nonequilibirum quantum phase transitions~\cite{Filippone2011,DelRe2016,Munoz-Arias2020,Correale2021}.

Our first goal is to explore whether a notion of ``critical coupling'' exists for the family of $p$TaT Hamiltonians. From our discussion of the $2$TaT in Sec.~\ref{sec:all_twist_turn} we saw that critical coupling is given by the parameter regime at which conditions $(1)$-$(4)$ in Sec.~\ref{subsec:separatrix_geo_tat} are satisfied simultaneously. In this subsection we study each of these conditions for the whole family of $p$TaT Hamiltonians. We will see that such a strong notion of critical coupling cannot be extended to the other $p$TaT models with $p>2$. However, a less stringent notion can still be satisfied by at least one more $p$TaT Hamiltonian: we will show that the only other case for which more than one of the conditions can be satisfied simultaneously is $p=3$. 

Condition (1) defines critical coupling as the parameter choice such that opposite ends of the separatrix are diametrically opposed. This statement holds true only for the $p$TaT with even values of $p$. This can be seen in the following way. Recall that $p$TaT Hamiltonians with even $p$ have a parity symmetry, given by the operator $\hat{\Pi} = e^{i\pi\hat{J}_x}$ and $[\hat{H}_{p{\rm TaT}}, \hat{\Pi}] = 0$. Thus, pairs of points which are diametrically opposed can have the same energy, as for instance the north and south poles with respect to the twisting axis, represented by the two stretched states $|J,\pm J\rangle$, with mean energy $\langle \hat{H}_{p{\rm TaT}}\rangle = \frac{\tilde{\chi}}{p}J$. Given that the separatrix is an isoenergetic curve, the classical flow associated with the $p$TaT Hamiltonian for even values of $p$ admits a regime of parameters where the separatrix has points diametrically opposed. On the contrary the lack of parity symmetry for the models with odd values of $p$ rules out this possibility.

We now turn to condition (3). We found that there is no parameter regime for which the separatrix branches are orthogonal in the vicinity of the saddle point, for any $p$TaT with $p>2$. This is formalized in the following theorem, whose proof we give in Appendix~\ref{app:ptat_stuff}.
\begin{theorem}[Absence of local optimality in \texorpdfstring{$p$}{\textit{p}}TaT]
\label{theo:no_local_ptat}
Given the real valued  control parameter $\frac{\Omega}{\tilde{\chi}}>0$, there is no value of this parameter for which the $p{\rm TaT}$ dynamics, with $p>2$, is locally optimal.
\end{theorem}

Let us now consider the other two conditions. Condition (4) demands the energy wells on both sides of the saddle point to have equal depth~\footnote{This is a central point in the phase space approach presented in this work. Approximating the time scales to certain metrologically useful states, via the motion of points along separatrix branches, requires this motion to take place symmetrically with respect to the saddle point.}. This property can be investigated using the classical energy $E(\mathbf{X}) = \langle \hat{H}_{p{\rm TaT}} \rangle/J$ where the expectation value is taken in a spin coherent state, in the limit $J\to\infty$. All the $p$TaT models have a parameter regime where the classical energy is a single well, and a parameter regime where the classical energy has a double ($p=2$, $p>2$ and odd) or triple well structure ($p>2$ and even). These different regimes are separated by bifurcation points and ground state critical points (for a in-detail discussion see Refs.~\cite{Bapst2012,Munoz-Arias2020}). For the $2$TaT, parity symmetry guarantees the individual wells of the double well to have equal depth for all parameter regimes. For the $p$TaT with odd values of $p$, the condition of having wells with the same depth defines the ground state critical point; as such, this condition is satisfied only at one specific parameter value, $\left. \frac{\tilde{\chi}}{\Omega}\right|_{\rm GS}$. For the $p$TaT with $p>2$ and even, parity symmetry guarantees the two outermost wells to have the same depth for all parameter regimes. However, these lie at the right and left side of two different saddle points. Similar to the case of odd values of $p$, the outer and central wells will have the same depth at a single parameter value, given also by $\left. \frac{\tilde{\chi}}{\Omega}\right|_{\rm GS}$. An immediate consequence is that for $p$TaT with $p>2$ and even, conditions (1) and (4) in Sec.~\ref{subsec:separatrix_geo_tat} are satisfied at two completely different parameter regimes, and thus they cannot be considered to define a notion of critical coupling.

The previous analysis implies that for the $p$TaT with $p>2$ there is no parameter regime for which at least three of the conditions in Sec.~\ref{subsec:separatrix_geo_tat} are satisfied simultaneously. In fact, we will see that for the models with $p>2$ and even, all four conditions occur at different parameter values, and thus a notion of critical coupling cannot be introduced. We are then left with the question of whether, for models with $p>2$ and odd, conditions (2) and (4) in Sec.~\ref{subsec:separatrix_geo_tat} can be satisfied simultaneously. In the following we show that this is true only for $p=3$, and thus, this is the only other $p$TaT Hamiltonian admitting a notion of critical coupling.

In order to show this, we need explicit expressions for $\left.\frac{\tilde{\chi}}{\Omega}\right|_{\rm GS}$, the ground state critical point, and $\left.\frac{\tilde{\chi}}{\Omega}\right|_{{\rm LE}}$, the maximum of the saddle point Lyapunov exponent, both as function of $p$. An explicit computation of the ground state critical point for this family of models was presented in Appendix B of Ref.~\cite{Munoz2022}. Here, we only mention the explicit expression. The ground state critical point is given by
\begin{equation}
\label{eqn:gsqpt_solutions}
Z_{\rm GS} = \sqrt{\frac{p(p-2)}{(p-1)^2}}, \enspace \left.\frac{\tilde{\chi}}{\Omega}\right|_{\rm GS} = \frac{(p-1)^{p-1}}{\sqrt{(p(p-2))^{p-2}}},
\end{equation}
where $Z_{\rm GS}$ is the $z$-coordinate of the new global minimum of the classical energy density. An explicit expression for the maximum of the saddle point Lyapunov exponent is given by 
\begin{subequations}
\label{eqn:max_lyap_sols}
\begin{align}
Z_{\rm LE} &= \frac{p-2}{p-1}, \\
\left.\frac{\tilde{\chi}}{\Omega}\right|_{\rm LE} &= \frac{(p-1)^{p-1}}{(p-2)^{p-2}\sqrt{(p-1)^2 - (p-2)^2}},
\end{align}
\end{subequations}
where $Z_{\rm LE}$, is the $z$-coordinate of the saddle point for the parameter values at which the Laypunov exponent is maximum. We give the explicit derivation of Eq.~(\ref{eqn:max_lyap_sols}) in Appendix~\ref{app:ptat_stuff}.

The form of Eq.~(\ref{eqn:gsqpt_solutions}) and Eq.~(\ref{eqn:max_lyap_sols}) imply that 
\begin{equation}
\left.\frac{\tilde{\chi}}{\Omega}\right|_{\rm GS}  \ge \left.\frac{\tilde{\chi}}{\Omega}\right|_{\rm LE},    
\end{equation}
with equality only when $p=3$. Hence, the only other $p$TaT Hamiltonian which admits a notion of critical coupling, similar to that of the TaT, is the $3$TaT, with critical coupling defined as the ground state critical point.

Before closing this subsection let us mention that the different coupling regimes of the $p$TaT family, defining the bifurcation (spinodal) point, ground state critical point, dynamical critical point, and maximum of the saddle point Lyapunov exponent, establish a classification of this family of models. In fact, one encounters four different types of inequality chains. If $p=2$, these parameter values satisfy
\begin{equation}
\left.\frac{\tilde{\chi}}{\Omega}\right|_{\rm spino} = \left.\frac{\tilde{\chi}}{\Omega}\right|_{\rm GS} < \left.\frac{\tilde{\chi}}{\Omega}\right|_{\rm DQPT} = \left.\frac{\tilde{\chi}}{\Omega}\right|_{\rm LE},
\end{equation}
where $\left.\frac{\tilde{\chi}}{\Omega}\right|_{\rm DQPT}$ indicates the critical point of the dynamical quantum phase transition. If $p=3$, these parameter values satisfy 
\begin{equation}
\left.\frac{\tilde{\chi}}{\Omega}\right|_{\rm spino} < \left.\frac{\tilde{\chi}}{\Omega}\right|_{\rm GS} = \left.\frac{\tilde{\chi}}{\Omega}\right|_{\rm LE} < \left.\frac{\tilde{\chi}}{\Omega}\right|_{\rm DQPT}.
\end{equation}
If $p=4$, these parameter values satisfy
\begin{equation}
\left.\frac{\tilde{\chi}}{\Omega}\right|_{\rm spino} <
\left.\frac{\tilde{\chi}}{\Omega}\right|_{\rm LE} <
\left.\frac{\tilde{\chi}}{\Omega}\right|_{\rm GS}  < \left.\frac{\tilde{\chi}}{\Omega}\right|_{\rm DQPT}.
\end{equation}
For all other values of $p>4$, these parameter values satisfy
\begin{equation}
\left.\frac{\tilde{\chi}}{\Omega}\right|_{\rm spino} <
\left.\frac{\tilde{\chi}}{\Omega}\right|_{\rm LE} <
\left.\frac{\tilde{\chi}}{\Omega}\right|_{\rm DQPT} <
\left.\frac{\tilde{\chi}}{\Omega}\right|_{\rm GS}.
\end{equation}
We gave explicit expressions for the spinodal, ground state critical and dynamical critical points of this family of models in Appendix B of Ref.~\cite{Munoz2022}.

\begin{figure}[t!]
\centering{\includegraphics[width=0.43\textwidth]{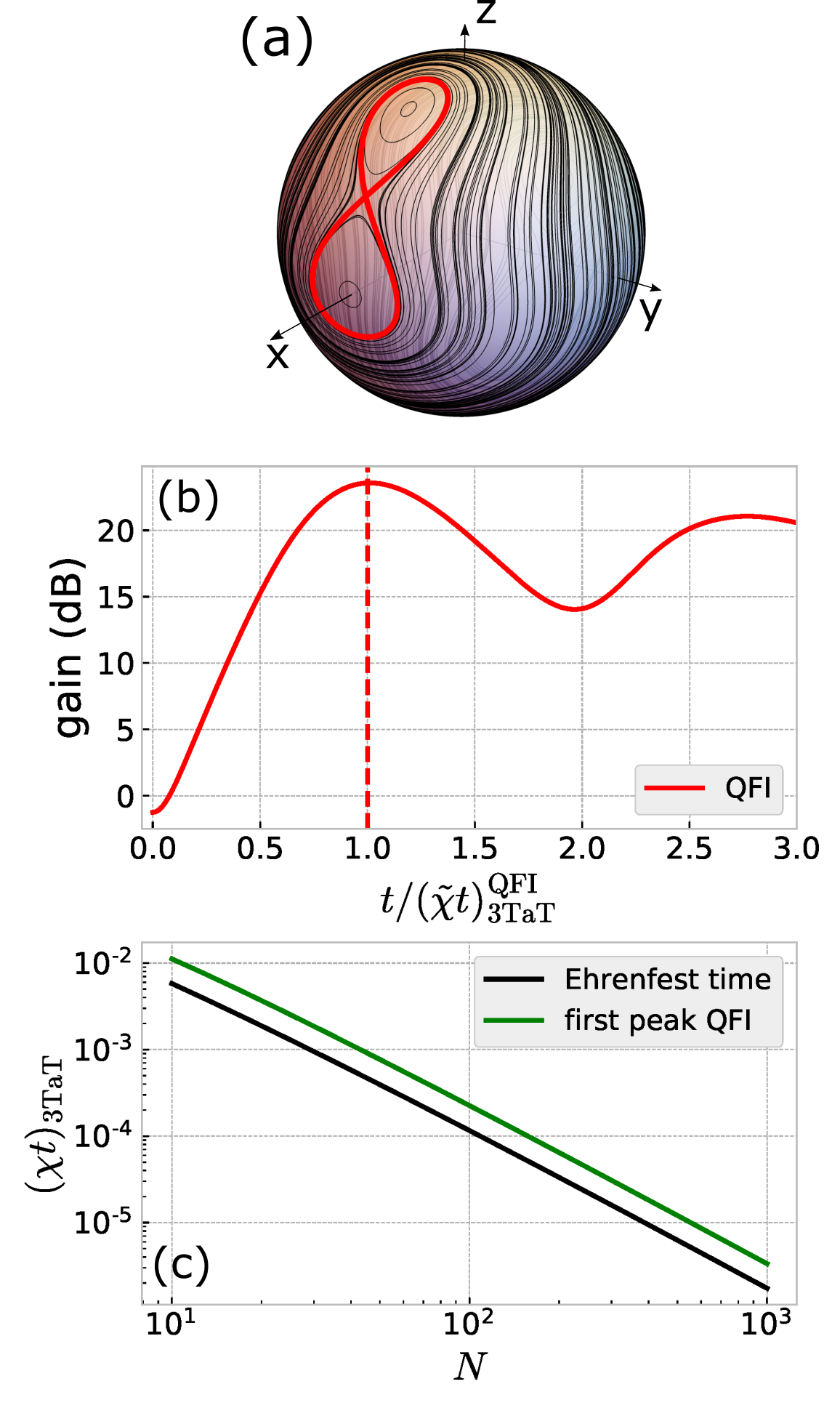}}
\caption{\textbf{(a)} Phase portrait of the phase space flow of the $3$TaT Hamiltonian at critical coupling. The separatrix is shown in solid red. \textbf{(b)} Metrological gain based the QFI (solid red), at critical coupling, for an initial SCS centered at the position of the saddle point. The dashed line show our result for the time to the first peak. Results are for a system with $N = 1024$. \textbf{(c)} Time scale to first peak of QFI with $3$TaT at critical coupling, contrasted against the Ehrenfest time.}
\label{fig:3tat_gain}
\end{figure}

\subsection{Time scale to first peak of QFI with the \texorpdfstring{$3$}{\textit{3}}TaT}
Our second goal in this section is to give an explicit expression for the time scale to the first peak of the QFI, for the dynamics with the $3$TaT. The fact that at critical coupling the semiclassical energy is a double well with wells of equal depth, indicates that one can prepare a ``cat-like'' state where the peaks of each lobe lied at the bottom of each of the wells, and the time scale to this state is estimated similarly as that of the TaT in Sec.~\ref{sec:all_twist_turn}. We show the separatrix of the $3$TaT at critical coupling in Fig.~\ref{fig:3tat_gain} (a).

Given that the position of the saddle at critical coupling is $Z_{\rm sd} = \frac{1}{2}$,  we consider an initial SCS centered at the saddle, that is, $|\theta_0,\varphi_0\rangle = |\frac{\pi}{3},0\rangle$. Since the separatrix is extended in the $z$-direction, the cat-like state being prepared will be extend in this direction as well. We then consider $\hat{G} = \hat{J}_z$, and will look at the QFI $F_{Q}[|\psi\rangle, \hat{J}_z]$. We then approximate the desired time scale as the traveling time of a point starting at the border of the SCS uncertainty patch and moving, along the separatrix, all the way to its end. This procedure can be tracked analytically, but we include details in Appendix \ref{app:ptat_stuff}. The initial point $Z(0)$ turns out to be 
\begin{equation}
Z(0) = \frac{1}{2} + \frac{1}{2\sqrt{\left(1 + \frac{1}{3}\left(\cos^{-1}\left( \sqrt{3/5}\right)\right)^{-2} \right)N}},
\end{equation}
and the end point is given by one of the two points at the intersection of the great circle in the $x$-$z$ plane, and the separatrix, which have the same energy as the saddle point. This leads to $Z(t_f) \approx 0.96$. The desired time scale is then given by
\begin{equation}
\label{eqn:3tat_time_to_cat}
(\tilde{\chi}t)_{3{\rm TaT}}^{\rm QFI} = \int_{Z(0)}^{Z(t_f)} \frac{12 dZ}{\sqrt{27(1-Z^2) - (5 - 4Z^3)^2}}.
\end{equation}
We compare this time scale with the respective Ehrenfest time in Fig.~\ref{fig:3tat_gain} (c). In this case the saddle point Lyapunov exponent at critical coupling is given by 
\begin{equation}
\left.\Lambda_{\rm sd}^{3{\rm TaT}}\right|_{\rm CC} = \frac{\sqrt{7}}{4}\tilde{\chi},
\end{equation}
and thus, following from Eq.~(\ref{eqn:ehrenfest_saddle}), the Ehrenfest time for the $3$TaT at critical coupling is given by
\begin{equation}
(\tilde{\chi} t)^{\rm Erfst}_{3{\rm TaT}} = \frac{2}{\sqrt{7}}\ln[N].
\end{equation}
The time to the first peak of QFI and the Ehrenfest time for the $3$TAT at critical coupling are compared in Fig.~\ref{fig:3tat_gain} (c). We immediately see that $(\tilde{\chi}t)_{3{\rm TAT}}^{\rm QFI} > (\tilde{\chi} t)_{3\rm{TaT}}^{\rm Erfst}$ for all values of $N$. As such , one would have expected a semiclassical analysis to give inaccurate predictions. However, we see this is not the case, pointing at a deeper connection between the geometry of separatrices and the generation of nonclassical states.

\section{Summary and outlook}
\label{sec:outlook}
In this work we have studied the preparation of nonclassical states for use in quantum metrology with collective spin systems based on the geometry of classical phase space. This picture is completely general for states generated by an arbitrary collective spin Hamiltonian. Using this, we have framed the quantum dynamics problem of state preparation as a problem involving the geometry of curves on the surface of the unit sphere. We have established a notion of local optimality for state preparation, which is given by the existence of a saddle point together with the orthogonality of separatrix branches in its vicinity, giving a geometrical interpretation to a pure squeezing Hamiltonian. Correspondingly, this framework also leads to a natural notion of global optimality, given by separatrix branches which are aligned with geodesics on the phase space, which for collective spins correspond to great circles on the unit sphere.

With these geometric tools in hand, we analyzed two paradigmatic examples of collective spin Hamiltonians that generate metrologically relevant quantum states at an exponential rate: the two-axis counter-twisting, and the twisting and turning Hamiltonians. We gave a geometrical interpretation to the parameter regime known as ``critical coupling'' in the TaT Hamiltonian composed of four main properties, and proved the local optimality of this parameter regime. Furthermore, we extended the TaT Hamiltonian to a large family of models, the $p$-twisting and turning Hamiltonians, by allowing ourselves to consider arbitrary $p$-body collective interaction terms. We proved a no-go theorem for the local optimality of $p$TaT dynamics with $p>2$, and showed that the only model which still admits a notion of critical coupling involving more than one of its geometric properties is the $3$TaT. 

It is thus the geometry of the classical separatrix that controls these exponentially short time scales. We focused on initial spin coherent states centered at the saddle points, but our conclusions are general, in the sense that \emph{any} initial spin coherent state with considerable overlap with the separatrix line, will have an exponentially short evolution time to a metrologically relevant state. The physical reason behind this fact can be traced back to the structure of the eigenstates which in the classical limit correspond to trajectories within a small neighborhood of the sepatarix~\footnote{One can define this neighborhood by taking a small energy window centered at the separatrix energy.}. It is known that these states live at the border of the principle of correspondence and behave as WKB states~\cite{Schubert2012wave}, thus leading to a dynamics of wave packed spreading which inherits many of the characteristics of the underlying classical dynamics. Furthermore, it was noted that these states lead to ``saddle point scrambling''~\cite{Xu2020,Kidd2021}, a process where out-of-time-order correlators grow exponentially but the system does not equilibrate, pointing at an interesting connection between this type of scrambling and metrological advantage.



The analysis done here is not only restricted to the standard approach to local quantum metrology involving state preparation, sensing, and then measurement. In fact, it can be readily applied to other quantum metrology protocols.  For example, consider the spin amplification protocol studied in~\cite{Koppenhofer2021}. The key mechanism behind the amplification procedure in this protocol can be understood by considering the phase space flow associated with collective dissipative dynamics, that is, by Lindbladian dynamics with the jump operator $\hat{L} = \sqrt{\Gamma} \hat{J}_-$. The motion described by its phase space flow is that of an overdamped pendulum (see for instance~\cite{Grobe1987}), given by the classical equations
\begin{equation}
\label{eqn:overdamped_pendulum}
\frac{d\theta}{dt} = k\sin(\theta), \enspace \frac{d\phi}{dt} = 0
\end{equation}
with $k$ some positive number, and $(\theta, \phi)$ are the two angular coordinates on the unit sphere. As discussed in~\cite{Koppenhofer2021}, one has $k = \Gamma$. Notice that canonical equations of motion for the 2ACT in Eq.~(\ref{eqn:canonical_eqs_ct}) map exactly to those in Eq.~(\ref{eqn:overdamped_pendulum}) when we take $k=\tilde{\chi}$, and write in the angular variables and constrained to the separatrices. In other words, the separatrices of the 2ACT are spin amplifiers, and the protocol in Ref.~~\cite{Koppenhofer2021} can be executed in a fully coherent manner by replacing the collective dissipation with the 2ACT dynamics. A more detailed study of this latter fact is the object of ongoing work.

Beyond its application to spin amplification, the present analysis can be used in other types of quantum metrology protocols, where one expect the separatrix line to play a role. For instance, in certain tasks of critical metrology, or in situations where one is interested in refining the knowledge about some Hamiltonian parameter. In the latter case, the best strategy will always be to chose an initial spin coherent state placed on top of the separatrix. Furthermore, some types of dynamical sensors, as for instance Floquet time crystals~\cite{Munoz2022}, or chaotic sensors~\cite{Fiderer2018}. There, the idea of a separatrix generalizes to that of a classical border between regions of distinct macroscopic motion, and spin coherent states placed at these borders are known to provide the best strategy~\cite{Fiderer2018}.

\acknowledgements
The authors are grateful to Tyler J. Volkoff and Jason Twamley for engaging discussions. This work was supported by NSF Grant No. PHY-1606989, and Quantum Leap Challenge Institutes program, Award No. 2016244. In the last stages of the manuscript conception, MHMA was supported in part via funding from NSERC, the Canada First Research Excellence Fund, and the Ministère de l’Économie et de l’Innovation du Québec.

\appendix
\section{Derivations of some results with the two-axis counter-twisting Hamiltonian}
\label{app:ct_stuff}
\subsection{Derivation of phase space flow, fixed points and their stability}
\label{subsec:flow_details}
In order to compute the phase space flow, we need to introduce a rescaled version of Eq.~(\ref{eqn:ct_hamil_2}), so that its energy will be extensive in the thermodynamic limit. This can be achieved by introducing the rescaled counter-twisting strength $\chi = \frac{\tilde{\chi}}{N}$, then 
\begin{equation}
\label{eqn:ct_hamil_3}
\hat{H}_{\rm CT} = \frac{\tilde{\chi}}{N}(\hat{J}_x\hat{J}_y + \hat{J}_y\hat{J}_x).
\end{equation}
Using the Hamiltonian in Eq.~(\ref{eqn:ct_hamil_3}) we can write the Heisenberg equations of motion, $\frac{d\hat{J}_\gamma}{dt} = i[\hat{H}, \hat{J}_\gamma]$ with $\gamma = x,y,z$, for the components of the collective spin. They are 
\begin{subequations}
\label{eqn:heisenberg_ct}
\begin{align}
\frac{d\hat{J}_x}{dt} &= \frac{\tilde{\chi}}{N}\left(\hat{J}_x\hat{J}_z + \hat{J}_z\hat{J}_x\right), \\
\frac{d\hat{J}_x}{dt} &= -\frac{\tilde{\chi}}{N}\left(\hat{J}_y\hat{J}_z + \hat{J}_z\hat{J}_y\right), \\
\frac{d\hat{J}_z}{dt} &= -\frac{2\tilde{\chi}}{N}\left(\hat{J}_x^2 - \hat{J}_y^2\right).
\end{align}
\end{subequations}
In the thermodynamic limit, $J\to\infty$, Eq.~(\ref{eqn:heisenberg_ct}) lead to the phase space flow of the classical variables $\mathbf{X} = \frac{\langle \mathbf{\hat{J}}\rangle}{J}$, which after neglecting correlations $\langle \hat{A}\hat{B} \rangle = \langle\hat{A} \rangle \langle \hat{B}\rangle$, is given by 
\begin{subequations}
\label{eqn:flow_ct_app}
\begin{align}
\frac{dX}{dt} &= -\tilde{\chi}XZ, \\
\frac{dY}{dt} &= -\tilde{\chi}YZ, \\
\frac{dZ}{dt} &= -\tilde{\chi}(X^2 - Y^2). 
\end{align}
\end{subequations}
Fixed points of a phase space flow are those initial conditions with a trivial evolution, that is, solutions of $\frac{d\mathbf{X}}{dt} = 0$. The phase space flow in Eq.~(\ref{eqn:flow_ct}) has six different fixed points given by 
\begin{subequations}
\label{eqn:fixed_points_ct_app}
\begin{align}
(X,Y,Z) &= (0,0,\pm1), \\
(X,Y,Z) &= (\frac{1}{\sqrt{2}},\mp \frac{1}{\sqrt{2}},0), \\
(X,Y,Z) &= (-\frac{1}{\sqrt{2}}, \pm\frac{1}{\sqrt{2}},0).
\end{align}
\end{subequations}
Thus, the phase space flow has fixed points in the north and south poles of the unit sphere, and in the poles of each of the twisting axis.

The stability of fixed points can be analyzed using the eigenvalues of the Jacobi matrix, $\mathbb{M}[\mathbf{X}] = \frac{\partial}{\partial\mathbf X} \frac{d\mathbf{X}}{dt}$, evaluated at the fixed point. For the phase space flow in Eq.~(\ref{eqn:flow_ct}) this matrix is given by 
\begin{equation}
\label{eqn:tangent_map_ct_app}
\mathbb{M}[\mathbf{X}] = 
\begin{pmatrix}
\tilde{\chi}Z && 0 && \tilde{\chi}X \\ 
0 && -\tilde{\chi}Z && -\tilde{\chi}Y \\
-4\tilde{\chi}X && 4\tilde{\chi}Y && 0
\end{pmatrix}.
\end{equation}
When evaluated at the fixed point $(X,Y,Z) = (0,0,\pm1)$, Eq.~(\ref{eqn:tangent_map_ct_app}) is diagonal, with diagonal equal to ${\rm Diag}[\mathbb{M}] = (\pm\tilde{\chi}, \mp\tilde{\chi},0)$. Thus the two eigenvalues are real and equal to $\mathcal{M}_\pm = \pm\tilde{\chi}$. This implies the fixed point is a saddle point, and that the principal directions of the separatrix emerging from it are orthogonal and aligned with the $x$- and $y$-axis, respectively. In other words, they define great circles in the $x-z$ and $y-z$ planes, respectively. Furthermore, we know that in the vicinity of the saddle point, an initial condition on one of the separatrix branches evolves according to $\frac{dP_{\pm}}{dt} = \pm\tilde{\chi}P_\pm$, where $P_\pm = X,Y$.

The other four fixed points in Eq.~(\ref{eqn:fixed_points_ct_app}) are stable centers. This can be easily verified by looking at the eigenvalues of the Jacobi matrix evaluated at the fixed points, which are given by $\mathcal{M}_{\rm} = \pm i2\tilde{\chi}$. 

\subsection{Classical energy and separatrix}
Using the form of the 2ACT Hamiltonian in Eq.~(\ref{eqn:ct_hamil_3}) we can write down an expression of the classical energy by taking the thermodynamic limit of $\frac{\langle\hat{H}_{\rm CT}\rangle}{J}$, then after neglecting correlation and in terms of the classical variables $\mathbf{X}$, the classical energy density reads 
\begin{equation}
\label{eqn:classical_energy_ct_1}
E(X,Y;\tilde{\chi}) = \frac{\langle\hat{H}_{\rm CT}\rangle}{J} = \tilde{\chi} X Y.
\end{equation}
Introducing spherical coordinates $(X,Y,Z) = (\sin(\theta)\cos(\phi), \sin(\theta)\sin(\phi), \cos{\theta})$, and noticing that the variables $\phi$ and $Z$ are classical conjugate variables playing the role of ``position" and ``momentum", respectively. We can write Eq.~(\ref{eqn:classical_energy_ct_1}) as 
\begin{equation}
\label{eqn:classical_energy_ct_2}
E(Z,\phi;\tilde{\chi}) = \frac{\tilde{\chi}}{2}(1-Z^2)\sin(2\phi).
\end{equation}
We can use the expression of the classical energy in Eq.~(\ref{eqn:classical_energy_ct_2}) to compute an equation for the classical separatrix and subsequently, the time required for points to travel along sections of this curve.

First of all, recall that the separatrix connects the two saddles at $(0,0,\pm1)$, thus, conservation of energy say that any point along the curve has energy $E_{\rm sepa} = 0$. Thus, the separatrix equation is $E(Z,\phi;\tilde{\chi}) = 0$, this is true if $Z=\pm1$, an also for the range of allowed values of $\phi\in[0,2\pi]$, then the separatrix is define by all the pairs $(Z,\phi)$ with $\phi =0,\frac{\pi}{2}$, defining the two great circles on the $x$-$z$ plane and the $y$-$z$ plane, respectively.

The equations of motion for the classical conjugate variables are given by 
\begin{subequations}
\label{eqn:canonical_eqs_ct}
\begin{align}
\frac{dZ}{dt} &= -\frac{\partial E}{\partial \phi} = -\tilde{\chi}(1-Z^2)\cos(2\phi), \\
\frac{d\phi}{dt} &= \frac{\partial E}{\partial Z} = -\tilde{\chi}Z\sin(2\phi).
\end{align}
\end{subequations}
For the saddle at $(X,Y,Z) = (0,0,1)$ the $x$-direction is the unstable manifold of the separatrix, thus motion of points along the separatrix satisfy $\phi(t) = 0$ for all times. Constrained to the separatrix, then we have 
\begin{equation}
\frac{dZ}{dt} = -\tilde{\chi}(1-Z^2),    
\end{equation}
and the desired time scale is given by 
\begin{equation}
\label{eqn:int_time_scale_ct}
\tilde{\chi} t = -\int_{Z(0)}^{Z(t_f)}\frac{dZ}{1-Z^2}.    
\end{equation}
After solving this integral, and recalling that $Z(0) = \sqrt{1-\frac{1}{N}}$, as explained in the main text, we obtain
\begin{equation}
\tilde{\chi}t = \ln\left[\frac{(1-Z(t_f))(\sqrt{N} + \sqrt{N-1})}{\sqrt{1 - Z^2(t_f)}} \right]
\end{equation}
which is the same as Eq.~(\ref{eqn:time_scale_ct}) in the main text.

\subsection{Explicit expressions for the spin squeezing parameter and quantum Fisher information}
\label{subsec:explicit_ct}
Our starting point in this section is the fact that the solution to the integral in Eq.~(\ref{eqn:int_time_scale_ct}) allows us to write a explicit expression for $Z(t)$. It is given by 
\begin{equation}
\label{eqn:z_evol_ct}
Z(t) = \frac{-\sinh(\tilde{\chi}t) + Z(0)\cosh(\tilde{\chi}t)}{\cosh(\tilde{\chi}t) - Z(0)\sinh(\tilde{\chi}t)} = \frac{Z(0) - \tanh(\tilde{\chi}t)}{1 -Z(0)\tanh(\tilde{\chi}t)}.
\end{equation}
With this expression at hand we can go forward and write explicit expressions for the squeezing parameter and quantum Fisher information. 
\subsubsection{Explicit expression for the squeezing parameter}
Let us write Eq.~(\ref{eqn:squeezing_ct}) of the main text in terms of $Z$ instead of the polar angle $\theta$. It is given by 
\begin{equation}
\label{eqn:squeezing_z_ct}
\xi^2_{\rm CT} (Z) = \frac{1}{NZ^2(t)(1-Z^2(t))}.
\end{equation}
Then by plugging Eq.~(\ref{eqn:z_evol_ct}) into Eq.~(\ref{eqn:squeezing_z_ct}) we obtain 
\begin{equation}
\xi^2_{\rm CT} = \frac{(1-Z(0)\tanh(\tilde{\chi}t))^4}{N(1-Z^2(0))(1-\tanh^2(\tilde{\chi}t))(Z(0) - \tanh(\tilde{\chi}t))^2},
\end{equation}
recalling that our initial state of interest provides $Z(0) = \sqrt{1-\frac{1}{N}}$, and thus $N(1-Z^2(0))=1$. Plugging these into the last expression we obtain the explicit form of the squeezing parameter under evolution with the two-axis counter-twisting
\begin{equation}
\label{eqn:explicit_squeezing_ct}
\xi^2_{\rm CT} = \frac{\left[ \sqrt{N} -\sqrt{N-1}\tanh(\tilde{\chi}t) \right]^4}{N(1-\tanh(\tilde{\chi}t))\left[\sqrt{N-1} - \sqrt{N}\tanh(\tilde{\chi}t)\right]^2}.
\end{equation}

\subsubsection{Explicit expression for the quantum Fisher information}
Recall that the metrological gain based on the QFI is defined as 
\begin{equation}
\zeta^2 = \frac{N}{F_{Q}}.
\end{equation}
Furthermore recall that the variance in the $x$-direction goes as $\Delta J_x = J\sin(\theta)$. Thus, after taking $\hat{J}_x$ as generator, we can write the QFI as 
\begin{equation}
\label{eqn:QFI_ct_1}
F_{\rm Q} = N^2 \sin^2(\theta) = N^2(1-Z^2(t)),
\end{equation}
after plugging Eq.~(\ref{eqn:z_evol_ct}) into Eq.~(\ref{eqn:QFI_ct_1}) we obtain 
\begin{equation}
\label{eqn:QFI_ct_2}
F_{\rm Q} = N^2\frac{(1-\tanh^2(\tilde{\chi}t))(1-Z(0)^2)}{(1-Z(0)\tanh(\tilde{\chi}t))^2},
\end{equation}
recalling that in our setting $Z(0) = \sqrt{1-\frac{1}{N}}$, then we can write 
\begin{equation}
\label{eqn:QFI_ct_3}
F_{\rm Q} = N^2\frac{1-\tanh^2(\tilde{\chi})}{(\sqrt{N} - \sqrt{N-1}\tanh(\tilde{\chi}t))^2},
\end{equation}
this last expression allows us to write the classical estimate for the metrological gain based on the QFI as 
\begin{equation}
\zeta^2(t)_{\rm CT} = \frac{(\sqrt{N} - \sqrt{N-1}\tanh(\tilde{\chi}t))^2}{N(1-\tanh^2(\tilde{\chi}))}
\end{equation}

\subsection{Some other quantum states relevant for metrology accessible with the two-axis counter-twisting}
\label{subsec:other_states_ct}
In this section we illustrate how the time scales for to the BWS, the EWSS and YUS are computed using the traveling time for motion of points along sections of the separatrix. 

This calculation is based on the values of the variances of the states of interest. In order to obtain them we need a explicit expression for these states. Let us start with the Berry-Wiseman state. Initially introduced by Berry and Wiseman~\cite{Berry2000}, as the optimal state for phase encoding in an interferometric setting, it was latter considered by Combes and Wiseman~\cite{Combes2004} as the optimal state for phase estimation. It is given by
\begin{equation}
|{\rm BW}\rangle = \frac{1}{\sqrt{1 + N/2}}\sum_{M = -J}^{J}\cos\left[\frac{M\pi}{N + 2}\right]|J, M\rangle,    
\end{equation}
with variance, for $N \gg 1$, equal to $(\Delta J_z)^2 \approx 0.13 J^2$.

The EWSS is the pure state version of the maximally mixed state, that is, a superposition of all Dicke states with equal weight
\begin{equation}
|{\rm EWSS}\rangle = \frac{1}{\sqrt{N + 1}}\sum_{M = -J}^J |J, M\rangle,
\end{equation}
with a variance equal to $(\Delta J_z)^2 = \frac{J(J+1)}{3}$.

The family of Yurke states
\begin{equation}
|{\rm Yu}\rangle = \frac{\sin(\alpha)}{\sqrt{2}}|J, 1\rangle + \cos(\alpha)|J,0\rangle + \frac{\sin(\alpha)}{\sqrt{2}}|J,-1\rangle,
\end{equation}
with a variance equal to 
\begin{equation}
(\Delta J_z)^2 = \frac{J}{4}\left[(J+1)(2 - \sin^2(\alpha)) - 2\sin^2(\alpha)\right],
\end{equation}
We notice, however, that only Yurke states with relatively small values of $\alpha$ are prepared by the 2ACT dynamics.

Now that we know the structure of the states in the Dicke basis and their respective values of the variances, let us illustrate how the time scales are calculated. 

\subsubsection{Time scale to Berry-Wiseman state}
In the limit of large $N\gg1$, the variance of the state goes as $\sim 0.13J^2$, thus, the two values $\Delta J^{\rm low} = \frac{J}{\sqrt{8}}$ an $\Delta J^{\rm up} = \frac{J}{\sqrt{7}}$, are good lower and upper bounds, respectively. From the above observation we have
\begin{equation}
\sin(\theta^{\rm low}) = \frac{1}{\sqrt{8}}, \quad \sin(\theta^{\rm up}) = \frac{1}{\sqrt{7}}.
\end{equation}
Hence, we get the values of $Z(t_f)$ for the time scale to a BWS as 
\begin{equation}
Z^{\rm low}(t_f) = \sqrt{\frac{7}{8}}, \quad Z^{\rm up}(t_f) = \sqrt{\frac{6}{7}},
\end{equation}
giving the following lower and upper bounds for the time scale to the peak of the fidelity to the BWS state
\begin{subequations}
\begin{align}
(\tilde{\chi} t)^{\rm low}_{\rm CT} &= \ln\left[(\sqrt{8} - \sqrt{7})(\sqrt{N} + \sqrt{N-1})\right], \\
(\tilde{\chi} t)^{\rm up}_{\rm CT} &=  \ln\left[(\sqrt{7} - \sqrt{6})(\sqrt{N} + \sqrt{N-1})\right],
\end{align}
\end{subequations}
which are the same as Eq.~(\ref{eqn:time_BWS}) in the main text.

\subsubsection{Time scale to equally weighted superposition state}
The variance of this state is $\Delta J = \sqrt{J(J+1)}/3$. Thus, we have that at the time at which the fidelity to this state peaks, $\theta$ is given by 
\begin{equation}
\sin(\theta) = \frac{1}{\sqrt{3}}\sqrt{1 + \frac{1}{J}},
\end{equation}
giving the value of $Z(t_f)$
\begin{equation}
Z(t_f) = \cos(\theta) = \frac{\sqrt{2J - 1}}{\sqrt{3J}}.
\end{equation}
From here it is easy to get the time scale as
\begin{subequations}
\begin{align}
(\tilde{\chi} t)_{\rm CT}^{\rm EWSS} &= \ln\left[\left(\frac{\sqrt{3} - \sqrt{2 - \frac{2}{N}}}{\sqrt{1 + \frac{2}{N}}}\right)(\sqrt{N} + \sqrt{N - 1})\right] \\ 
&= \ln\left[(\sqrt{3} - \sqrt{2})(\sqrt{N} + \sqrt{N - 1})\right],
\end{align}
\end{subequations}
which is the same Eq.~(\ref{eqn:time_EWSS}) in the main text.

\subsubsection{Time scale to some Yurke states}
Using the expression for the variance we get
\begin{equation}
\sin(\theta) = \frac{1}{2}\sqrt{2-\sin^2(\alpha) + \frac{4 - 6\sin^2(\alpha)}{N}},
\end{equation}
from where we obtain $Z(t_f)$ as 
\begin{equation}
Z(t_f) = \cos(\theta) = \frac{1}{2}\sqrt{2 + \sin^2(\alpha) - \frac{4 - 6\sin^2(\alpha)}{N}},
\end{equation}
leading to the following expression for the time scale to Yurke states
\begin{widetext}
\begin{equation}
\label{eqn:time_yurke_full}
(\tilde{\chi} t)_{\rm CT}^{\rm YUS} = \ln\left[\left( \frac{2 - \sqrt{2 + \sin^2(\alpha) - \frac{4 - 6\sin^2(\alpha)}{N}}}{ \sqrt{2-\sin^2(\alpha) + \frac{4 - 6\sin^2(\alpha)}{N}} } \right) (\sqrt{N} + \sqrt{N - 1})\right],
\end{equation}
\end{widetext}
where Eq.~(\ref{eqn:time_yurke}) is recovered from Eq.~(\ref{eqn:time_yurke_full}) in the limit $N\to\infty$.

\section{Derivation of some results with the Twisting and Turning Hamiltonian at critical coupling}
\label{app:tat_stuff}

\subsection{Derivation of the phase space flow, fixed points and their stability}
In order to construct the phase space flow in the thermodynamic limit, we need to rescale the TaT Hamiltonian in Eq.~(\ref{eqn:tat_hamil}) so that in the thermodynamic limit, we can guarantee the extensivity of the energy density. This is achieved by introducing the normalized twisting strength $\chi = \frac{\tilde{\chi}}{N}$, then we write 
\begin{equation}
\label{eqn:tat_hamil_2}
\hat{H}_{\rm TaT} = \Omega \hat{J}_x + \frac{\tilde{\chi}}{N}\hat{J}_z^2.
\end{equation}
With this Hamiltonian we write the Heisenberg equations for the components of the collective spin, they are given by 
\begin{subequations}
\label{eqn:heisenberg_tat}
\begin{align}
\frac{d\hat{J}_x}{dt} &= -\frac{\tilde{\chi}}{N}(\hat{J}_y\hat{J}_z + \hat{J}_z\hat{J}_y), \\
\frac{d\hat{J}_y}{dt} &= -\Omega \hat{J}_z - \frac{\tilde{\chi}}{N}(\hat{J}_x\hat{J}_z + \hat{J}_z\hat{J}_x), \\
\frac{d\hat{J}_z}{dt} &= \Omega\hat{J}_y.   
\end{align}
\end{subequations}

In the thermodynamic limit, $J\to\infty$, Eq.~(\ref{eqn:heisenberg_tat}) lead to the phase space flow of the classical variables $\mathbf{X} = \frac{\langle \mathbf{\hat{J}}\rangle}{J}$, which after neglecting correlations $\langle \hat{A}\hat{B} \rangle = \langle\hat{A} \rangle \langle \hat{B}\rangle$, is given by
\begin{subequations}
\label{eqn:flow_tat_app}
\begin{align}
\frac{dX}{dt} &= \tilde{\chi}YZ, \\
\frac{dY}{dt} &= -\Omega Z - \tilde{\chi}XZ, \\
\frac{dX}{dt} &= \Omega Y.
\end{align}
\end{subequations}
The fixed points of the phase space flow in Eq.~(\ref{eqn:flow_tat_app}), which are solutions of $\frac{dX}{dt}=0$, define two different type of sets, depending on whether $\frac{\Omega}{\tilde{\chi}}>1$ or $\frac{\Omega}{\tilde{\chi}}<1$. In the case of the former there are only two fixed points at $(X,Y,Z) = (\pm1,0,0)$, \textit{i.e.}, the poles with respect to the turning axis, and phase space is filed with trajectories representing Larmor precessions of the mean spin. In the case of the latter parameter regime, there are four fixed points two given by $(X,Y,Z) = (\pm1,0,0)$, where the one at $X = -1$ is stable and the one at $X = 1$ is a saddle. The two additional fixed points are at 
\begin{equation}
(X,Y,Z) = \left(\frac{\Omega}{\tilde{\chi}},0,\pm\sqrt{1-\left(\frac{\Omega}{\tilde{\chi}}\right)^2}\right),
\end{equation}
The two parameter regimes are connected through a pitchfork bifurcation of the stable point at $(X,Y,Z) = (1,0,0)$ happening at $\frac{\Omega}{\tilde{\chi}}=1$.

The saddle point at $(X,Y,Z) = (1,0,0)$ existing for parameters such that $\frac{\Omega}{\tilde{\chi}}<1$, is of key importance for our analysis. We know that separatrix branches emerge from it and conservation of energy guarantees that all points on the separatrix have the same energy as the saddle. Furthermore, by evaluating the eigenvalues of the Jacobi matrix at this saddle, one obtains its Lyapunov exponent as 
\begin{equation}
\label{eqn:lyap_tat_app}
\Lambda^{\rm TaT}_{\rm sd} = \tilde{\chi} \sqrt{\frac{\Omega}{\tilde{\chi}}\left(1 - \frac{\Omega}{\tilde{\chi}} \right)}.
\end{equation}
Importantly, the Lyapunov exponent in Eq.~(\ref{eqn:lyap_tat}) has, as a function of the ratio $\Omega/\tilde{\chi}$ a maximum when $\frac{\Omega}{\tilde{\chi}}=\frac{1}{2}$, parameter regime which defines critical coupling.

\subsection{Classical energy and separatrix}
Similarly to how we obtained the classical energy density of the 2ACT Hamiltonian, we can obtain the energy density of the TaT Hamiltonian by computing $\frac{\langle\hat{H}_{\rm TaT}\rangle}{J}$ for Eq.~(\ref{eqn:tat_hamil_2}) in the thermodynamic limit. After neglecting correlations and in terms of the classical variables $\mathbf{X}$, this energy density reads
\begin{equation}
\label{eqn:tat_energy_1}
E(X,Z;\Omega, \tilde{\chi}) = \frac{\langle\hat{H}_{\rm TaT}\rangle}{J} = \Omega X + \frac{\tilde{\chi}}{2}Z^2,
\end{equation}
which we can write in terms of the classical conjugated variables $Z$ and $\phi$ as 
\begin{equation}
\label{eqn:tat_energy_2}
E(Z,\phi;\Omega, \tilde{\chi}) = \Omega \sqrt{1-Z^2}\cos(\phi) + \frac{\tilde{\chi}}{2}Z^2.
\end{equation}
The separatrix equation can be constructed by noticing that conservation of energy guarantees any point on the separatrix to have the same energy as the energy of the saddle, which is given by $E(0,0;\tilde{\chi},\Omega) = \Omega$, and thus the separatrix is defined by $E(Z,\phi;\tilde{\chi},\Omega) = \Omega$. We can use this last expression to write the following explicit equation for the separatrix line of the TaT phase space flow
\begin{equation}
\label{eqn:tat_separatrix}
\cos(\phi) = \frac{1 - \frac{\cos^2(\theta)}{\omega}}{\sin(\theta)},
\end{equation}
where we have explicitly written $Z = \cos(\theta)$, and have defined $\omega = \frac{2\Omega}{\tilde{\chi}}$.

\subsubsection{Angle between separatrix branches}
Let us now use Eq.~(\ref{eqn:tat_separatrix}) to derive the result in Eq.~(\ref{eqn:angle_separatrix_tat}) of the main text. The first step is to write a parametric equation for the separatrix. Equation~(\ref{eqn:tat_separatrix}) allows us to write $X$ and $Y$ in terms of $Z$ as 
\begin{equation}
X = 1 - \frac{\chi}{2\Omega}Z^2, \enspace Y = \pm Z\sqrt{(\frac{\chi}{\Omega} - 1) - (\frac{\chi}{2\Omega})^2 Z^2},    
\end{equation}
then the parametric equation for the separatrix is given by 
\begin{equation}
\label{eqn:tat_separatrix_para_1}
\vec{S}_{\pm}(Z;\tilde{\chi}, \Omega) = \left( 1 - \frac{\chi}{2\Omega}Z^2,\enspace \pm Z\sqrt{(\frac{\chi}{\Omega} - 1) - (\frac{\chi}{2\Omega})^2 Z^2},\enspace Z \right),
\end{equation}
where $\pm$ denotes the two different separatrix branches. From Eq.~(\ref{eqn:tat_separatrix_para_1}) we take its projection onto the $y$-$z$ plane, given by 
\begin{equation}
\label{eqn:tat_separatrix_para_2}
\vec{S}^{xy}_{\pm}(Z;\tilde{\chi}, \Omega) = \left( 1 - \frac{\chi}{2\Omega}Z^2,\enspace \pm Z\sqrt{(\frac{\chi}{\Omega} - 1) - (\frac{\chi}{2\Omega})^2 Z^2} \right),
\end{equation}
from this last expression the angle between the $(+)$ and $(-)$ separatrix branches can be readily computed, giving
\begin{equation}
\label{eqn:angle_separatrix_tat_full}
\cos(\upsilon) = \frac{\vec{S}_+^{(yz)} . \vec{S}_-^{(yz)}}{|\vec{S}_+^{(yz)}| |\vec{S}_-^{(yz)}|} = \frac{2 - \frac{\chi}{\Omega} + (\frac{\chi}{2\Omega})^2 Z^2}{\frac{\chi}{\Omega} - (\frac{\chi}{2\Omega})^2 Z^2},
\end{equation}
which recovers Eq.~(\ref{eqn:angle_separatrix_tat}) in the main text.

\subsubsection{Effective 2ATC Hamiltonian at critical coupling}
Let us now turn our attention to the derivation of the effective 2ACT Hamiltonian for the TaT Hamiltonian. Our starting point is to write the TaT Hamiltonian in Eq.~(\ref{eqn:tat_hamil_2}) using the new collective operators defined in Eq.~(\ref{eqn:new_opes}). After this procedure we find
\begin{equation}
\label{eqn:tat_hamil_eff_app_1}
\hat{H}_{\rm TaT}^{\rm eff} = \Omega \hat{J}_x + \frac{\tilde{\chi}}{2N}\left(\hat{J}_1\hat{J}_2 + \hat{J}_2  \hat{J}_1\right) + \frac{\tilde{\chi}}{2N}\left(J^2 - \hat{J}_x^2\right),
\end{equation}
we then write $\hat{J}_x = J - \hat{o}$, that is, its mean-field minus fluctuations, and keep terms up to linear in the fluctuations $\hat{o}$, to obtain 
\begin{equation}
\label{eqn:tat_hamil_eff_app_2}
\hat{H}_{\rm TaT}^{\rm eff} = \Omega J - \left(\Omega - \frac{\tilde{\chi}}{2}\right)\hat{o} + \frac{\tilde{\chi}}{2N}\left(\hat{J}_1\hat{J}_2 + \hat{J}_2  \hat{J_1}\right),
\end{equation}
this last expression recovers, at critical coupling $\tilde{\chi} = 2\Omega$, Eq.~(\ref{eqn:tat_hamil_eff_2}) in the main text.

\subsubsection{Details of the computation of time scales at critical coupling}
Let us now look into the details of the calculation of the time scales to the peak spin squeezing and the first peak of the QFI for the TaT dynamics at critical coupling. Our starting point are the equation of motion for the classical conjugated variables $Z$ and $\phi$, they are 
\begin{subequations}
\label{eqn:tat_canonical_eqs}
\begin{align}
\frac{dZ}{dt} &= -\frac{\partial E}{\partial \phi} = \Omega \sqrt{1-Z^2}\sin(\phi),\\
\frac{d\phi}{dt} &= \frac{\partial E}{\partial Z} = -\frac{\Omega Z}{\sqrt{1-Z^2}}\cos(\phi) + \tilde{\chi}Z.
\end{align}
\end{subequations}
Noticing that at critical coupling Eq.~(\ref{eqn:tat_separatrix}) simplifies to $\cos(\phi) = \sin(\theta)$, and thus $\sin(\phi) = |\cos(\theta)|$, we can write the equation for $Z$ in Eq.~(\ref{eqn:tat_canonical_eqs}) constrained to motion along the separatrix as 
\begin{equation}
\frac{dZ}{dt} = \Omega Z \sqrt{1-Z^2} = \frac{\tilde{\chi}}{2} Z \sqrt{1-Z^2},    
\end{equation}
and thus the desired time scale is given by 
\begin{equation}
\label{eqn:tat_time_scale_int}
\tilde{\chi}t = 2\int_{Z(0)}^{Z(t_f)} \frac{dZ}{Z\sqrt{1-Z^2}},
\end{equation}
recalling that our initial state is a SCS along the positive $x$-direction, then $Z(0)\approx\frac{1}{\sqrt{2N}}$. Using this value, and after solving the integral we find 
\begin{equation}
\label{eqn:tat_time_scale_app}
\tilde{\chi}t = 2\ln\left[ \frac{Z(t_f)(\sqrt{2N} + \sqrt{2N-1})}{1 + \sqrt{1 - Z^2(t_f)}} \right].
\end{equation}

For the time scale to peak spin squeezing we will use the fact that, locally, the TaT separatrices at critical coupling must be aligned with great circles on the unit sphere. We then identify $Z(t_f)$ as the limit value for which this statement holds true. 

We then exploit the fundamental theorem of the local theory of curves~\cite{Do1976}, and compute the curvature $\kappa$ and torsion $\tau$ for both the 2ACT separatrices and the TaT separatrices at critical coupling. For a great circle in the unit sphere the values of these two quantities are 
\begin{equation}
\label{eqn:ct_curv_tor_app}
\kappa_{\rm CT} = 1, \enspace \tau_{\rm CT} = 0.
\end{equation}
To compute the curvature and torsion of the TaT separatrix we will consider the following parametric expression for the separatrix
\begin{equation}
\vec{S}_{+}(\theta) = \left(\sin^2(\theta),\enspace \sin(\theta)\cos(\theta),\enspace \cos(\theta)\right),
\end{equation}
where $\theta$ is the polar angle of spherical coordinates, and we only need to consider the positive branch. The curvature and torsion for this curve are given by 
\begin{subequations}
\label{eqn:tat_curv_tor}
\begin{align}
\kappa_{\rm TaT}(\theta) &= \frac{2\sqrt{13 - 3\cos(2\theta)}}{(3 - \cos(2\theta))^\frac{3}{2}}, \\
\tau_{\rm TaT}(\theta) &= \frac{3}{4} - \frac{12\sin(\theta)}{13 - 3\cos(2\theta)},
\end{align}
\end{subequations}
where we have translated the ``origin" to the position of the saddle point. We look for the value of $\theta$ in $[0,\pi/2]$ such that these two quantities differ from those of the CT dynamics at most by $1\%$. For the curvature we solve the equation
\begin{equation}
\label{eqn:solve_curv}
\kappa_{\rm TaT}(\theta) = 1.01,
\end{equation}
in the desired range, Eq.~(\ref{eqn:solve_curv}) has solution given by $\theta^* \approx 1.43772$, thus $Z^*(t_f) = \cos(\theta^*) \approx 0.132684$. At the same time $\theta^*$ gives a torsion $\tau_{\rm TaT}(\theta^*) \approx 10^{-3} \approx 0$. Which satisfy our requirement and provide the right hand side of Eq.~(\ref{eqn:z_squeezing_bound}) in the main text.

\subsubsection{Length difference between the separatrices of 2ACT and TaT at critical coupling}
For the separatrix length of the 2ACT dynamics we have 
\begin{equation}
l_{\rm CT} = \int_{\theta_0}^{\frac{\pi}{2}}d\theta\sqrt{\sin^2(\theta) + \cos^2(\theta)} = \frac{\pi}{2} - \theta_0,
\end{equation}
where $\theta_0 = \cos^{-1}\left(\sqrt{1 - \frac{1}{N}}\right)$. 

For the TaT at critical coupling, the separatrix length is given by 
\begin{equation}
l_{\rm TaT} = \int_{0}^{\theta_f} d\theta \sqrt{1 + \sin^2(\theta)} = {\rm E}(\theta_f,i),
\end{equation}
where $\theta_f = \cos~^{-1}\left(\frac{1}{\sqrt{2N}}\right)$, E$(\varphi,k)$ is the incomplete elliptic integral of the second kind, and $i$ the imaginary unit.

Finally in the limit $N\to\infty$, we find the difference in lengths to be  
\begin{equation*}
l_{\rm TaT} - l_{\rm CT} = {\rm E}(i) - \frac{\pi}{2}\approx 0.3393 \approx \frac{\ln(2)}{2},
\end{equation*}
which recovers Eq.~(\ref{eqn:length_diff}) in the main text.

\section{Details of the quantum speed limit section}
\label{app:speed_limit}
We introduced the Holstein-Primakoff approximation with respect to the positive $x$-direction, that is 
\begin{equation}
\label{eqn:hp_along_x}
\hat{J}_x = J - \hat{a}^\dagger\hat{a}, \enspace \hat{J}_y = \sqrt{J}\hat{q}, \enspace \hat{J}_y = \sqrt{J}\hat{p},    
\end{equation}
where $\hat{a}^\dagger$ ($\hat{a}$) are creation (annihilation) operators for a single bosonic mode, and $\hat{q}$, $\hat{p}$ are bosonic quadrature operators. Using Eq.~(\ref{eqn:hp_along_x}) we can write Eq.~(\ref{eqn:tat_hamil_2}) as 
\begin{equation}
\label{eqn:tat_hamil_hp_app}
\hat{H}_{\rm TaT} = -\Omega \hat{n} + \frac{\tilde{\chi}}{2}\hat{p}^2,
\end{equation}
where $\hat{n} = \hat{a}^\dagger\hat{a}$ is the bosonic number operator. Recalling that $\hat{q}^2 + \hat{p}^2 = \hat{n} + \frac{1}{2}$, we can rewrite Eq.~(\ref{eqn:tat_hamil_hp_app}) as 
\begin{equation}
\label{eqn:tat_hamil_hp_app_2}
\hat{H}_{\rm TaT} = \Omega\left(J + \frac{1}{2}\right) - \frac{\Omega}{2}\hat{q}^2 + \frac{1}{2}(\tilde{\chi} - \Omega)\hat{p}^2,
\end{equation}
which after dropping a constant factor, recovers Eq.~(\ref{eqn:tat_hamil_hp}) in the main text.

We want to compare the speed of evolution of our initial state under unitary dynamics generated by $\hat{H}_{\rm TaT}$, $V_0^2$, with the maximum speed of evolution optimized over all quantum states, $V_{\rm max}^2$. Restricted to Gaussian initial states, the speed is given by \cite{Poggi2021} 
\begin{equation}
\label{eqn:speed_gauss}
V_{\rm Gauss}^2 = \frac{1}{8}\left({\rm Tr}\left[(\mathbb{G}\Sigma)^2 \right] + {\rm Tr}\left[(\mathbb{G}\mathcal{J})^2\right]\right),
\end{equation}
which holds for undisplaced states, and where $\mathbb{G}$ is defined in Eq.~(\ref{eqn:g_matrix}), $\mathcal{J}$ is the symplectic form matrix, given by
\begin{equation}
\mathcal{J} = \begin{pmatrix}
0 && 1\\
-1 && 0
\end{pmatrix},
\end{equation}
and $\Sigma$ is the covariance matrix, which for a single mode Gaussian state is given by 
\begin{equation}
\Sigma = \mathcal{R}\mathcal{D}\mathcal{R}^{T},
\end{equation}
where the matrices $\mathcal{R}$ and $\mathcal{D}$ are given by
\begin{equation}
\mathcal{R} = \begin{pmatrix}
\cos(\beta) && \sin(\beta) \\
-\sin(\beta) && \cos(\beta)
\end{pmatrix}, \enspace \mathcal{D} = \begin{pmatrix}
e^{r} && 0 \\
0 && e^{-r}
\end{pmatrix}
\end{equation}
with $\beta$ an angle of rotation, and $r\in\mathbb{R}$ the squeezing parameter. 
 
The speed in Eq. (\ref{eqn:speed_gauss}) is in general unbounded, and so $V_{\rm max}^2\to\infty$. We then restrict ourselves to consider the speed of evolution under Gaussian optimized over all states with a given squeezing $r$. Under this consideration, Eq.~(\ref{eqn:speed_gauss}) can be written as
\begin{equation}
\label{eqn:max_speed_gauss_sq}
V_{\rm max}^2 = \frac{\tilde{\chi}^2}{8} + \frac{1}{4}\tilde{\chi}(\tilde{\chi} - 2\Omega)\cos(2\beta)r + \mathcal{O}(r^2).
\end{equation}
By noticing that, when $\tilde{\chi} < 2\Omega$ one has $\cos(2\beta) = -1$, and when $\tilde{\chi}>2\Omega$ one has $\cos(2\beta) = 1$, we can write Eq.~(\ref{eqn:max_speed_gauss_sq}) as 
\begin{equation}
\label{eqn:max_speed}
V^2_{\rm max}(r) = \frac{\tilde{\chi}^2}{8} + \frac{1}{4}\tilde{\chi}\left| \tilde{\chi} - 2\Omega \right|r. 
\end{equation}
On the other hand, the speed of unitary evolution for our initial state is given by 
\begin{equation}
\label{eqn:speed_initial}
V_0^2 =  \langle \Delta^2 \hat{H}_{\rm TaT} \rangle_0 = \frac{\tilde{\chi}^2}{8}.
\end{equation}
Thus, we can explore parameter regime where the quantum speed limit is saturated by considering the ratio of Eq.~(\ref{eqn:speed_initial}) and Eq.~(\ref{eqn:max_speed}), given by
\begin{equation}
\frac{V_0^2}{V_{\rm max}^2(r)} = \frac{\frac{\tilde{\chi}^2}{\Omega^2}}{\frac{\tilde{\chi}^2}{\Omega^2} + 2\frac{\tilde{\chi}}{\Omega}\left| \frac{\tilde{\chi}}{\Omega} - 2\right|r},
\end{equation}
which is recovers Eq.~(\ref{eqn:ratio_speed_limit}) in the main text.\\

\section{Derivation of some results with the \texorpdfstring{$p$}{\textit{p}}-twisting Hamiltonians}
\label{app:ptat_stuff}

\subsection{2ACT with $p$-order twisting}

Starting from classical flow of Eqs.~\ref{eqn:flow_ct_p}), we can compute the Jacobi matrix, which reads
\begin{widetext}
\begin{equation}
    \mathbb{M}[\mbf{X}]=\tilde{\chi}\left( \begin{array}{c c c}
    0 & -(p-1) Y^{p-2}Z & - Y^{p-1} \\
    -(p-1)X^{p-2}Z & 0 & -X^{p-1} \\
    (p-1)X^{p-2}Y+Y^{p-1} & (p-1)Y^{p-2}X + X^{p-1} & 0 
    \end{array}
    \right).
\end{equation}
\end{widetext}
We focus on the case of $p$ even for this model, and the case of $p=2$ was covered before. For $p>2$, the Jacobi matrix evaluated at the fixed points the equator gives
\begin{equation}
    \mathbb{M}[(\pm 1,0,0)]= \tilde{\chi}\left(\begin{array}{c c c}
    0 & 0 & 0 \\
    0 & 0 & \mp 1 \\
    0 & \pm 1 & 0
    \end{array}\right),
\end{equation}
\noindent and 
\begin{equation}
    \mathbb{M}[(0,\pm 1,0)]= \tilde{\chi}\left(\begin{array}{c c c}
    0 & 0 & \mp 1 \\
    0 & 0 & 0 \\
    \pm 1 & 0 & 0
    \end{array}\right).
\end{equation}

All these cases lead to purely imaginary eigenvalues $\pm i \tilde{\chi}$, meaning that the fixed points are elliptic or stable. For the fixed points at the pole $(0,0,\pm 1)$ however, we obtain that $\mathbb{M}=0$, meaning that we need to study the system beyond linearization, as done in the main text. 

\subsection{Phase pace flow and classical energy for $p$TaT}
In order to construct the phase space flow in the thermodynamic limit, we need to rescale the $p$TaT Hamiltonian in Eq.~(\ref{eqn:ptat_hamil}) so that in the thermodynamic limit, we can guarantee the extensivity of the energy density. This is achieved by introducing the normalized $p$-twisting strength $\chi = \frac{\tilde{\chi}}{pJ^{p-1}}$, then we write 
\begin{equation}
\label{eqn:ptat_hamil_2}
\hat{H}_{p{\rm TaT}} = \Omega \hat{J}_x + \frac{\tilde{\chi}}{pJ^{p-1}}\hat{J}_z^p.
\end{equation}
Using the Heisenberg equations of motion for the collective angular momentum $\mathbf{\hat{J}}$, and defining the classical variables $\mathbf{X} = \frac{\langle\mathbf{\hat{J}} \rangle}{J}$, in the thermodynamic limit, we can write the $p$TaT phase space flow as
\begin{subequations}
\label{eqn:ptat_flow}
\begin{align}
\frac{dX}{dt} &= \tilde{\chi}YZ^{p-2}, \\
\frac{dY}{dt} &= \Omega Z - \tilde{\chi}XZ^{p-2}, \\
\frac{dZ}{dt} &= -\Omega Y.
\end{align}
\end{subequations}
Fixed points of this flow are defined as solutions of $\frac{d\mathbf{X}}{dt} = 0$. Then, solving for the fixed points of Eq.~(\ref{eqn:ptat_flow}) gives the following condition 
\begin{equation}
\label{eqn:fixed_point_cond_ptat}
Y = 0, \enspace \text{and}\enspace X = \left(\frac{\Omega}{\tilde{\chi}}\right)\frac{1}{Z^{p-2}}.
\end{equation}
This condition, together with $|\mathbf{X}|^2 = 1$, give the same algebraic equation 
\begin{equation}
Z^{2p-2} - Z^{2p-4} + \left(\frac{\Omega}{\tilde{\chi}}\right)^2 = 0,
\end{equation}
whose solution is the $Z$ coordinate of the fixed points.

The Jacobi matrix associated with the phase space flow in Eq.~(\ref{eqn:ptat_flow}) is given by
\begin{equation}
\label{eqn:jacobi_mat_ptat}
\mathbb{M}[\mathbf{X}] = \begin{pmatrix}
0 && \tilde{\chi}Z^{p-1} && (p-1)\tilde{\chi}Z^{p-1}Y \\
-\tilde{\chi}Z^{p-1} && 0 && \Omega - (p-1)\tilde{\chi}Z^{p-1}X \\ 
0 && -\Omega && 0
\end{pmatrix}.
\end{equation}
We can obtain the energy density of the $p$TaT Hamiltonian by computing $\frac{\langle\hat{H}_{p{\rm TaT}}\rangle}{J}$ for Eq.~(\ref{eqn:ptat_hamil_2}) in the thermodynamic limit. After neglecting correlations and in terms of the classical variables $\mathbf{X}$, this energy density reads
\begin{equation}
\label{eqn:ptat_energy_1}
E(X,Z;\Omega, \tilde{\chi}) = \frac{\langle\hat{H}_{\rm TaT}\rangle}{J} = \Omega X + \frac{\tilde{\chi}}{p}Z^p,
\end{equation}
which we can write in terms of the classical conjugated variables $Z$ and $\phi$ as 
\begin{equation}
\label{eqn:ptat_energy_2}
E(Z,\phi;\Omega, \tilde{\chi}) = \Omega \sqrt{1-Z^2}\cos(\phi) + \frac{\tilde{\chi}}{p}Z^p.
\end{equation}

\subsection{Proof of Theorem~\ref{theo:no_local_ptat}}
The proof of Theorem~\ref{theo:no_local_ptat} follows from the results in the previous subsection. To see this, we need to rewrite the statement of the theorem. Local optimality is a geometric statement, it implies that the phase space flow has a saddle point, and that the separatrix branches are orthogonal in the vicinity of the saddle. As such, lack of local optimality implies that there is no parameter regime for which the separatrix branches are orthogonal. Furthermore, this implies that there is no parameter regime for which the principal directions of the saddle point are orthogonal. This last statement is the one we use to build the proof.

Proving the theorem then reduces to study the principal directions of the Jacobi matrix in Eq.~(\ref{eqn:jacobi_mat_ptat}) evaluated at the saddle point. Furthermore, we know these principal directions will be described by orthogonal vectors if and only if the Jacobi matrix evaluated at the saddle point is real and symmetric. Noticing that the saddle point is a fixed point of Eq.~(\ref{eqn:ptat_flow}), then it is true that it should satisfy the condition in Eq.~(\ref{eqn:fixed_point_cond_ptat}). 

We then evaluate the Jacobi matrix, Eq.~(\ref{eqn:jacobi_mat_ptat}), at the saddle via the condition in Eq.~(\ref{eqn:fixed_point_cond_ptat}), and recognized two different scenarios. First, if $p=2$, \textit{i.e.}, TaT, then one has
\begin{equation}
\mathbb{M}[\mathbf{X}_{\rm sd}] = \begin{pmatrix}
0 && 0 && 0 \\
0 && 0 && \Omega - \tilde{\chi} \\ 
0 && -\Omega && 0
\end{pmatrix},    
\end{equation}
which is real an symmetric whenever $\tilde{\chi} = 2\Omega$, \textit{i.e.}, at critical coupling. Second, for all $p>2$, one has 
\begin{equation}
\label{eqn:ptat_tangent_map}
\mathbb{M}[\mathbf{X}_{\rm sd}] = \begin{pmatrix}
0 && \tilde{\chi}Z^{p-1}_{\rm sd} && 0 \\
-\tilde{\chi}Z^{p-1}_{\rm sd} && 0 && -(p-2)\Omega \\ 
0 && -\Omega && 0
\end{pmatrix},    
\end{equation}
which is only real and symmetric when $p=3$ and $Z_{\rm sd}\to0$. The latter only happens in the limit $\frac{\tilde{\chi}}{\Omega}\to\infty$, and thus there is no parameter regime at which $p$TaT, with $p>2$, is locally optimal.

\subsection{Maximum saddle point Lyapunov exponent for the \texorpdfstring{$p$}{\textit{p}}TaT family}
The first step in this calculation is to considered a reduced form of the classical energy density, obtained from Eq.~(\ref{eqn:ptat_energy_2}) by setting $\phi = 0$, given by 
\begin{equation}
\label{eqn:ptat_energy_3}
E(Z;\Omega, \tilde{\chi}) = \Omega \sqrt{1-Z^2} + \frac{\tilde{\chi}}{p}Z^p.
\end{equation}
This reduced energy only looks at the great circle on the $x$-$z$ plane, along which extreme points of the classical energy, \textit{i.e.}, fixed points of the phase space flow, might emerge. In order to find the value of $\frac{\tilde{\chi}}{\Omega}$ at which the saddle point Lyapunov exponent we need an auxiliary equation, such that the system of equations can be solve for the pair
\begin{equation}
\left( Z_{\rm sd}^{@{{\rm max}[\Lambda_{\rm sd}]}}, \left.\frac{\tilde{\chi}}{\Omega}\right|_{{\rm max}[\Lambda_{\rm sd}]}\right).
\end{equation}
To find this auxiliary equation we solve for the eigenvalues of Eq.~(\ref{eqn:ptat_tangent_map}), and write an expression of the Lyapunov exponent of the saddle point
\begin{equation}
\label{eqn:ptat_lyap}
\Lambda_{\rm sd} = \tilde{\chi}\sqrt{\frac{(p-2)}{U^2} - Z_{\rm sd}^{2p-2}},
\end{equation}
where $U = \frac{\tilde{\chi}}{\Omega}$. In order to proceed further, we use the fact that the saddle point is an extreme point of the classical energy density, and thus it satisfies
\begin{equation}
\frac{dE(Z)}{dZ} = 0 \rightarrow Z = U Z^{p-1}\sqrt{1 - Z^2},
\end{equation}
where $E(Z)$ is define in Eq.~(\ref{eqn:ptat_energy_3}). From this expression we can write the ratio $\frac{\tilde{\chi}}{\Omega}$ as
\begin{equation}
\label{eqn:implicit_eq}
U = \frac{1}{Z^{p-2}\sqrt{1-Z^2}}.
\end{equation}
By plugging Eq.~(\ref{eqn:implicit_eq}) into Eq.~(\ref{eqn:ptat_lyap}), we obtain
\begin{equation}
\label{eqn:ptat_lyap_2}
\Lambda_{\rm sd} = \tilde{\chi}Z^{p-2}\sqrt{(p-1) - (p-2)Z^2},
\end{equation}
then by solving $\frac{d\Lambda_{\rm sd}}{dZ} = 0$, we obtain the position of the saddle point when the Lyapunov exponent is maximum, given by 
\begin{equation}
\label{eqn:z_saddle_pos_max_lyap}
Z_{\rm sd}^{{{\rm max}[\Lambda_{\rm sd}]}} = \frac{p-2}{p-1},
\end{equation}
by plugging Eq.~(\ref{eqn:z_saddle_pos_max_lyap}) into Eq.~(\ref{eqn:implicit_eq}), we obtain
\begin{equation}
\label{eqn:max_lyap}
\left.\frac{\tilde{\chi}}{\Omega}\right|_{{\rm max}[\Lambda_{\rm sd}]} = \frac{(p-1)^{p-1}}{(p-2)^{p-2}\sqrt{(p-1)^2 - (p-2)^2}}.
\end{equation}
Together Eq.~(\ref{eqn:z_saddle_pos_max_lyap}) and Eq.~(\ref{eqn:max_lyap}) recovers Eq.~(\ref{eqn:max_lyap_sols}) in the main text.

\subsection{Angle between separatrix branches at critical coupling for the \texorpdfstring{$3$}{\textit{3}}TaT}
For the $3$TaT at critical coupling the position of the saddle and the control parameter are given by 
\begin{equation}
\label{eqn:3tat_critcal_coupling}
Z_{\rm sd}^{@{{\rm max}[\Lambda_{\rm sd}]}} = \frac{1}{2}, \enspace\left.\frac{\tilde{\chi}}{\Omega}\right|_{{\rm max}[\Lambda_{\rm sd}]} = \frac{4}{\sqrt{3}}.
\end{equation}
Given these two values, it is easy to see that the energy of the saddle and thus the energy of the separatrix is $E_{\rm sep} = \frac{5}{3\sqrt{3}}\Omega$. With this we can write an equation for the separatrix line of the $3$TaT at critical coupling, by writing $E(Z,\phi;\tilde{\chi},\Omega) = \frac{5}{3\sqrt{3}}\Omega$. In terms of the angular variables we find
\begin{equation}
\label{eqn:sep_3tat_cc}
\cos(\phi) = \frac{1}{3\sqrt{3}\sin(\theta)}\left(4 - 5\cos^3(\theta)\right).
\end{equation}
We can use Eq.~(\ref{eqn:sep_3tat_cc}) to write the separatrix in parametric form, as 
\begin{equation}
\label{eqn:sep_3tat_cc_para}
\vec{S}_{\pm}(\theta) = \left(X(\theta),\enspace Y_{\pm}(\theta),\enspace Z(\theta)\right),
\end{equation}
whose components are given by 
\begin{subequations}
\label{eqn:sep_3tat_cc_components}
\begin{align}
X(\theta) &= \frac{1}{3\sqrt{3}}\left(4 - 5\cos^3(\theta)\right), \\
Y_\pm(\theta) &= \pm\sqrt{\sin^2(\theta) - \frac{1}{27}\left(4 - 5\cos^3(\theta) \right)^2},\\
Z(\theta) &= \cos(\theta).
\end{align}
\end{subequations}
To compute the angle between separatrix branches we take the projection of the separatrix onto the $y$-$z$ plane and translate the saddle to the origin, that is, we take $Z(\theta) = \cos(\theta) - \frac{1}{2}$, then the angle is calculated as the dot product between the two branches
\begin{equation}
\cos(\upsilon) = \frac{\vec{S}^{(YZ)}_+ . {\vec{S}^{(YZ)}_-}}{|\vec{S}^{(YZ)}_+| |\vec{S}^{(YZ)}_-|},
\end{equation}
where
\begin{equation}
\vec{S}^{(YZ)}_\pm = \left(\pm\sqrt{\sin^2(\theta) - \frac{1}{27}\left(4 - 5\cos^3(\theta) \right)^2}, \enspace \cos(\theta) - \frac{1}{2} \right).
\end{equation}
In the vicinity of the saddle point, we find $\cos(\upsilon) = \frac{1}{5}$.

\subsection{Time scale to first peak of the QFI with the \texorpdfstring{$3$}{\textit{3}}TaT}
The fact that at critical coupling the semiclassical energy is a double well with wells of equal depth indicates that one can prepare a "cat-like" state where the peaks of each love lie at the bottom of each of the wells, and the time scale to this state can, in principle, be estimated via a similar argument as the one used of the $p=2$ spin.

Our starting point are the canonical equations of motion obtained from the classical energy density in Eq.~(\ref{eqn:ptat_energy_2}), given by
\begin{subequations}
\label{eqn:3tat_canonical_eqs}
\begin{align}
\frac{dZ}{dt} &= -\frac{\partial E}{\partial \phi} = \Omega\sqrt{1 - Z^2}\sin(\phi), \\
\frac{d\phi}{dt} &= \frac{\partial E}{\partial Z} = -\Omega\frac{\cos(\phi)Z}{\sqrt{1 - Z^2}} + \tilde{\chi}Z^2.    
\end{align}
\end{subequations}
Using the equation of the separatrix at critical coupling we can write an equation of motion for $Z$ constrained to the separatrix branches, it reads
\begin{equation}
\frac{dZ}{dt} = \frac{\tilde{\chi}}{12}\sqrt{27(1-Z^2) - (5 - 4Z^3)^2},
\end{equation}
and thus, the desired time scale is given by 
\begin{equation}
\label{eqn:3tat_time_scale_QFI_app}
\tilde{\chi} t = \int_{Z(0)}^{Z(t_t)} \frac{12 dZ}{\sqrt{27(1-Z^2) - (5 - 4Z^3)^2}}.
\end{equation}
In order to complete this computation we need to specify the limits of integration in Eq.~(\ref{eqn:3tat_time_scale_QFI_app}). Specifying these limits of integration requires us to think a little bit more about the separatrix geometry. 

The value of $Z(t_f)$ is either one of the two ``end'' points of the separatrix, that is either of the two points on the intersection of the separatrix with the great circle on the $x$-$z$ plane, having the same energy as the saddle point. We can find them by equating the energy of the saddle with the semiclassical energy and solving for the roots of the resulting algebraic equation. Such algebraic equation is given by
\begin{equation}
16Z^6 - 40Z^3 + 27Z^2 - 2 = 0,
\end{equation}
which has three solutions, $Z = 1/2$, \textit{i.e.}, the saddle point, and 
\begin{equation}
Z_u = 0.9590789, \qquad Z_l = -0.23433406,
\end{equation}
for the ``upper'' and ``lower'' end points respectively. For the time scale calculation let me assume $Z(t_f) = Z_u$, that is, motion from the saddle to the ``upper'' end point (notice that motion happens simmetrycally towards $Z_l$ as well).

For the initial condition, it is not hard to realized that it should have the form 
\begin{equation}
Z_0 = \frac{1}{2} + d_{\rm sp},
\end{equation}
that is the position of the saddle plus a small increment $d_{\rm sp}$. The latter is the distance from the center of the SCS distribution to the edge measured along the separatrix and projected back to the $z$-axis. Given that the angle between separatrix branches is $\cos(\upsilon) = \frac{1}{5}$, we can get the angle between the separatrix branch and the $z$-axis. It is given by 
\begin{equation}
\label{eqn:angle_sep_and_z}
\cos(\tilde{\upsilon}) = \sqrt{\frac{3}{5}}.
\end{equation}
From Eq.~(\ref{eqn:angle_sep_and_z}) we can compute $d_{\rm sp}$ in the following way. Consider the initial uncertainty patch of the SCS projected onto the $y$-$z$ plane. It corresponds with an ellipse whose semi-major and semi-minor axis, on the $y$- and $z$-axis, respectively, are given by 
\begin{equation}
a = \frac{\sqrt{3}}{2\sqrt{N}}, \qquad b = \frac{1}{2\sqrt{N}},
\end{equation}
and thus, the ellipse is described by the equation 
\begin{equation}
\label{eqn:ellipse_eq}
4NZ^2 + \frac{4N}{3}Y^2 = 1.
\end{equation}
On the other hand, the point at the intersection between the ellipse and the separatrix branch satisfies 
\begin{equation}
\tan \left( \frac{Z_{\rm sp}}{Y_{\rm sp}}\right) = \frac{\sin(\tilde{\upsilon})}{\cos(\tilde{\upsilon})} = \sqrt{\frac{2}{3}},
\end{equation}
thus
\begin{equation}
\label{eqn:y_eq}
Y_{\rm sp} = \frac{Z_{\rm sp}}{\tilde{\upsilon}} = \tan^{-1}\left(\sqrt{\frac{2}{3}}\right).
\end{equation}
After plugging Eq.~(\ref{eqn:y_eq}) into Eq.~(\ref{eqn:ellipse_eq}) we find 
\begin{equation}
Z_{\rm sp} = \frac{1}{2\sqrt{\left(1 + \frac{1}{3\tilde{\upsilon}^2} \right)N}},
\end{equation}
and thus
\begin{equation}
Z(0) = \frac{1}{2} + \frac{1}{2\sqrt{\left(1 + \frac{1}{3\tilde{\upsilon}^2} \right)N}}.
\end{equation}
After which the time scale is completely determined, and it completes the derivation of Eq.~(\ref{eqn:3tat_time_to_cat}) in the main text.

\bibliography{references_metrology}

\begin{thebibliography}{75}%
\makeatletter
\providecommand \@ifxundefined [1]{%
 \@ifx{#1\undefined}
}%
\providecommand \@ifnum [1]{%
 \ifnum #1\expandafter \@firstoftwo
 \else \expandafter \@secondoftwo
 \fi
}%
\providecommand \@ifx [1]{%
 \ifx #1\expandafter \@firstoftwo
 \else \expandafter \@secondoftwo
 \fi
}%
\providecommand \natexlab [1]{#1}%
\providecommand \enquote  [1]{``#1''}%
\providecommand \bibnamefont  [1]{#1}%
\providecommand \bibfnamefont [1]{#1}%
\providecommand \citenamefont [1]{#1}%
\providecommand \href@noop [0]{\@secondoftwo}%
\providecommand \href [0]{\begingroup \@sanitize@url \@href}%
\providecommand \@href[1]{\@@startlink{#1}\@@href}%
\providecommand \@@href[1]{\endgroup#1\@@endlink}%
\providecommand \@sanitize@url [0]{\catcode `\\12\catcode `\$12\catcode
  `\&12\catcode `\#12\catcode `\^12\catcode `\_12\catcode `\%12\relax}%
\providecommand \@@startlink[1]{}%
\providecommand \@@endlink[0]{}%
\providecommand \url  [0]{\begingroup\@sanitize@url \@url }%
\providecommand \@url [1]{\endgroup\@href {#1}{\urlprefix }}%
\providecommand \urlprefix  [0]{URL }%
\providecommand \Eprint [0]{\href }%
\providecommand \doibase [0]{https://doi.org/}%
\providecommand \selectlanguage [0]{\@gobble}%
\providecommand \bibinfo  [0]{\@secondoftwo}%
\providecommand \bibfield  [0]{\@secondoftwo}%
\providecommand \translation [1]{[#1]}%
\providecommand \BibitemOpen [0]{}%
\providecommand \bibitemStop [0]{}%
\providecommand \bibitemNoStop [0]{.\EOS\space}%
\providecommand \EOS [0]{\spacefactor3000\relax}%
\providecommand \BibitemShut  [1]{\csname bibitem#1\endcsname}%
\let\auto@bib@innerbib\@empty
\bibitem [{\citenamefont {Giovannetti}\ \emph {et~al.}(2006)\citenamefont
  {Giovannetti}, \citenamefont {Lloyd},\ and\ \citenamefont
  {Maccone}}]{Giovannetti2006}%
  \BibitemOpen
  \bibfield  {author} {\bibinfo {author} {\bibfnamefont {V.}~\bibnamefont
  {Giovannetti}}, \bibinfo {author} {\bibfnamefont {S.}~\bibnamefont {Lloyd}},\
  and\ \bibinfo {author} {\bibfnamefont {L.}~\bibnamefont {Maccone}},\
  }\bibfield  {title} {\bibinfo {title} {Quantum metrology},\ }\href
  {https://doi.org/10.1103/PhysRevLett.96.010401} {\bibfield  {journal}
  {\bibinfo  {journal} {Phys. Rev. Lett.}\ }\textbf {\bibinfo {volume} {96}},\
  \bibinfo {pages} {010401} (\bibinfo {year} {2006})}\BibitemShut {NoStop}%
\bibitem [{\citenamefont {Giovannetti}\ \emph {et~al.}(2011)\citenamefont
  {Giovannetti}, \citenamefont {Lloyd},\ and\ \citenamefont
  {Maccone}}]{Giovannetti2011}%
  \BibitemOpen
  \bibfield  {author} {\bibinfo {author} {\bibfnamefont {V.}~\bibnamefont
  {Giovannetti}}, \bibinfo {author} {\bibfnamefont {S.}~\bibnamefont {Lloyd}},\
  and\ \bibinfo {author} {\bibfnamefont {L.}~\bibnamefont {Maccone}},\
  }\bibfield  {title} {\bibinfo {title} {Advances in quantum metrology},\
  }\href@noop {} {\bibfield  {journal} {\bibinfo  {journal} {Nature photonics}\
  }\textbf {\bibinfo {volume} {5}},\ \bibinfo {pages} {222} (\bibinfo {year}
  {2011})}\BibitemShut {NoStop}%
\bibitem [{\citenamefont {Pezz\`e}\ \emph {et~al.}(2018)\citenamefont
  {Pezz\`e}, \citenamefont {Smerzi}, \citenamefont {Oberthaler}, \citenamefont
  {Schmied},\ and\ \citenamefont {Treutlein}}]{Pezze2018}%
  \BibitemOpen
  \bibfield  {author} {\bibinfo {author} {\bibfnamefont {L.}~\bibnamefont
  {Pezz\`e}}, \bibinfo {author} {\bibfnamefont {A.}~\bibnamefont {Smerzi}},
  \bibinfo {author} {\bibfnamefont {M.~K.}\ \bibnamefont {Oberthaler}},
  \bibinfo {author} {\bibfnamefont {R.}~\bibnamefont {Schmied}},\ and\ \bibinfo
  {author} {\bibfnamefont {P.}~\bibnamefont {Treutlein}},\ }\bibfield  {title}
  {\bibinfo {title} {Quantum metrology with nonclassical states of atomic
  ensembles},\ }\href {https://doi.org/10.1103/RevModPhys.90.035005} {\bibfield
   {journal} {\bibinfo  {journal} {Rev. Mod. Phys.}\ }\textbf {\bibinfo
  {volume} {90}},\ \bibinfo {pages} {035005} (\bibinfo {year}
  {2018})}\BibitemShut {NoStop}%
\bibitem [{\citenamefont {Degen}\ \emph {et~al.}(2017)\citenamefont {Degen},
  \citenamefont {Reinhard},\ and\ \citenamefont {Cappellaro}}]{Degen2017}%
  \BibitemOpen
  \bibfield  {author} {\bibinfo {author} {\bibfnamefont {C.~L.}\ \bibnamefont
  {Degen}}, \bibinfo {author} {\bibfnamefont {F.}~\bibnamefont {Reinhard}},\
  and\ \bibinfo {author} {\bibfnamefont {P.}~\bibnamefont {Cappellaro}},\
  }\bibfield  {title} {\bibinfo {title} {Quantum sensing},\ }\href
  {https://doi.org/10.1103/RevModPhys.89.035002} {\bibfield  {journal}
  {\bibinfo  {journal} {Rev. Mod. Phys.}\ }\textbf {\bibinfo {volume} {89}},\
  \bibinfo {pages} {035002} (\bibinfo {year} {2017})}\BibitemShut {NoStop}%
\bibitem [{\citenamefont {Kitagawa}\ and\ \citenamefont
  {Ueda}(1993)}]{KitagawaUeda1993}%
  \BibitemOpen
  \bibfield  {author} {\bibinfo {author} {\bibfnamefont {M.}~\bibnamefont
  {Kitagawa}}\ and\ \bibinfo {author} {\bibfnamefont {M.}~\bibnamefont
  {Ueda}},\ }\bibfield  {title} {\bibinfo {title} {Squeezed spin states},\
  }\href {https://doi.org/10.1103/PhysRevA.47.5138} {\bibfield  {journal}
  {\bibinfo  {journal} {Phys. Rev. A}\ }\textbf {\bibinfo {volume} {47}},\
  \bibinfo {pages} {5138} (\bibinfo {year} {1993})}\BibitemShut {NoStop}%
\bibitem [{\citenamefont {S\o{}rensen}\ and\ \citenamefont
  {M\o{}lmer}(2001)}]{Sorensen2001}%
  \BibitemOpen
  \bibfield  {author} {\bibinfo {author} {\bibfnamefont {A.~S.}\ \bibnamefont
  {S\o{}rensen}}\ and\ \bibinfo {author} {\bibfnamefont {K.}~\bibnamefont
  {M\o{}lmer}},\ }\bibfield  {title} {\bibinfo {title} {Entanglement and
  extreme spin squeezing},\ }\href
  {https://doi.org/10.1103/PhysRevLett.86.4431} {\bibfield  {journal} {\bibinfo
   {journal} {Phys. Rev. Lett.}\ }\textbf {\bibinfo {volume} {86}},\ \bibinfo
  {pages} {4431} (\bibinfo {year} {2001})}\BibitemShut {NoStop}%
\bibitem [{\citenamefont {Ma}\ \emph {et~al.}(2011)\citenamefont {Ma},
  \citenamefont {Wang}, \citenamefont {Sun},\ and\ \citenamefont
  {Nori}}]{Ma2011}%
  \BibitemOpen
  \bibfield  {author} {\bibinfo {author} {\bibfnamefont {J.}~\bibnamefont
  {Ma}}, \bibinfo {author} {\bibfnamefont {X.}~\bibnamefont {Wang}}, \bibinfo
  {author} {\bibfnamefont {C.-P.}\ \bibnamefont {Sun}},\ and\ \bibinfo {author}
  {\bibfnamefont {F.}~\bibnamefont {Nori}},\ }\bibfield  {title} {\bibinfo
  {title} {Quantum spin squeezing},\ }\href@noop {} {\bibfield  {journal}
  {\bibinfo  {journal} {Physics Reports}\ }\textbf {\bibinfo {volume} {509}},\
  \bibinfo {pages} {89} (\bibinfo {year} {2011})}\BibitemShut {NoStop}%
\bibitem [{\citenamefont {Gross}(2012)}]{Gross2012}%
  \BibitemOpen
  \bibfield  {author} {\bibinfo {author} {\bibfnamefont {C.}~\bibnamefont
  {Gross}},\ }\bibfield  {title} {\bibinfo {title} {Spin squeezing,
  entanglement and quantum metrology with bose--einstein condensates},\
  }\href@noop {} {\bibfield  {journal} {\bibinfo  {journal} {Journal of Physics
  B: Atomic, Molecular and Optical Physics}\ }\textbf {\bibinfo {volume}
  {45}},\ \bibinfo {pages} {103001} (\bibinfo {year} {2012})}\BibitemShut
  {NoStop}%
\bibitem [{\citenamefont {Wineland}\ \emph {et~al.}(1992)\citenamefont
  {Wineland}, \citenamefont {Bollinger}, \citenamefont {Itano}, \citenamefont
  {Moore},\ and\ \citenamefont {Heinzen}}]{Wineland1992}%
  \BibitemOpen
  \bibfield  {author} {\bibinfo {author} {\bibfnamefont {D.~J.}\ \bibnamefont
  {Wineland}}, \bibinfo {author} {\bibfnamefont {J.~J.}\ \bibnamefont
  {Bollinger}}, \bibinfo {author} {\bibfnamefont {W.~M.}\ \bibnamefont
  {Itano}}, \bibinfo {author} {\bibfnamefont {F.~L.}\ \bibnamefont {Moore}},\
  and\ \bibinfo {author} {\bibfnamefont {D.~J.}\ \bibnamefont {Heinzen}},\
  }\bibfield  {title} {\bibinfo {title} {Spin squeezing and reduced quantum
  noise in spectroscopy},\ }\href {https://doi.org/10.1103/PhysRevA.46.R6797}
  {\bibfield  {journal} {\bibinfo  {journal} {Phys. Rev. A}\ }\textbf {\bibinfo
  {volume} {46}},\ \bibinfo {pages} {R6797} (\bibinfo {year}
  {1992})}\BibitemShut {NoStop}%
\bibitem [{\citenamefont {T{\'o}th}\ and\ \citenamefont
  {Apellaniz}(2014)}]{Toth2014}%
  \BibitemOpen
  \bibfield  {author} {\bibinfo {author} {\bibfnamefont {G.}~\bibnamefont
  {T{\'o}th}}\ and\ \bibinfo {author} {\bibfnamefont {I.}~\bibnamefont
  {Apellaniz}},\ }\bibfield  {title} {\bibinfo {title} {Quantum metrology from
  a quantum information science perspective},\ }\href@noop {} {\bibfield
  {journal} {\bibinfo  {journal} {Journal of Physics A: Mathematical and
  Theoretical}\ }\textbf {\bibinfo {volume} {47}},\ \bibinfo {pages} {424006}
  (\bibinfo {year} {2014})}\BibitemShut {NoStop}%
\bibitem [{\citenamefont {Huang}\ \emph {et~al.}(2015)\citenamefont {Huang},
  \citenamefont {Qin}, \citenamefont {Zhong}, \citenamefont {Ke},\ and\
  \citenamefont {Lee}}]{Huang2015}%
  \BibitemOpen
  \bibfield  {author} {\bibinfo {author} {\bibfnamefont {J.}~\bibnamefont
  {Huang}}, \bibinfo {author} {\bibfnamefont {X.}~\bibnamefont {Qin}}, \bibinfo
  {author} {\bibfnamefont {H.}~\bibnamefont {Zhong}}, \bibinfo {author}
  {\bibfnamefont {Y.}~\bibnamefont {Ke}},\ and\ \bibinfo {author}
  {\bibfnamefont {C.}~\bibnamefont {Lee}},\ }\bibfield  {title} {\bibinfo
  {title} {Quantum metrology with spin cat states under dissipation},\
  }\href@noop {} {\bibfield  {journal} {\bibinfo  {journal} {Scientific
  reports}\ }\textbf {\bibinfo {volume} {5}},\ \bibinfo {pages} {1} (\bibinfo
  {year} {2015})}\BibitemShut {NoStop}%
\bibitem [{\citenamefont {Dicke}(1954)}]{Dicke1954}%
  \BibitemOpen
  \bibfield  {author} {\bibinfo {author} {\bibfnamefont {R.~H.}\ \bibnamefont
  {Dicke}},\ }\bibfield  {title} {\bibinfo {title} {Coherence in spontaneous
  radiation processes},\ }\href {https://doi.org/10.1103/PhysRev.93.99}
  {\bibfield  {journal} {\bibinfo  {journal} {Phys. Rev.}\ }\textbf {\bibinfo
  {volume} {93}},\ \bibinfo {pages} {99} (\bibinfo {year} {1954})}\BibitemShut
  {NoStop}%
\bibitem [{\citenamefont {Pezz\'e}\ and\ \citenamefont
  {Smerzi}(2009)}]{Pezze2009}%
  \BibitemOpen
  \bibfield  {author} {\bibinfo {author} {\bibfnamefont {L.}~\bibnamefont
  {Pezz\'e}}\ and\ \bibinfo {author} {\bibfnamefont {A.}~\bibnamefont
  {Smerzi}},\ }\bibfield  {title} {\bibinfo {title} {Entanglement, nonlinear
  dynamics, and the heisenberg limit},\ }\href
  {https://doi.org/10.1103/PhysRevLett.102.100401} {\bibfield  {journal}
  {\bibinfo  {journal} {Phys. Rev. Lett.}\ }\textbf {\bibinfo {volume} {102}},\
  \bibinfo {pages} {100401} (\bibinfo {year} {2009})}\BibitemShut {NoStop}%
\bibitem [{Note1()}]{Note1}%
  \BibitemOpen
  \bibinfo {note} {Here we emphasise that the types of quantum metrology
  protocol which exploit the quantum states we are about to mention is often
  referred as \protect \emph {local quantum metrology}. This with the aim of
  making clear the fact that we would not be thinking about Bayesian quantum
  metrology.}\BibitemShut {Stop}%
\bibitem [{\citenamefont {Micheli}\ \emph {et~al.}(2003)\citenamefont
  {Micheli}, \citenamefont {Jaksch}, \citenamefont {Cirac},\ and\ \citenamefont
  {Zoller}}]{Micheli2003}%
  \BibitemOpen
  \bibfield  {author} {\bibinfo {author} {\bibfnamefont {A.}~\bibnamefont
  {Micheli}}, \bibinfo {author} {\bibfnamefont {D.}~\bibnamefont {Jaksch}},
  \bibinfo {author} {\bibfnamefont {J.~I.}\ \bibnamefont {Cirac}},\ and\
  \bibinfo {author} {\bibfnamefont {P.}~\bibnamefont {Zoller}},\ }\bibfield
  {title} {\bibinfo {title} {Many-particle entanglement in two-component
  bose-einstein condensates},\ }\href
  {https://doi.org/10.1103/PhysRevA.67.013607} {\bibfield  {journal} {\bibinfo
  {journal} {Phys. Rev. A}\ }\textbf {\bibinfo {volume} {67}},\ \bibinfo
  {pages} {013607} (\bibinfo {year} {2003})}\BibitemShut {NoStop}%
\bibitem [{\citenamefont {Sorelli}\ \emph {et~al.}(2019)\citenamefont
  {Sorelli}, \citenamefont {Gessner}, \citenamefont {Smerzi},\ and\
  \citenamefont {Pezz\`e}}]{Sorelli2019}%
  \BibitemOpen
  \bibfield  {author} {\bibinfo {author} {\bibfnamefont {G.}~\bibnamefont
  {Sorelli}}, \bibinfo {author} {\bibfnamefont {M.}~\bibnamefont {Gessner}},
  \bibinfo {author} {\bibfnamefont {A.}~\bibnamefont {Smerzi}},\ and\ \bibinfo
  {author} {\bibfnamefont {L.}~\bibnamefont {Pezz\`e}},\ }\bibfield  {title}
  {\bibinfo {title} {Fast and optimal generation of entanglement in bosonic
  josephson junctions},\ }\href {https://doi.org/10.1103/PhysRevA.99.022329}
  {\bibfield  {journal} {\bibinfo  {journal} {Phys. Rev. A}\ }\textbf {\bibinfo
  {volume} {99}},\ \bibinfo {pages} {022329} (\bibinfo {year}
  {2019})}\BibitemShut {NoStop}%
\bibitem [{\citenamefont {Yukawa}\ \emph {et~al.}(2014)\citenamefont {Yukawa},
  \citenamefont {Milburn}, \citenamefont {Holmes}, \citenamefont {Ueda},\ and\
  \citenamefont {Nemoto}}]{Yukawa2014}%
  \BibitemOpen
  \bibfield  {author} {\bibinfo {author} {\bibfnamefont {E.}~\bibnamefont
  {Yukawa}}, \bibinfo {author} {\bibfnamefont {G.~J.}\ \bibnamefont {Milburn}},
  \bibinfo {author} {\bibfnamefont {C.~A.}\ \bibnamefont {Holmes}}, \bibinfo
  {author} {\bibfnamefont {M.}~\bibnamefont {Ueda}},\ and\ \bibinfo {author}
  {\bibfnamefont {K.}~\bibnamefont {Nemoto}},\ }\bibfield  {title} {\bibinfo
  {title} {Precision measurements using squeezed spin states via two-axis
  countertwisting interactions},\ }\href
  {https://doi.org/10.1103/PhysRevA.90.062132} {\bibfield  {journal} {\bibinfo
  {journal} {Phys. Rev. A}\ }\textbf {\bibinfo {volume} {90}},\ \bibinfo
  {pages} {062132} (\bibinfo {year} {2014})}\BibitemShut {NoStop}%
\bibitem [{\citenamefont {Kajtoch}\ and\ \citenamefont
  {Witkowska}(2015)}]{Kajtoch2015}%
  \BibitemOpen
  \bibfield  {author} {\bibinfo {author} {\bibfnamefont {D.}~\bibnamefont
  {Kajtoch}}\ and\ \bibinfo {author} {\bibfnamefont {E.}~\bibnamefont
  {Witkowska}},\ }\bibfield  {title} {\bibinfo {title} {Quantum dynamics
  generated by the two-axis countertwisting hamiltonian},\ }\href
  {https://doi.org/10.1103/PhysRevA.92.013623} {\bibfield  {journal} {\bibinfo
  {journal} {Phys. Rev. A}\ }\textbf {\bibinfo {volume} {92}},\ \bibinfo
  {pages} {013623} (\bibinfo {year} {2015})}\BibitemShut {NoStop}%
\bibitem [{\citenamefont {Stockton}\ \emph {et~al.}(2003)\citenamefont
  {Stockton}, \citenamefont {Geremia}, \citenamefont {Doherty},\ and\
  \citenamefont {Mabuchi}}]{Stockton2003}%
  \BibitemOpen
  \bibfield  {author} {\bibinfo {author} {\bibfnamefont {J.~K.}\ \bibnamefont
  {Stockton}}, \bibinfo {author} {\bibfnamefont {J.}~\bibnamefont {Geremia}},
  \bibinfo {author} {\bibfnamefont {A.~C.}\ \bibnamefont {Doherty}},\ and\
  \bibinfo {author} {\bibfnamefont {H.}~\bibnamefont {Mabuchi}},\ }\bibfield
  {title} {\bibinfo {title} {Characterizing the entanglement of symmetric
  many-particle spin-1 2 systems},\ }\href@noop {} {\bibfield  {journal}
  {\bibinfo  {journal} {Physical Review A}\ }\textbf {\bibinfo {volume} {67}},\
  \bibinfo {pages} {022112} (\bibinfo {year} {2003})}\BibitemShut {NoStop}%
\bibitem [{\citenamefont {Salvatori}\ \emph {et~al.}(2014)\citenamefont
  {Salvatori}, \citenamefont {Mandarino},\ and\ \citenamefont
  {Paris}}]{Salvatori2014}%
  \BibitemOpen
  \bibfield  {author} {\bibinfo {author} {\bibfnamefont {G.}~\bibnamefont
  {Salvatori}}, \bibinfo {author} {\bibfnamefont {A.}~\bibnamefont
  {Mandarino}},\ and\ \bibinfo {author} {\bibfnamefont {M.~G.~A.}\ \bibnamefont
  {Paris}},\ }\bibfield  {title} {\bibinfo {title} {Quantum metrology in
  lipkin-meshkov-glick critical systems},\ }\href
  {https://doi.org/10.1103/PhysRevA.90.022111} {\bibfield  {journal} {\bibinfo
  {journal} {Phys. Rev. A}\ }\textbf {\bibinfo {volume} {90}},\ \bibinfo
  {pages} {022111} (\bibinfo {year} {2014})}\BibitemShut {NoStop}%
\bibitem [{\citenamefont {Fr\'erot}\ and\ \citenamefont
  {Roscilde}(2018)}]{Frerot2018}%
  \BibitemOpen
  \bibfield  {author} {\bibinfo {author} {\bibfnamefont {I.}~\bibnamefont
  {Fr\'erot}}\ and\ \bibinfo {author} {\bibfnamefont {T.}~\bibnamefont
  {Roscilde}},\ }\bibfield  {title} {\bibinfo {title} {Quantum critical
  metrology},\ }\href {https://doi.org/10.1103/PhysRevLett.121.020402}
  {\bibfield  {journal} {\bibinfo  {journal} {Phys. Rev. Lett.}\ }\textbf
  {\bibinfo {volume} {121}},\ \bibinfo {pages} {020402} (\bibinfo {year}
  {2018})}\BibitemShut {NoStop}%
\bibitem [{\citenamefont {Fiderer}\ and\ \citenamefont
  {Braun}(2018)}]{Fiderer2018}%
  \BibitemOpen
  \bibfield  {author} {\bibinfo {author} {\bibfnamefont {L.~J.}\ \bibnamefont
  {Fiderer}}\ and\ \bibinfo {author} {\bibfnamefont {D.}~\bibnamefont
  {Braun}},\ }\bibfield  {title} {\bibinfo {title} {Quantum metrology with
  quantum-chaotic sensors},\ }\href
  {https://doi.org/10.1038/s41467-018-03623-z} {\bibfield  {journal} {\bibinfo
  {journal} {Nature communications}\ }\textbf {\bibinfo {volume} {9}},\
  \bibinfo {pages} {1} (\bibinfo {year} {2018})}\BibitemShut {NoStop}%
\bibitem [{\citenamefont {Boixo}\ \emph
  {et~al.}(2008{\natexlab{a}})\citenamefont {Boixo}, \citenamefont {Datta},
  \citenamefont {Davis}, \citenamefont {Flammia}, \citenamefont {Shaji},\ and\
  \citenamefont {Caves}}]{Boixo2008a}%
  \BibitemOpen
  \bibfield  {author} {\bibinfo {author} {\bibfnamefont {S.}~\bibnamefont
  {Boixo}}, \bibinfo {author} {\bibfnamefont {A.}~\bibnamefont {Datta}},
  \bibinfo {author} {\bibfnamefont {M.~J.}\ \bibnamefont {Davis}}, \bibinfo
  {author} {\bibfnamefont {S.~T.}\ \bibnamefont {Flammia}}, \bibinfo {author}
  {\bibfnamefont {A.}~\bibnamefont {Shaji}},\ and\ \bibinfo {author}
  {\bibfnamefont {C.~M.}\ \bibnamefont {Caves}},\ }\bibfield  {title} {\bibinfo
  {title} {Quantum metrology: Dynamics versus entanglement},\ }\href
  {https://doi.org/10.1103/PhysRevLett.101.040403} {\bibfield  {journal}
  {\bibinfo  {journal} {Phys. Rev. Lett.}\ }\textbf {\bibinfo {volume} {101}},\
  \bibinfo {pages} {040403} (\bibinfo {year} {2008}{\natexlab{a}})}\BibitemShut
  {NoStop}%
\bibitem [{\citenamefont {Boixo}\ \emph
  {et~al.}(2008{\natexlab{b}})\citenamefont {Boixo}, \citenamefont {Datta},
  \citenamefont {Flammia}, \citenamefont {Shaji}, \citenamefont {Bagan},\ and\
  \citenamefont {Caves}}]{Boixo2008b}%
  \BibitemOpen
  \bibfield  {author} {\bibinfo {author} {\bibfnamefont {S.}~\bibnamefont
  {Boixo}}, \bibinfo {author} {\bibfnamefont {A.}~\bibnamefont {Datta}},
  \bibinfo {author} {\bibfnamefont {S.~T.}\ \bibnamefont {Flammia}}, \bibinfo
  {author} {\bibfnamefont {A.}~\bibnamefont {Shaji}}, \bibinfo {author}
  {\bibfnamefont {E.}~\bibnamefont {Bagan}},\ and\ \bibinfo {author}
  {\bibfnamefont {C.~M.}\ \bibnamefont {Caves}},\ }\bibfield  {title} {\bibinfo
  {title} {Quantum-limited metrology with product states},\ }\href
  {https://doi.org/10.1103/PhysRevA.77.012317} {\bibfield  {journal} {\bibinfo
  {journal} {Phys. Rev. A}\ }\textbf {\bibinfo {volume} {77}},\ \bibinfo
  {pages} {012317} (\bibinfo {year} {2008}{\natexlab{b}})}\BibitemShut
  {NoStop}%
\bibitem [{\citenamefont {Beau}\ and\ \citenamefont {del
  Campo}(2017)}]{Beau2017}%
  \BibitemOpen
  \bibfield  {author} {\bibinfo {author} {\bibfnamefont {M.}~\bibnamefont
  {Beau}}\ and\ \bibinfo {author} {\bibfnamefont {A.}~\bibnamefont {del
  Campo}},\ }\bibfield  {title} {\bibinfo {title} {Nonlinear quantum metrology
  of many-body open systems},\ }\href
  {https://doi.org/10.1103/PhysRevLett.119.010403} {\bibfield  {journal}
  {\bibinfo  {journal} {Phys. Rev. Lett.}\ }\textbf {\bibinfo {volume} {119}},\
  \bibinfo {pages} {010403} (\bibinfo {year} {2017})}\BibitemShut {NoStop}%
\bibitem [{\citenamefont {Macr\`{\i}}\ \emph {et~al.}(2016)\citenamefont
  {Macr\`{\i}}, \citenamefont {Smerzi},\ and\ \citenamefont
  {Pezz\`e}}]{Macri2016}%
  \BibitemOpen
  \bibfield  {author} {\bibinfo {author} {\bibfnamefont {T.}~\bibnamefont
  {Macr\`{\i}}}, \bibinfo {author} {\bibfnamefont {A.}~\bibnamefont {Smerzi}},\
  and\ \bibinfo {author} {\bibfnamefont {L.}~\bibnamefont {Pezz\`e}},\
  }\bibfield  {title} {\bibinfo {title} {Loschmidt echo for quantum
  metrology},\ }\href {https://doi.org/10.1103/PhysRevA.94.010102} {\bibfield
  {journal} {\bibinfo  {journal} {Phys. Rev. A}\ }\textbf {\bibinfo {volume}
  {94}},\ \bibinfo {pages} {010102} (\bibinfo {year} {2016})}\BibitemShut
  {NoStop}%
\bibitem [{\citenamefont {Davis}\ \emph {et~al.}(2016)\citenamefont {Davis},
  \citenamefont {Bentsen},\ and\ \citenamefont {Schleier-Smith}}]{Davis2016}%
  \BibitemOpen
  \bibfield  {author} {\bibinfo {author} {\bibfnamefont {E.}~\bibnamefont
  {Davis}}, \bibinfo {author} {\bibfnamefont {G.}~\bibnamefont {Bentsen}},\
  and\ \bibinfo {author} {\bibfnamefont {M.}~\bibnamefont {Schleier-Smith}},\
  }\bibfield  {title} {\bibinfo {title} {Approaching the heisenberg limit
  without single-particle detection},\ }\href
  {https://doi.org/10.1103/PhysRevLett.116.053601} {\bibfield  {journal}
  {\bibinfo  {journal} {Phys. Rev. Lett.}\ }\textbf {\bibinfo {volume} {116}},\
  \bibinfo {pages} {053601} (\bibinfo {year} {2016})}\BibitemShut {NoStop}%
\bibitem [{\citenamefont {Anders}\ \emph {et~al.}(2018)\citenamefont {Anders},
  \citenamefont {Pezz\`e}, \citenamefont {Smerzi},\ and\ \citenamefont
  {Klempt}}]{Anders2018}%
  \BibitemOpen
  \bibfield  {author} {\bibinfo {author} {\bibfnamefont {F.}~\bibnamefont
  {Anders}}, \bibinfo {author} {\bibfnamefont {L.}~\bibnamefont {Pezz\`e}},
  \bibinfo {author} {\bibfnamefont {A.}~\bibnamefont {Smerzi}},\ and\ \bibinfo
  {author} {\bibfnamefont {C.}~\bibnamefont {Klempt}},\ }\bibfield  {title}
  {\bibinfo {title} {Phase magnification by two-axis countertwisting for
  detection-noise robust interferometry},\ }\href
  {https://doi.org/10.1103/PhysRevA.97.043813} {\bibfield  {journal} {\bibinfo
  {journal} {Phys. Rev. A}\ }\textbf {\bibinfo {volume} {97}},\ \bibinfo
  {pages} {043813} (\bibinfo {year} {2018})}\BibitemShut {NoStop}%
\bibitem [{\citenamefont {Nolan}\ \emph {et~al.}(2017)\citenamefont {Nolan},
  \citenamefont {Szigeti},\ and\ \citenamefont {Haine}}]{Nolan2017}%
  \BibitemOpen
  \bibfield  {author} {\bibinfo {author} {\bibfnamefont {S.~P.}\ \bibnamefont
  {Nolan}}, \bibinfo {author} {\bibfnamefont {S.~S.}\ \bibnamefont {Szigeti}},\
  and\ \bibinfo {author} {\bibfnamefont {S.~A.}\ \bibnamefont {Haine}},\
  }\bibfield  {title} {\bibinfo {title} {Optimal and robust quantum metrology
  using interaction-based readouts},\ }\href
  {https://doi.org/10.1103/PhysRevLett.119.193601} {\bibfield  {journal}
  {\bibinfo  {journal} {Phys. Rev. Lett.}\ }\textbf {\bibinfo {volume} {119}},\
  \bibinfo {pages} {193601} (\bibinfo {year} {2017})}\BibitemShut {NoStop}%
\bibitem [{\citenamefont {Volkoff}\ and\ \citenamefont
  {Martin}(2022)}]{Volkoff2022}%
  \BibitemOpen
  \bibfield  {author} {\bibinfo {author} {\bibfnamefont {T.~J.}\ \bibnamefont
  {Volkoff}}\ and\ \bibinfo {author} {\bibfnamefont {M.~J.}\ \bibnamefont
  {Martin}},\ }\bibfield  {title} {\bibinfo {title} {Asymptotic optimality of
  twist-untwist protocols for heisenberg scaling in atom-based sensing},\
  }\href {https://doi.org/10.1103/PhysRevResearch.4.013236} {\bibfield
  {journal} {\bibinfo  {journal} {Phys. Rev. Research}\ }\textbf {\bibinfo
  {volume} {4}},\ \bibinfo {pages} {013236} (\bibinfo {year}
  {2022})}\BibitemShut {NoStop}%
\bibitem [{Note2()}]{Note2}%
  \BibitemOpen
  \bibinfo {note} {This is the well know results of Heisenberg limited spin
  squeezing achieved with 2ACT~\cite {KitagawaUeda1993}, for $N=1024$, this
  gives an squeezing of $80\protect \qopname \relax o{log}_{10}(2)\approx
  25$dB}\BibitemShut {NoStop}%
\bibitem [{\citenamefont {Berry}\ and\ \citenamefont
  {Wiseman}(2000)}]{Berry2000}%
  \BibitemOpen
  \bibfield  {author} {\bibinfo {author} {\bibfnamefont {D.~W.}\ \bibnamefont
  {Berry}}\ and\ \bibinfo {author} {\bibfnamefont {H.~M.}\ \bibnamefont
  {Wiseman}},\ }\bibfield  {title} {\bibinfo {title} {Optimal states and almost
  optimal adaptive measurements for quantum interferometry},\ }\href
  {https://doi.org/10.1103/PhysRevLett.85.5098} {\bibfield  {journal} {\bibinfo
   {journal} {Phys. Rev. Lett.}\ }\textbf {\bibinfo {volume} {85}},\ \bibinfo
  {pages} {5098} (\bibinfo {year} {2000})}\BibitemShut {NoStop}%
\bibitem [{\citenamefont {Combes}\ and\ \citenamefont
  {Wiseman}(2004)}]{Combes2004}%
  \BibitemOpen
  \bibfield  {author} {\bibinfo {author} {\bibfnamefont {J.}~\bibnamefont
  {Combes}}\ and\ \bibinfo {author} {\bibfnamefont {H.~M.}\ \bibnamefont
  {Wiseman}},\ }\bibfield  {title} {\bibinfo {title} {States for phase
  estimation in quantum interferometry},\ }\href
  {https://doi.org/10.1088/1464-4266/7/1/004} {\bibfield  {journal} {\bibinfo
  {journal} {Journal of Optics B: Quantum and Semiclassical Optics}\ }\textbf
  {\bibinfo {volume} {7}},\ \bibinfo {pages} {14} (\bibinfo {year}
  {2004})}\BibitemShut {NoStop}%
\bibitem [{\citenamefont {Schubert}\ \emph {et~al.}(2012)\citenamefont
  {Schubert}, \citenamefont {Vallejos},\ and\ \citenamefont
  {Toscano}}]{Schubert2012wave}%
  \BibitemOpen
  \bibfield  {author} {\bibinfo {author} {\bibfnamefont {R.}~\bibnamefont
  {Schubert}}, \bibinfo {author} {\bibfnamefont {R.~O.}\ \bibnamefont
  {Vallejos}},\ and\ \bibinfo {author} {\bibfnamefont {F.}~\bibnamefont
  {Toscano}},\ }\bibfield  {title} {\bibinfo {title} {How do wave packets
  spread? time evolution on ehrenfest time scales},\ }\href@noop {} {\bibfield
  {journal} {\bibinfo  {journal} {Journal of Physics A: Mathematical and
  Theoretical}\ }\textbf {\bibinfo {volume} {45}},\ \bibinfo {pages} {215307}
  (\bibinfo {year} {2012})}\BibitemShut {NoStop}%
\bibitem [{\citenamefont {Pappalardi}\ \emph {et~al.}(2018)\citenamefont
  {Pappalardi}, \citenamefont {Russomanno}, \citenamefont {\ifmmode
  \check{Z}\else \v{Z}\fi{}unkovi\ifmmode~\check{c}\else \v{c}\fi{}},
  \citenamefont {Iemini}, \citenamefont {Silva},\ and\ \citenamefont
  {Fazio}}]{Pappalardi2018}%
  \BibitemOpen
  \bibfield  {author} {\bibinfo {author} {\bibfnamefont {S.}~\bibnamefont
  {Pappalardi}}, \bibinfo {author} {\bibfnamefont {A.}~\bibnamefont
  {Russomanno}}, \bibinfo {author} {\bibfnamefont {B.}~\bibnamefont {\ifmmode
  \check{Z}\else \v{Z}\fi{}unkovi\ifmmode~\check{c}\else \v{c}\fi{}}}, \bibinfo
  {author} {\bibfnamefont {F.}~\bibnamefont {Iemini}}, \bibinfo {author}
  {\bibfnamefont {A.}~\bibnamefont {Silva}},\ and\ \bibinfo {author}
  {\bibfnamefont {R.}~\bibnamefont {Fazio}},\ }\bibfield  {title} {\bibinfo
  {title} {Scrambling and entanglement spreading in long-range spin chains},\
  }\href {https://doi.org/10.1103/PhysRevB.98.134303} {\bibfield  {journal}
  {\bibinfo  {journal} {Phys. Rev. B}\ }\textbf {\bibinfo {volume} {98}},\
  \bibinfo {pages} {134303} (\bibinfo {year} {2018})}\BibitemShut {NoStop}%
\bibitem [{\citenamefont {Strobel}\ \emph {et~al.}(2014)\citenamefont
  {Strobel}, \citenamefont {Muessel}, \citenamefont {Linnemann}, \citenamefont
  {Zibold}, \citenamefont {Hume}, \citenamefont {Pezzè}, \citenamefont
  {Smerzi},\ and\ \citenamefont {Oberthaler}}]{Strobel2014}%
  \BibitemOpen
  \bibfield  {author} {\bibinfo {author} {\bibfnamefont {H.}~\bibnamefont
  {Strobel}}, \bibinfo {author} {\bibfnamefont {W.}~\bibnamefont {Muessel}},
  \bibinfo {author} {\bibfnamefont {D.}~\bibnamefont {Linnemann}}, \bibinfo
  {author} {\bibfnamefont {T.}~\bibnamefont {Zibold}}, \bibinfo {author}
  {\bibfnamefont {D.~B.}\ \bibnamefont {Hume}}, \bibinfo {author}
  {\bibfnamefont {L.}~\bibnamefont {Pezzè}}, \bibinfo {author} {\bibfnamefont
  {A.}~\bibnamefont {Smerzi}},\ and\ \bibinfo {author} {\bibfnamefont {M.~K.}\
  \bibnamefont {Oberthaler}},\ }\bibfield  {title} {\bibinfo {title} {Fisher
  information and entanglement of non-gaussian spin states},\ }\href
  {https://doi.org/10.1126/science.1250147} {\bibfield  {journal} {\bibinfo
  {journal} {Science}\ }\textbf {\bibinfo {volume} {345}},\ \bibinfo {pages}
  {424} (\bibinfo {year} {2014})}\BibitemShut {NoStop}%
\bibitem [{\citenamefont {Muessel}\ \emph {et~al.}(2015)\citenamefont
  {Muessel}, \citenamefont {Strobel}, \citenamefont {Linnemann}, \citenamefont
  {Zibold}, \citenamefont {Juli\'a-D\'{\i}az},\ and\ \citenamefont
  {Oberthaler}}]{Muessel2015}%
  \BibitemOpen
  \bibfield  {author} {\bibinfo {author} {\bibfnamefont {W.}~\bibnamefont
  {Muessel}}, \bibinfo {author} {\bibfnamefont {H.}~\bibnamefont {Strobel}},
  \bibinfo {author} {\bibfnamefont {D.}~\bibnamefont {Linnemann}}, \bibinfo
  {author} {\bibfnamefont {T.}~\bibnamefont {Zibold}}, \bibinfo {author}
  {\bibfnamefont {B.}~\bibnamefont {Juli\'a-D\'{\i}az}},\ and\ \bibinfo
  {author} {\bibfnamefont {M.~K.}\ \bibnamefont {Oberthaler}},\ }\bibfield
  {title} {\bibinfo {title} {Twist-and-turn spin squeezing in bose-einstein
  condensates},\ }\href {https://doi.org/10.1103/PhysRevA.92.023603} {\bibfield
   {journal} {\bibinfo  {journal} {Phys. Rev. A}\ }\textbf {\bibinfo {volume}
  {92}},\ \bibinfo {pages} {023603} (\bibinfo {year} {2015})}\BibitemShut
  {NoStop}%
\bibitem [{\citenamefont {Mirkhalaf}\ \emph {et~al.}(2018)\citenamefont
  {Mirkhalaf}, \citenamefont {Nolan},\ and\ \citenamefont
  {Haine}}]{Mirkhalaf2018}%
  \BibitemOpen
  \bibfield  {author} {\bibinfo {author} {\bibfnamefont {S.~S.}\ \bibnamefont
  {Mirkhalaf}}, \bibinfo {author} {\bibfnamefont {S.~P.}\ \bibnamefont
  {Nolan}},\ and\ \bibinfo {author} {\bibfnamefont {S.~A.}\ \bibnamefont
  {Haine}},\ }\bibfield  {title} {\bibinfo {title} {Robustifying twist-and-turn
  entanglement with interaction-based readout},\ }\href
  {https://doi.org/10.1103/PhysRevA.97.053618} {\bibfield  {journal} {\bibinfo
  {journal} {Phys. Rev. A}\ }\textbf {\bibinfo {volume} {97}},\ \bibinfo
  {pages} {053618} (\bibinfo {year} {2018})}\BibitemShut {NoStop}%
\bibitem [{\citenamefont {Braverman}\ \emph {et~al.}(2019)\citenamefont
  {Braverman}, \citenamefont {Kawasaki}, \citenamefont {Pedrozo-Pe\~nafiel},
  \citenamefont {Colombo}, \citenamefont {Shu}, \citenamefont {Li},
  \citenamefont {Mendez}, \citenamefont {Yamoah}, \citenamefont {Salvi},
  \citenamefont {Akamatsu}, \citenamefont {Xiao},\ and\ \citenamefont
  {Vuleti\ifmmode~\acute{c}\else \'{c}\fi{}}}]{Braverman2019}%
  \BibitemOpen
  \bibfield  {author} {\bibinfo {author} {\bibfnamefont {B.}~\bibnamefont
  {Braverman}}, \bibinfo {author} {\bibfnamefont {A.}~\bibnamefont {Kawasaki}},
  \bibinfo {author} {\bibfnamefont {E.}~\bibnamefont {Pedrozo-Pe\~nafiel}},
  \bibinfo {author} {\bibfnamefont {S.}~\bibnamefont {Colombo}}, \bibinfo
  {author} {\bibfnamefont {C.}~\bibnamefont {Shu}}, \bibinfo {author}
  {\bibfnamefont {Z.}~\bibnamefont {Li}}, \bibinfo {author} {\bibfnamefont
  {E.}~\bibnamefont {Mendez}}, \bibinfo {author} {\bibfnamefont
  {M.}~\bibnamefont {Yamoah}}, \bibinfo {author} {\bibfnamefont
  {L.}~\bibnamefont {Salvi}}, \bibinfo {author} {\bibfnamefont
  {D.}~\bibnamefont {Akamatsu}}, \bibinfo {author} {\bibfnamefont
  {Y.}~\bibnamefont {Xiao}},\ and\ \bibinfo {author} {\bibfnamefont
  {V.}~\bibnamefont {Vuleti\ifmmode~\acute{c}\else \'{c}\fi{}}},\ }\bibfield
  {title} {\bibinfo {title} {Near-unitary spin squeezing in
  $^{171}\mathrm{Yb}$},\ }\href
  {https://doi.org/10.1103/PhysRevLett.122.223203} {\bibfield  {journal}
  {\bibinfo  {journal} {Phys. Rev. Lett.}\ }\textbf {\bibinfo {volume} {122}},\
  \bibinfo {pages} {223203} (\bibinfo {year} {2019})}\BibitemShut {NoStop}%
\bibitem [{\citenamefont {Li}\ \emph {et~al.}(2022)\citenamefont {Li},
  \citenamefont {Braverman}, \citenamefont {Colombo}, \citenamefont {Shu},
  \citenamefont {Kawasaki}, \citenamefont {Adiyatullin}, \citenamefont
  {Pedrozo-Pe\~nafiel}, \citenamefont {Mendez},\ and\ \citenamefont
  {Vuleti\ifmmode~\acute{c}\else \'{c}\fi{}}}]{Li2022}%
  \BibitemOpen
  \bibfield  {author} {\bibinfo {author} {\bibfnamefont {Z.}~\bibnamefont
  {Li}}, \bibinfo {author} {\bibfnamefont {B.}~\bibnamefont {Braverman}},
  \bibinfo {author} {\bibfnamefont {S.}~\bibnamefont {Colombo}}, \bibinfo
  {author} {\bibfnamefont {C.}~\bibnamefont {Shu}}, \bibinfo {author}
  {\bibfnamefont {A.}~\bibnamefont {Kawasaki}}, \bibinfo {author}
  {\bibfnamefont {A.~F.}\ \bibnamefont {Adiyatullin}}, \bibinfo {author}
  {\bibfnamefont {E.}~\bibnamefont {Pedrozo-Pe\~nafiel}}, \bibinfo {author}
  {\bibfnamefont {E.}~\bibnamefont {Mendez}},\ and\ \bibinfo {author}
  {\bibfnamefont {V.}~\bibnamefont {Vuleti\ifmmode~\acute{c}\else
  \'{c}\fi{}}},\ }\bibfield  {title} {\bibinfo {title} {Collective spin-light
  and light-mediated spin-spin interactions in an optical cavity},\ }\href
  {https://doi.org/10.1103/PRXQuantum.3.020308} {\bibfield  {journal} {\bibinfo
   {journal} {PRX Quantum}\ }\textbf {\bibinfo {volume} {3}},\ \bibinfo {pages}
  {020308} (\bibinfo {year} {2022})}\BibitemShut {NoStop}%
\bibitem [{\citenamefont {Lipkin}\ \emph {et~al.}(1965)\citenamefont {Lipkin},
  \citenamefont {Meshkov},\ and\ \citenamefont {Glick}}]{Lipkin1965}%
  \BibitemOpen
  \bibfield  {author} {\bibinfo {author} {\bibfnamefont {H.}~\bibnamefont
  {Lipkin}}, \bibinfo {author} {\bibfnamefont {N.}~\bibnamefont {Meshkov}},\
  and\ \bibinfo {author} {\bibfnamefont {A.}~\bibnamefont {Glick}},\ }\bibfield
   {title} {\bibinfo {title} {{Validity of many-body approximation methods for
  a solvable model: (I). Exact solutions and perturbation theory}},\ }\href
  {https://doi.org/10.1016/0029-5582(65)90862-X} {\bibfield  {journal}
  {\bibinfo  {journal} {Nuclear Physics}\ }\textbf {\bibinfo {volume} {62}},\
  \bibinfo {pages} {188} (\bibinfo {year} {1965})}\BibitemShut {NoStop}%
\bibitem [{\citenamefont {Vidal}\ \emph {et~al.}(2004)\citenamefont {Vidal},
  \citenamefont {Palacios},\ and\ \citenamefont {Aslangul}}]{Vidal2004}%
  \BibitemOpen
  \bibfield  {author} {\bibinfo {author} {\bibfnamefont {J.}~\bibnamefont
  {Vidal}}, \bibinfo {author} {\bibfnamefont {G.}~\bibnamefont {Palacios}},\
  and\ \bibinfo {author} {\bibfnamefont {C.}~\bibnamefont {Aslangul}},\
  }\bibfield  {title} {\bibinfo {title} {Entanglement dynamics in the
  lipkin-meshkov-glick model},\ }\href
  {https://doi.org/10.1103/PhysRevA.70.062304} {\bibfield  {journal} {\bibinfo
  {journal} {Phys. Rev. A}\ }\textbf {\bibinfo {volume} {70}},\ \bibinfo
  {pages} {062304} (\bibinfo {year} {2004})}\BibitemShut {NoStop}%
\bibitem [{\citenamefont {Dusuel}\ and\ \citenamefont
  {Vidal}(2004)}]{Dusuel2004}%
  \BibitemOpen
  \bibfield  {author} {\bibinfo {author} {\bibfnamefont {S.}~\bibnamefont
  {Dusuel}}\ and\ \bibinfo {author} {\bibfnamefont {J.}~\bibnamefont {Vidal}},\
  }\bibfield  {title} {\bibinfo {title} {Finite-size scaling exponents of the
  lipkin-meshkov-glick model},\ }\href
  {https://doi.org/10.1103/PhysRevLett.93.237204} {\bibfield  {journal}
  {\bibinfo  {journal} {Phys. Rev. Lett.}\ }\textbf {\bibinfo {volume} {93}},\
  \bibinfo {pages} {237204} (\bibinfo {year} {2004})}\BibitemShut {NoStop}%
\bibitem [{\citenamefont {Latorre}\ \emph {et~al.}(2005)\citenamefont
  {Latorre}, \citenamefont {Or\'us}, \citenamefont {Rico},\ and\ \citenamefont
  {Vidal}}]{Latorre2005}%
  \BibitemOpen
  \bibfield  {author} {\bibinfo {author} {\bibfnamefont {J.~I.}\ \bibnamefont
  {Latorre}}, \bibinfo {author} {\bibfnamefont {R.}~\bibnamefont {Or\'us}},
  \bibinfo {author} {\bibfnamefont {E.}~\bibnamefont {Rico}},\ and\ \bibinfo
  {author} {\bibfnamefont {J.}~\bibnamefont {Vidal}},\ }\bibfield  {title}
  {\bibinfo {title} {Entanglement entropy in the lipkin-meshkov-glick model},\
  }\href {https://doi.org/10.1103/PhysRevA.71.064101} {\bibfield  {journal}
  {\bibinfo  {journal} {Phys. Rev. A}\ }\textbf {\bibinfo {volume} {71}},\
  \bibinfo {pages} {064101} (\bibinfo {year} {2005})}\BibitemShut {NoStop}%
\bibitem [{\citenamefont {Dusuel}\ and\ \citenamefont
  {Vidal}(2005)}]{Dusuel2005}%
  \BibitemOpen
  \bibfield  {author} {\bibinfo {author} {\bibfnamefont {S.}~\bibnamefont
  {Dusuel}}\ and\ \bibinfo {author} {\bibfnamefont {J.}~\bibnamefont {Vidal}},\
  }\bibfield  {title} {\bibinfo {title} {Continuous unitary transformations and
  finite-size scaling exponents in the lipkin-meshkov-glick model},\ }\href
  {https://doi.org/10.1103/PhysRevB.71.224420} {\bibfield  {journal} {\bibinfo
  {journal} {Phys. Rev. B}\ }\textbf {\bibinfo {volume} {71}},\ \bibinfo
  {pages} {224420} (\bibinfo {year} {2005})}\BibitemShut {NoStop}%
\bibitem [{\citenamefont {Heiss}\ \emph {et~al.}(2005)\citenamefont {Heiss},
  \citenamefont {Scholtz},\ and\ \citenamefont {Geyer}}]{Heiss2005}%
  \BibitemOpen
  \bibfield  {author} {\bibinfo {author} {\bibfnamefont {W.~D.}\ \bibnamefont
  {Heiss}}, \bibinfo {author} {\bibfnamefont {F.~G.}\ \bibnamefont {Scholtz}},\
  and\ \bibinfo {author} {\bibfnamefont {H.~B.}\ \bibnamefont {Geyer}},\
  }\bibfield  {title} {\bibinfo {title} {The large $n$ behaviour of the lipkin
  model and exceptional points},\ }\href
  {https://doi.org/10.1088/0305-4470/38/9/002} {\bibfield  {journal} {\bibinfo
  {journal} {Journal of Physics A: Mathematical and General}\ }\textbf
  {\bibinfo {volume} {38}},\ \bibinfo {pages} {1843} (\bibinfo {year}
  {2005})}\BibitemShut {NoStop}%
\bibitem [{\citenamefont {Guti\'errez-Ruiz}\ \emph {et~al.}(2021)\citenamefont
  {Guti\'errez-Ruiz}, \citenamefont {Gonzalez}, \citenamefont
  {Ch\'avez-Carlos}, \citenamefont {Hirsch},\ and\ \citenamefont
  {Vergara}}]{Gutierrez2021}%
  \BibitemOpen
  \bibfield  {author} {\bibinfo {author} {\bibfnamefont {D.}~\bibnamefont
  {Guti\'errez-Ruiz}}, \bibinfo {author} {\bibfnamefont {D.}~\bibnamefont
  {Gonzalez}}, \bibinfo {author} {\bibfnamefont {J.}~\bibnamefont
  {Ch\'avez-Carlos}}, \bibinfo {author} {\bibfnamefont {J.~G.}\ \bibnamefont
  {Hirsch}},\ and\ \bibinfo {author} {\bibfnamefont {J.~D.}\ \bibnamefont
  {Vergara}},\ }\bibfield  {title} {\bibinfo {title} {Quantum geometric tensor
  and quantum phase transitions in the lipkin-meshkov-glick model},\ }\href
  {https://doi.org/10.1103/PhysRevB.103.174104} {\bibfield  {journal} {\bibinfo
   {journal} {Phys. Rev. B}\ }\textbf {\bibinfo {volume} {103}},\ \bibinfo
  {pages} {174104} (\bibinfo {year} {2021})}\BibitemShut {NoStop}%
\bibitem [{Note3()}]{Note3}%
  \BibitemOpen
  \bibinfo {note} {This fact implies that opposite ends of the separatrix touch
  the points $(X,Y,Z) = (0,0,\pm 1)$. In other words, the state $|\psi \rangle
  = |J,J\rangle $ has energy equal to that of the separatrix line. This defines
  the dynamical critical point of the dynamical quantum phase transition of
  Hamiltonian in Eq.~(\ref {eqn:tat_hamil}). Interestingly, the metrological
  relevance of this dynamical quantum phase transition was investigated
  recently in Ref.~\cite {Guan2021}}\BibitemShut {NoStop}%
\bibitem [{\citenamefont {do~Carmo}(1976)}]{Do1976}%
  \BibitemOpen
  \bibfield  {author} {\bibinfo {author} {\bibfnamefont {M.~P.}\ \bibnamefont
  {do~Carmo}},\ }\href {https://books.google.com/books?id=1v0YAQAAIAAJ} {\emph
  {\bibinfo {title} {Differential Geometry of Curves and Surfaces}}}\ (\bibinfo
   {publisher} {Prentice-Hall},\ \bibinfo {year} {1976})\BibitemShut {NoStop}%
\bibitem [{\citenamefont {Pires}\ \emph {et~al.}(2016)\citenamefont {Pires},
  \citenamefont {Cianciaruso}, \citenamefont {C{\'e}leri}, \citenamefont
  {Adesso},\ and\ \citenamefont {Soares-Pinto}}]{Pires2016}%
  \BibitemOpen
  \bibfield  {author} {\bibinfo {author} {\bibfnamefont {D.~P.}\ \bibnamefont
  {Pires}}, \bibinfo {author} {\bibfnamefont {M.}~\bibnamefont {Cianciaruso}},
  \bibinfo {author} {\bibfnamefont {L.~C.}\ \bibnamefont {C{\'e}leri}},
  \bibinfo {author} {\bibfnamefont {G.}~\bibnamefont {Adesso}},\ and\ \bibinfo
  {author} {\bibfnamefont {D.~O.}\ \bibnamefont {Soares-Pinto}},\ }\bibfield
  {title} {\bibinfo {title} {Generalized geometric quantum speed limits},\
  }\href@noop {} {\bibfield  {journal} {\bibinfo  {journal} {Physical Review
  X}\ }\textbf {\bibinfo {volume} {6}},\ \bibinfo {pages} {021031} (\bibinfo
  {year} {2016})}\BibitemShut {NoStop}%
\bibitem [{\citenamefont {Deffner}\ and\ \citenamefont
  {Campbell}(2017)}]{Deffner2017}%
  \BibitemOpen
  \bibfield  {author} {\bibinfo {author} {\bibfnamefont {S.}~\bibnamefont
  {Deffner}}\ and\ \bibinfo {author} {\bibfnamefont {S.}~\bibnamefont
  {Campbell}},\ }\bibfield  {title} {\bibinfo {title} {Quantum speed limits:
  from heisenberg’s uncertainty principle to optimal quantum control},\
  }\href@noop {} {\bibfield  {journal} {\bibinfo  {journal} {Journal of Physics
  A: Mathematical and Theoretical}\ }\textbf {\bibinfo {volume} {50}},\
  \bibinfo {pages} {453001} (\bibinfo {year} {2017})}\BibitemShut {NoStop}%
\bibitem [{\citenamefont {Poggi}\ \emph {et~al.}(2021)\citenamefont {Poggi},
  \citenamefont {Campbell},\ and\ \citenamefont {Deffner}}]{Poggi2021}%
  \BibitemOpen
  \bibfield  {author} {\bibinfo {author} {\bibfnamefont {P.~M.}\ \bibnamefont
  {Poggi}}, \bibinfo {author} {\bibfnamefont {S.}~\bibnamefont {Campbell}},\
  and\ \bibinfo {author} {\bibfnamefont {S.}~\bibnamefont {Deffner}},\
  }\bibfield  {title} {\bibinfo {title} {Diverging quantum speed limits: A
  herald of classicality},\ }\href
  {https://doi.org/10.1103/PRXQuantum.2.040349} {\bibfield  {journal} {\bibinfo
   {journal} {PRX Quantum}\ }\textbf {\bibinfo {volume} {2}},\ \bibinfo {pages}
  {040349} (\bibinfo {year} {2021})}\BibitemShut {NoStop}%
\bibitem [{\citenamefont {Holstein}\ and\ \citenamefont
  {Primakoff}(1940)}]{Holstein1940}%
  \BibitemOpen
  \bibfield  {author} {\bibinfo {author} {\bibfnamefont {T.}~\bibnamefont
  {Holstein}}\ and\ \bibinfo {author} {\bibfnamefont {H.}~\bibnamefont
  {Primakoff}},\ }\bibfield  {title} {\bibinfo {title} {{Field Dependence of
  the Intrinsic Domain Magnetization of a Ferromagnet}},\ }\href
  {https://doi.org/10.1103/PhysRev.58.1098} {\bibfield  {journal} {\bibinfo
  {journal} {Physical Review}\ }\textbf {\bibinfo {volume} {58}},\ \bibinfo
  {pages} {1098} (\bibinfo {year} {1940})}\BibitemShut {NoStop}%
\bibitem [{\citenamefont {Trail}\ \emph {et~al.}(2010)\citenamefont {Trail},
  \citenamefont {Jessen},\ and\ \citenamefont {Deutsch}}]{Trail2010}%
  \BibitemOpen
  \bibfield  {author} {\bibinfo {author} {\bibfnamefont {C.~M.}\ \bibnamefont
  {Trail}}, \bibinfo {author} {\bibfnamefont {P.~S.}\ \bibnamefont {Jessen}},\
  and\ \bibinfo {author} {\bibfnamefont {I.~H.}\ \bibnamefont {Deutsch}},\
  }\bibfield  {title} {\bibinfo {title} {Strongly enhanced spin squeezing via
  quantum control},\ }\href {https://doi.org/10.1103/PhysRevLett.105.193602}
  {\bibfield  {journal} {\bibinfo  {journal} {Phys. Rev. Lett.}\ }\textbf
  {\bibinfo {volume} {105}},\ \bibinfo {pages} {193602} (\bibinfo {year}
  {2010})}\BibitemShut {NoStop}%
\bibitem [{\citenamefont {Wang}\ and\ \citenamefont
  {P\'erez-Bernal}(2019)}]{Wang2019}%
  \BibitemOpen
  \bibfield  {author} {\bibinfo {author} {\bibfnamefont {Q.}~\bibnamefont
  {Wang}}\ and\ \bibinfo {author} {\bibfnamefont {F.}~\bibnamefont
  {P\'erez-Bernal}},\ }\bibfield  {title} {\bibinfo {title} {Probing an
  excited-state quantum phase transition in a quantum many-body system via an
  out-of-time-order correlator},\ }\href
  {https://doi.org/10.1103/PhysRevA.100.062113} {\bibfield  {journal} {\bibinfo
   {journal} {Phys. Rev. A}\ }\textbf {\bibinfo {volume} {100}},\ \bibinfo
  {pages} {062113} (\bibinfo {year} {2019})}\BibitemShut {NoStop}%
\bibitem [{\citenamefont {Stránský}\ \emph {et~al.}(2014)\citenamefont
  {Stránský}, \citenamefont {Macek},\ and\ \citenamefont
  {Cejnar}}]{Stransky2014}%
  \BibitemOpen
  \bibfield  {author} {\bibinfo {author} {\bibfnamefont {P.}~\bibnamefont
  {Stránský}}, \bibinfo {author} {\bibfnamefont {M.}~\bibnamefont {Macek}},\
  and\ \bibinfo {author} {\bibfnamefont {P.}~\bibnamefont {Cejnar}},\
  }\bibfield  {title} {\bibinfo {title} {Excited-state quantum phase
  transitions in systems with two degrees of freedom: level density, level
  dynamics, thermal properties},\ }\href
  {https://doi.org/http://dx.doi.org/10.1016/j.aop.2014.03.006} {\bibfield
  {journal} {\bibinfo  {journal} {Ann. Phys. (N. Y.)}\ }\textbf {\bibinfo
  {volume} {345}},\ \bibinfo {pages} {73 } (\bibinfo {year}
  {2014})}\BibitemShut {NoStop}%
\bibitem [{\citenamefont {Cejnar}\ \emph {et~al.}(2021)\citenamefont {Cejnar},
  \citenamefont {Str{\'{a}}nsk{\'{y}}}, \citenamefont {Macek},\ and\
  \citenamefont {Kloc}}]{Cejnar2021}%
  \BibitemOpen
  \bibfield  {author} {\bibinfo {author} {\bibfnamefont {P.}~\bibnamefont
  {Cejnar}}, \bibinfo {author} {\bibfnamefont {P.}~\bibnamefont
  {Str{\'{a}}nsk{\'{y}}}}, \bibinfo {author} {\bibfnamefont {M.}~\bibnamefont
  {Macek}},\ and\ \bibinfo {author} {\bibfnamefont {M.}~\bibnamefont {Kloc}},\
  }\bibfield  {title} {\bibinfo {title} {Excited-state quantum phase
  transitions},\ }\href {https://doi.org/10.1088/1751-8121/abdfe8} {\bibfield
  {journal} {\bibinfo  {journal} {Journal of Physics A: Mathematical and
  Theoretical}\ }\textbf {\bibinfo {volume} {54}},\ \bibinfo {pages} {133001}
  (\bibinfo {year} {2021})}\BibitemShut {NoStop}%
\bibitem [{\citenamefont {Corps}\ and\ \citenamefont
  {Rela{\~n}o}(2022)}]{corps2022a}%
  \BibitemOpen
  \bibfield  {author} {\bibinfo {author} {\bibfnamefont {{\'A}.~L.}\
  \bibnamefont {Corps}}\ and\ \bibinfo {author} {\bibfnamefont
  {A.}~\bibnamefont {Rela{\~n}o}},\ }\bibfield  {title} {\bibinfo {title}
  {Theory of dynamical phase transitions in collective quantum systems},\
  }\href@noop {} {\bibfield  {journal} {\bibinfo  {journal} {arXiv preprint
  arXiv:2205.03443}\ } (\bibinfo {year} {2022})}\BibitemShut {NoStop}%
\bibitem [{\citenamefont {Corps}\ and\ \citenamefont
  {Rela\~no}(2022)}]{Corps2022b}%
  \BibitemOpen
  \bibfield  {author} {\bibinfo {author} {\bibfnamefont {A.~L.}\ \bibnamefont
  {Corps}}\ and\ \bibinfo {author} {\bibfnamefont {A.}~\bibnamefont
  {Rela\~no}},\ }\bibfield  {title} {\bibinfo {title} {Dynamical and
  excited-state quantum phase transitions in collective systems},\ }\href
  {https://doi.org/10.1103/PhysRevB.106.024311} {\bibfield  {journal} {\bibinfo
   {journal} {Phys. Rev. B}\ }\textbf {\bibinfo {volume} {106}},\ \bibinfo
  {pages} {024311} (\bibinfo {year} {2022})}\BibitemShut {NoStop}%
\bibitem [{\citenamefont {{Zhou}}\ \emph {et~al.}(2022)\citenamefont {{Zhou}},
  \citenamefont {{Kong}}, \citenamefont {{Lan}},\ and\ \citenamefont
  {{Zhang}}}]{Zhou2022}%
  \BibitemOpen
  \bibfield  {author} {\bibinfo {author} {\bibfnamefont {L.}~\bibnamefont
  {{Zhou}}}, \bibinfo {author} {\bibfnamefont {J.}~\bibnamefont {{Kong}}},
  \bibinfo {author} {\bibfnamefont {Z.}~\bibnamefont {{Lan}}},\ and\ \bibinfo
  {author} {\bibfnamefont {W.}~\bibnamefont {{Zhang}}},\ }\bibfield  {title}
  {\bibinfo {title} {{Dynamical quantum phase transitions in a spinor
  Bose-Einstein condensate and criticality enhanced quantum sensing}},\
  }\href@noop {} {\bibfield  {journal} {\bibinfo  {journal} {arXiv preprints
  arXiv:2209.11415}\ } (\bibinfo {year} {2022})}\BibitemShut {NoStop}%
\bibitem [{\citenamefont {J{\"{o}}rg}\ \emph {et~al.}(2010)\citenamefont
  {J{\"{o}}rg}, \citenamefont {Krzakala}, \citenamefont {Kurchan},
  \citenamefont {Maggs},\ and\ \citenamefont {Pujos}}]{Jorg2010}%
  \BibitemOpen
  \bibfield  {author} {\bibinfo {author} {\bibfnamefont {T.}~\bibnamefont
  {J{\"{o}}rg}}, \bibinfo {author} {\bibfnamefont {F.}~\bibnamefont
  {Krzakala}}, \bibinfo {author} {\bibfnamefont {J.}~\bibnamefont {Kurchan}},
  \bibinfo {author} {\bibfnamefont {A.~C.}\ \bibnamefont {Maggs}},\ and\
  \bibinfo {author} {\bibfnamefont {J.}~\bibnamefont {Pujos}},\ }\bibfield
  {title} {\bibinfo {title} {{Energy gaps in quantum first-order
  mean-field–like transitions: The problems that quantum annealing cannot
  solve}},\ }\href {https://doi.org/10.1209/0295-5075/89/40004} {\bibfield
  {journal} {\bibinfo  {journal} {EPL (Europhysics Letters)}\ }\textbf
  {\bibinfo {volume} {89}},\ \bibinfo {pages} {40004} (\bibinfo {year}
  {2010})}\BibitemShut {NoStop}%
\bibitem [{\citenamefont {Bapst}\ and\ \citenamefont
  {Semerjian}(2012)}]{Bapst2012}%
  \BibitemOpen
  \bibfield  {author} {\bibinfo {author} {\bibfnamefont {V.}~\bibnamefont
  {Bapst}}\ and\ \bibinfo {author} {\bibfnamefont {G.}~\bibnamefont
  {Semerjian}},\ }\bibfield  {title} {\bibinfo {title} {On quantum mean-field
  models and their quantum annealing},\ }\href
  {https://doi.org/10.1088/1742-5468/2012/06/p06007} {\bibfield  {journal}
  {\bibinfo  {journal} {Journal of Statistical Mechanics: Theory and
  Experiment}\ }\textbf {\bibinfo {volume} {2012}},\ \bibinfo {pages} {P06007}
  (\bibinfo {year} {2012})}\BibitemShut {NoStop}%
\bibitem [{\citenamefont {Matsuura}\ \emph {et~al.}(2017)\citenamefont
  {Matsuura}, \citenamefont {Nishimori}, \citenamefont {Vinci}, \citenamefont
  {Albash},\ and\ \citenamefont {Lidar}}]{Matsuura2017}%
  \BibitemOpen
  \bibfield  {author} {\bibinfo {author} {\bibfnamefont {S.}~\bibnamefont
  {Matsuura}}, \bibinfo {author} {\bibfnamefont {H.}~\bibnamefont {Nishimori}},
  \bibinfo {author} {\bibfnamefont {W.}~\bibnamefont {Vinci}}, \bibinfo
  {author} {\bibfnamefont {T.}~\bibnamefont {Albash}},\ and\ \bibinfo {author}
  {\bibfnamefont {D.~A.}\ \bibnamefont {Lidar}},\ }\bibfield  {title} {\bibinfo
  {title} {{Quantum-annealing correction at finite temperature: Ferromagnetic p
  -spin models}},\ }\href {https://doi.org/10.1103/PhysRevA.95.022308}
  {\bibfield  {journal} {\bibinfo  {journal} {Physical Review A}\ }\textbf
  {\bibinfo {volume} {95}},\ \bibinfo {pages} {022308} (\bibinfo {year}
  {2017})}\BibitemShut {NoStop}%
\bibitem [{\citenamefont {Filippone}\ \emph {et~al.}(2011)\citenamefont
  {Filippone}, \citenamefont {Dusuel},\ and\ \citenamefont
  {Vidal}}]{Filippone2011}%
  \BibitemOpen
  \bibfield  {author} {\bibinfo {author} {\bibfnamefont {M.}~\bibnamefont
  {Filippone}}, \bibinfo {author} {\bibfnamefont {S.}~\bibnamefont {Dusuel}},\
  and\ \bibinfo {author} {\bibfnamefont {J.}~\bibnamefont {Vidal}},\ }\bibfield
   {title} {\bibinfo {title} {Quantum phase transitions in fully connected spin
  models: An entanglement perspective},\ }\href
  {https://doi.org/10.1103/PhysRevA.83.022327} {\bibfield  {journal} {\bibinfo
  {journal} {Phys. Rev. A}\ }\textbf {\bibinfo {volume} {83}},\ \bibinfo
  {pages} {022327} (\bibinfo {year} {2011})}\BibitemShut {NoStop}%
\bibitem [{\citenamefont {Del~Re}\ \emph {et~al.}(2016)\citenamefont {Del~Re},
  \citenamefont {Fabrizio},\ and\ \citenamefont {Tosatti}}]{DelRe2016}%
  \BibitemOpen
  \bibfield  {author} {\bibinfo {author} {\bibfnamefont {L.}~\bibnamefont
  {Del~Re}}, \bibinfo {author} {\bibfnamefont {M.}~\bibnamefont {Fabrizio}},\
  and\ \bibinfo {author} {\bibfnamefont {E.}~\bibnamefont {Tosatti}},\
  }\bibfield  {title} {\bibinfo {title} {Nonequilibrium and nonhomogeneous
  phenomena around a first-order quantum phase transition},\ }\href
  {https://doi.org/10.1103/PhysRevB.93.125131} {\bibfield  {journal} {\bibinfo
  {journal} {Phys. Rev. B}\ }\textbf {\bibinfo {volume} {93}},\ \bibinfo
  {pages} {125131} (\bibinfo {year} {2016})}\BibitemShut {NoStop}%
\bibitem [{\citenamefont {Mu\~noz Arias}\ \emph {et~al.}(2020)\citenamefont
  {Mu\~noz Arias}, \citenamefont {Deutsch}, \citenamefont {Jessen},\ and\
  \citenamefont {Poggi}}]{Munoz-Arias2020}%
  \BibitemOpen
  \bibfield  {author} {\bibinfo {author} {\bibfnamefont {M.~H.}\ \bibnamefont
  {Mu\~noz Arias}}, \bibinfo {author} {\bibfnamefont {I.~H.}\ \bibnamefont
  {Deutsch}}, \bibinfo {author} {\bibfnamefont {P.~S.}\ \bibnamefont
  {Jessen}},\ and\ \bibinfo {author} {\bibfnamefont {P.~M.}\ \bibnamefont
  {Poggi}},\ }\bibfield  {title} {\bibinfo {title} {Simulation of the complex
  dynamics of mean-field $p$-spin models using measurement-based quantum
  feedback control},\ }\href {https://doi.org/10.1103/PhysRevA.102.022610}
  {\bibfield  {journal} {\bibinfo  {journal} {Phys. Rev. A}\ }\textbf {\bibinfo
  {volume} {102}},\ \bibinfo {pages} {022610} (\bibinfo {year}
  {2020})}\BibitemShut {NoStop}%
\bibitem [{\citenamefont {Correale}\ and\ \citenamefont
  {Silva}(2021)}]{Correale2021}%
  \BibitemOpen
  \bibfield  {author} {\bibinfo {author} {\bibfnamefont {L.}~\bibnamefont
  {Correale}}\ and\ \bibinfo {author} {\bibfnamefont {A.}~\bibnamefont
  {Silva}},\ }\bibfield  {title} {\bibinfo {title} {Changing the order of a
  dynamical phase transition through fluctuations in a quantum p-spin model},\
  }\href@noop {} {\bibfield  {journal} {\bibinfo  {journal} {arXiv preprint
  arXiv:2110.13524}\ } (\bibinfo {year} {2021})}\BibitemShut {NoStop}%
\bibitem [{Note4()}]{Note4}%
  \BibitemOpen
  \bibinfo {note} {This is a central point in the phase space approach
  presented in this work. Approximating the time scales to certain
  metrologically useful states, via the motion of points along separatrix
  branches, requires this motion to take place symmetrically with respect to
  the saddle point.}\BibitemShut {Stop}%
\bibitem [{\citenamefont {Mu\~noz Arias}\ \emph {et~al.}(2022)\citenamefont
  {Mu\~noz Arias}, \citenamefont {Chinni},\ and\ \citenamefont
  {Poggi}}]{Munoz2022}%
  \BibitemOpen
  \bibfield  {author} {\bibinfo {author} {\bibfnamefont {M.~H.}\ \bibnamefont
  {Mu\~noz Arias}}, \bibinfo {author} {\bibfnamefont {K.}~\bibnamefont
  {Chinni}},\ and\ \bibinfo {author} {\bibfnamefont {P.~M.}\ \bibnamefont
  {Poggi}},\ }\bibfield  {title} {\bibinfo {title} {Floquet time crystals in
  driven spin systems with all-to-all $p$-body interactions},\ }\href
  {https://doi.org/10.1103/PhysRevResearch.4.023018} {\bibfield  {journal}
  {\bibinfo  {journal} {Phys. Rev. Research}\ }\textbf {\bibinfo {volume}
  {4}},\ \bibinfo {pages} {023018} (\bibinfo {year} {2022})}\BibitemShut
  {NoStop}%
\bibitem [{Note5()}]{Note5}%
  \BibitemOpen
  \bibinfo {note} {One can define this neighborhood by taking a small energy
  window centered at the separatrix energy.}\BibitemShut {Stop}%
\bibitem [{\citenamefont {Xu}\ \emph {et~al.}(2020)\citenamefont {Xu},
  \citenamefont {Scaffidi},\ and\ \citenamefont {Cao}}]{Xu2020}%
  \BibitemOpen
  \bibfield  {author} {\bibinfo {author} {\bibfnamefont {T.}~\bibnamefont
  {Xu}}, \bibinfo {author} {\bibfnamefont {T.}~\bibnamefont {Scaffidi}},\ and\
  \bibinfo {author} {\bibfnamefont {X.}~\bibnamefont {Cao}},\ }\bibfield
  {title} {\bibinfo {title} {Does scrambling equal chaos?},\ }\href
  {https://doi.org/10.1103/PhysRevLett.124.140602} {\bibfield  {journal}
  {\bibinfo  {journal} {Phys. Rev. Lett.}\ }\textbf {\bibinfo {volume} {124}},\
  \bibinfo {pages} {140602} (\bibinfo {year} {2020})}\BibitemShut {NoStop}%
\bibitem [{\citenamefont {Kidd}\ \emph {et~al.}(2021)\citenamefont {Kidd},
  \citenamefont {Safavi-Naini},\ and\ \citenamefont {Corney}}]{Kidd2021}%
  \BibitemOpen
  \bibfield  {author} {\bibinfo {author} {\bibfnamefont {R.~A.}\ \bibnamefont
  {Kidd}}, \bibinfo {author} {\bibfnamefont {A.}~\bibnamefont {Safavi-Naini}},\
  and\ \bibinfo {author} {\bibfnamefont {J.~F.}\ \bibnamefont {Corney}},\
  }\bibfield  {title} {\bibinfo {title} {Saddle-point scrambling without
  thermalization},\ }\href {https://doi.org/10.1103/PhysRevA.103.033304}
  {\bibfield  {journal} {\bibinfo  {journal} {Phys. Rev. A}\ }\textbf {\bibinfo
  {volume} {103}},\ \bibinfo {pages} {033304} (\bibinfo {year}
  {2021})}\BibitemShut {NoStop}%
\bibitem [{\citenamefont {Koppenh{\"o}fer}\ \emph {et~al.}(2021)\citenamefont
  {Koppenh{\"o}fer}, \citenamefont {Groszkowski}, \citenamefont {Lau},\ and\
  \citenamefont {Clerk}}]{Koppenhofer2021}%
  \BibitemOpen
  \bibfield  {author} {\bibinfo {author} {\bibfnamefont {M.}~\bibnamefont
  {Koppenh{\"o}fer}}, \bibinfo {author} {\bibfnamefont {P.}~\bibnamefont
  {Groszkowski}}, \bibinfo {author} {\bibfnamefont {H.-K.}\ \bibnamefont
  {Lau}},\ and\ \bibinfo {author} {\bibfnamefont {A.~A.}\ \bibnamefont
  {Clerk}},\ }\bibfield  {title} {\bibinfo {title} {Dissipative superradiant
  spin amplifier for enhanced quantum sensing},\ }\href@noop {} {\bibfield
  {journal} {\bibinfo  {journal} {arXiv preprint arXiv:2111.15647}\ } (\bibinfo
  {year} {2021})}\BibitemShut {NoStop}%
\bibitem [{\citenamefont {Grobe}\ and\ \citenamefont
  {Haake}(1987)}]{Grobe1987}%
  \BibitemOpen
  \bibfield  {author} {\bibinfo {author} {\bibfnamefont {R.}~\bibnamefont
  {Grobe}}\ and\ \bibinfo {author} {\bibfnamefont {F.}~\bibnamefont {Haake}},\
  }\bibfield  {title} {\bibinfo {title} {Dissipative death of quantum
  coherences in a spin system},\ }\href {https://doi.org/10.1007/BF01471081}
  {\bibfield  {journal} {\bibinfo  {journal} {Zeitschrift für Physik B -
  Condensed Matter}\ }\textbf {\bibinfo {volume} {68}},\ \bibinfo {pages}
  {503–512} (\bibinfo {year} {1987})}\BibitemShut {NoStop}%
\bibitem [{\citenamefont {Guan}\ and\ \citenamefont
  {Lewis-Swan}(2021)}]{Guan2021}%
  \BibitemOpen
  \bibfield  {author} {\bibinfo {author} {\bibfnamefont {Q.}~\bibnamefont
  {Guan}}\ and\ \bibinfo {author} {\bibfnamefont {R.~J.}\ \bibnamefont
  {Lewis-Swan}},\ }\bibfield  {title} {\bibinfo {title} {Identifying and
  harnessing dynamical phase transitions for quantum-enhanced sensing},\ }\href
  {https://doi.org/10.1103/PhysRevResearch.3.033199} {\bibfield  {journal}
  {\bibinfo  {journal} {Phys. Rev. Research}\ }\textbf {\bibinfo {volume}
  {3}},\ \bibinfo {pages} {033199} (\bibinfo {year} {2021})}\BibitemShut
  {NoStop}%
\end{thebibliography}%
\end{document}